\documentclass[11pt]{article}
\pdfoutput=1
\usepackage{jcapmod}
\usepackage{shorthand}

\usepackage{mathtools}
\usepackage{booktabs}
\usepackage[english]{babel}
\usepackage{amsmath,amssymb,amsbsy,amstext, amsthm, simplewick, amsfonts}
\usepackage{hyperref}
\usepackage{graphicx}
\usepackage[small]{caption}
\usepackage{siunitx}
\usepackage{upgreek}
\usepackage{framed}
\usepackage{wrapfig}
\usepackage{multirow}
\usepackage{bbm}
\usepackage[svgnames,dvipsnames,x11names]{xcolor}
\usepackage[utf8x]{inputenc}
\usepackage{selinput}
\usepackage{bm}
\usepackage{float}
\usepackage{geometry}
\usepackage{yfonts}
\usepackage{caption}
\usepackage{subcaption}
\usepackage{sidecap}
\usepackage{longtable}
\usepackage{anyfontsize}

\usepackage{genyoungtabtikz}
\Yboxdim{11pt} 
\Ylinethick{.75pt}
\newcommand*{\medotimes}{\raisebox{-0.3ex}{\scalebox{1.25}{$\otimes$}}}
\newcommand*{\medoplus}{\raisebox{-0.3ex}{\scalebox{1.25}{$\oplus$}}}

\usepackage[framemethod=default]{mdframed}
\newmdenv[skipabove=7pt,
skipbelow=7pt,
rightline=false,
leftline=false,
topline=false,
bottomline=false,
backgroundcolor=gray!10,
linecolor=gray,
innerleftmargin=5pt,
innerrightmargin=5pt,
innertopmargin=5pt,
innerbottommargin=5pt,
leftmargin=0cm,
rightmargin=0cm,
linewidth=4pt]{eBox}
\newmdenv[skipabove=7pt,
skipbelow=7pt,
rightline=false,
leftline=false,
topline=false,
bottomline=false,
backgroundcolor=gray!10,
linecolor=gray,
innerleftmargin=5pt,
innerrightmargin=5pt,
innertopmargin=-5pt,
innerbottommargin=5pt,
leftmargin=0cm,
rightmargin=0cm,
linewidth=4pt]{eBox2}

\definecolor{blue3}{RGB}{31, 119, 180}
\definecolor{red3}{RGB}{	214, 39, 40}
\definecolor{orange3}{RGB}{255, 127, 14}
\definecolor{green3}{RGB}{44, 160, 44}

\usepackage{colortbl}
\definecolor{lightgreen}{cmyk}{0.2, 0, 0.2, 0.2}
\definecolor{lightgray}{cmyk}{0.1,0.2,0,0.1}
\definecolor{lightgray2}{cmyk}{0.1,0.1,0,0.1}

\makeatletter
\newlength{\apb@width}
\newcommand{\autoparbox}[2][c]{\settowidth{\apb@width}{#2}\parbox[#1]{\apb@width}{#2}}
\newcommand{\includegraphicsbox}[2][]{\autoparbox{\includegraphics[#1]{#2}}}
\makeatother


\def\hs{\hskip 1pt}

\def\k{{\vec k}}
\def\x{{\vec x}}
\def\z{{\vec z}}

\def\F{{\hat F}}
\def\a{{\hat \alpha}}
\def\b{{\hat \beta}}
\def\t{{\hat \tau}}

\def\kza{(\vec k_1\cdot \vec z_1)}
\def\kzb{(\vec k_2\cdot \vec z_2)}

\def\kdza{(\vec k_1\cdot \partial_{\vec z_1})}

\def\ddza{\nabla^2_{\vec z_1}}
\def\ddzb{\nabla^2_{\vec z_2}}

\def\beq{\begin{equation}}
\def\eeq{\end{equation}}

\allowdisplaybreaks[1]
\begin{document}



\begin{titlepage}
\setcounter{page}{1} \baselineskip=15.5pt 
\thispagestyle{empty}

\begin{center}
{\fontsize{20}{18} \bf The Cosmological Bootstrap:}\\[14pt]
{\fontsize{15.3}{18} \bf  Weight-Shifting Operators and Scalar Seeds}
\end{center}

\vskip 20pt
\begin{center}
\noindent
{\fontsize{12}{18}\selectfont Daniel Baumann,$^1$ Carlos Duaso Pueyo,$^{1}$
 Austin Joyce,$^{2}$\\[4pt] Hayden Lee,$^{3}$ and Guilherme L.~Pimentel$^{1,4}$}
\end{center}

\begin{center}
  \vskip8pt
\textit{$^1$ Institute of Physics, University of Amsterdam, Amsterdam, 1098 XH, The Netherlands}

  \vskip8pt
\textit{$^2$ Department of Physics, Columbia University, New York, NY 10027, USA}

\vskip 8pt
\textit{$^3$  Department of Physics, Harvard University, Cambridge, MA 02138, USA}

\vskip 8pt
\textit{$^4$ Lorentz Institute for Theoretical Physics, Leiden, 2333 CA, The Netherlands}
\end{center}

\vspace{0.4cm}
 \begin{center}{\bf Abstract}
 \end{center}

\noindent
A key insight of the bootstrap approach to cosmological correlations is the fact that all correlators of slow-roll inflation can be reduced to a unique building block---the four-point function of conformally coupled scalars, arising from the exchange of a massive scalar. Correlators corresponding to the exchange of particles with spin are then obtained by applying a spin-raising operator to the scalar-exchange solution. Similarly, the correlators of massless external fields can be derived by acting with a suitable weight-raising operator. In this paper, we present a systematic and highly streamlined derivation of these operators (and their generalizations) using tools of conformal field theory.  Our results greatly simplify the theoretical foundations of the cosmological bootstrap program.

\end{titlepage}

\restoregeometry

\newpage
\setcounter{tocdepth}{2}
\setcounter{page}{2}
\tableofcontents

\newpage
\section{Introduction}

It is remarkable how much physics can sometimes be derived from just a few basic principles.  
For example, the form of all consistent particle interactions is a nearly inevitable consequence of the twin pillars of quantum mechanics and special relativity. 
This is made manifest in the ``S-matrix bootstrap," where simple self-consistency requirements (such as Lorentz symmetry, locality and causality) completely fix the analytic structure of tree-level scattering amplitudes. Recently, a similar bootstrap philosophy was applied to 
cosmological correlation functions~\cite{Arkani-Hamed:2018kmz} (see also~\cite{Arkani-Hamed:2017fdk, Arkani-Hamed:2018bjr, Benincasa:2018ssx, Sleight:2019mgd, Sleight:2019hfp, Benincasa:2019vqr}). Working under the lamppost of slow-roll inflation with weak couplings to extra massive particles, this ``cosmological bootstrap" allowed for a complete classification of all scalar three- and four-point functions at tree level.

\vskip 4pt
The standard approach to computing inflationary correlation functions involves following the time-evolution of fields during inflation. From this viewpoint, locality is a fundamental input, and the interactions between particles lead to complicated time integrals that encode the manifestly local time evolution. The outputs of this procedure are late-time correlation functions that live on the (spacelike) future boundary of de Sitter space. These correlation functions are encoded in the statistics of late-time cosmological observables, and so form the fundamental observable output of inflation. The bootstrap philosophy is to focus on these final observable quantities and construct them directly, granting primacy to principles aside from manifest locality. In particular, the cosmological bootstrap exploits the approximate de Sitter symmetries---which act as conformal transformations on the boundary~\cite{Maldacena:2011nz,Antoniadis:2011ib,Creminelli:2011mw, Kehagias:2012pd, Arkani-Hamed:2015bza}---along with consistency requirements on the singularity structure of correlation functions to reconstruct the output of bulk time evolution without ever talking about time. Correlation functions instead arise as solutions (with particular singularities) of conformal Ward identities~\cite{Arkani-Hamed:2018kmz}. (See \cite{Raju:2012zr, Raju:2012zs, Mata:2012bx, Ghosh:2014kba, Kundu:2014gxa, Kundu:2015xta, Coriano:2013jba, Coriano:2018bbe, Maglio:2019grh, Bzowski:2013sza,Anninos:2014lwa, Bzowski:2015pba, Bzowski:2015yxv, Bzowski:2017poo, Isono:2018rrb, Isono:2019ihz, Isono:2019wex, Albayrak:2018tam, Albayrak:2019asr, Albayrak:2019yve, Bzowski:2019kwd, Pajer:2016ieg, Farrow:2018yni} for other studies on conformal correlators in momentum space.) Although these solutions describe static boundary correlators, they encode time-dependent processes in the bulk, including the production and decay of very massive particles~\cite{Chen:2009zp, Baumann:2011nk, Assassi:2012zq, Chen:2012ge, Pi:2012gf, Noumi:2012vr, Baumann:2012bc, Assassi:2013gxa, Gong:2013sma, Arkani-Hamed:2015bza, Lee:2016vti, Kehagias:2017cym, Kumar:2017ecc, An:2017hlx, An:2017rwo, Baumann:2017jvh, Goon:2018fyu, Kumar:2018jxz, Liu:2019fag, Kumar:2019ebj, Alexander:2019vtb, Kim:2019wjo}.

\begin{figure}[h!]
\centering
      \includegraphics{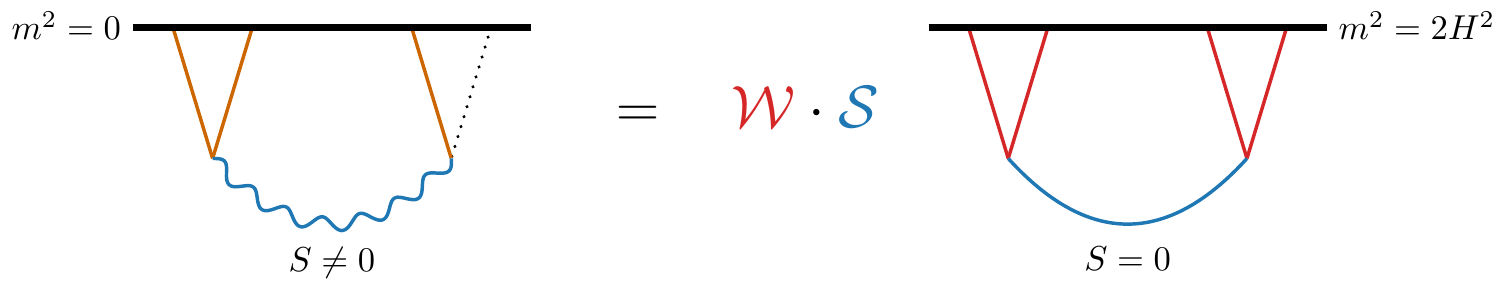}
           \caption{Schematic illustration of the spin-raising and weight-shifting procedure studied in this paper.} 
    \label{fig:schematic}
\end{figure}

\vskip 4pt
One of the main insights of the bootstrap approach is the fact that all correlators can be reduced to a unique building block---the de Sitter four-point function of {\it conformally coupled} scalars, mediated by the exchange of a massive scalar.  Solutions corresponding to the exchange of particles with spin are obtained by applying a spin-raising operator, ${\cal S}$, to the scalar-exchange solution. Similarly, solutions for {\it massless} external fields are derived by acting with a set of weight-raising operators, ${\cal W}$. Finally, the de Sitter four-point functions lead to inflationary three-point functions when one of the external legs is evaluated on the time-dependent background (see Fig.~\ref{fig:schematic}). 
The derivation of these spin-raising and weight-shifting operators in~\cite{Arkani-Hamed:2018kmz} was somewhat unsatisfactory, involving a combination of bulk reasoning, boundary considerations, and educated guesswork.  In this paper, we will present a more systematic derivation of these operators (and their generalizations) using tools of conformal field theory (CFT). This has the advantage of being purely intrinsic to the boundary, along with placing these operators in a broader context, opening up new avenues to the cosmological bootstrap~\cite{CosmoBoot2}. 

\vskip 4pt
Given that late-time correlation functions in de Sitter space are conformally invariant, it is natural 
 to try to connect the study of cosmological correlators to CFT techniques.
However, most of our understanding of CFTs has been developed in position space, while the natural habitat of cosmological correlators is Fourier space.
If standard CFT results could be translated to Fourier space, we would learn a great deal about the structure of inflationary correlators. Moreover, the momentum-space approach can also be useful in studying CFTs in Lorentzian signature~\cite{Gillioz:2016jnn, Gillioz:2018kwh, Gillioz:2018mto, Bautista:2019qxj, Gillioz:2019lgs}.
Unfortunately, taking the direct Fourier transform of position space CFT correlators is quite nontrivial. First of all, CFT correlators are singular at coincident points and need to be renormalized before the result can be Fourier transformed.
Second, even after renormalization, the explicit Fourier transforms are technically challenging.
In practice, it turns out to be easier to solve the conformal Ward identities directly in momentum space. 
However, even this approach quickly becomes intractable for operators with spin. Fortunately, the weight-shifting approach provides a more elegant way to proceed. This formalism allows us to generate new solutions to the conformal Ward identities by acting with differential operators on an initial seed solution.

\vskip 4pt
In this paper, we point out that the relevant spin-raising and weight-shifting operators used in cosmology are equivalent to similar operators used in the CFT literature~\cite{Costa:2011dw,Karateev:2017jgd,Costa:2018mcg}. The latter are defined most naturally in embedding space~\cite{Dirac:1936fq, Weinberg:2010fx, Costa:2011mg}. We show that the CFT weight-shifting operators are easily Fourier transformed, thereby bypassing the usual challenge of relating position space and Fourier space. Moreover, we show that the lift to embedding space provides an elegant way to derive and generalize the cosmological weight-shifting operators found in~\cite{Arkani-Hamed:2018kmz}. 
This new viewpoint clarifies the fact that all inflationary correlators can be obtained 
from a unique seed function corresponding to the exchange of a scalar particle and streamlines its derivation from the boundary perspective.
The weight-shifting approach also makes it clear how the spins of the external fields can be raised to obtain spinning solutions to the conformal Ward identities. 
We will present the details in a separate publication~\cite{CosmoBoot2}, where we show that inflationary tensor correlators can also be obtained from scalar seeds.
 
\paragraph{Outline} The plan of the paper is as follows: In Section~\ref{sec:CosmoBoot}, we recall a few relevant results of the cosmological bootstrap~\cite{Arkani-Hamed:2018kmz}. In particular, we present the de Sitter four-point function of conformally coupled scalars, arising from the tree-exchange of a generic scalar. This solution is the essential building block from which all other correlators will be derived via the action of suitable differential operators.
In Section~\ref{sec:EmbeddingSpace}, we briefly review the embedding space formalism for CFTs. Experts may skip this part without loss of continuity. 
In Section~\ref{sec:SpinExchange}, we work in embedding space to derive an operator which raises the spin of the particles exchanged in the correlators introduced in Section~\ref{sec:CosmoBoot}, and then translate this operator to Fourier space to obtain spin-exchange solutions. In Section~\ref{sec:Inflation}, we use the formalism to derive a weight-raising operator that transforms the four-point functions of conformally coupled scalars to those of massless scalars. We then show that the soft limit of these correlators with weakly perturbed scaling dimension leads to inflationary three-point functions. This allows for a compact expression for the inflationary bispectrum coming from the exchange of massive particles. We also present new results for the exchange of (partially) massless fields of arbitrary spin.
Our conclusions are presented in Section~\ref{sec:Conclusions}.
The appendices contain additional technical details and derivations: In Appendix~\ref{app:WeightEmbedding}, we provide a  systematic derivation of weight-shifting operators in embedding space. In Appendix~\ref{app:WeightFourier}, we transform these operators to Fourier space. In~Appendix~\ref{sec:PT}, we present explicit results for the polarization tensors used in the main text. Finally, Appendix~\ref{app:Notation} collects important variables used in this work. 

\paragraph{Notation and conventions} Unless stated otherwise, we will follow closely the notation and conventions used in~\cite{Arkani-Hamed:2018kmz}. 
Generic scalar operators (of dimension $\Delta$) will be denoted by $O$.
We will use $\varphi$ and $\phi$ for operators with $\Delta=2$ and $\Delta=3$, respectively. When we need to refer to the corresponding bulk fields, we will use $\upvarphi$ and $\upphi$.
The bulk de Sitter coordinates are $x^\mu$ and the coordinates on the spatial boundary are $x^i$, with conjugate momentum $k^i$. The boundary has $d$ spatial dimensions, and we often specialize to $d=3$, corresponding to the four-dimensional de Sitter space that seems to be relevant for our universe. 
Our convention for the $d$-dimensional Fourier transform is
\beq
O({\vec x}\hs) = \int\frac{{\rm d}^d k}{(2\pi)^d}\,e^{-i \k \cdot \x}\,O_{\k}\, .
\eeq
The coordinates in the embedding space formalism are $X^M$, with $M=-1,0,1, \ldots, d$, and the corresponding lightcone coordinates are $X^\pm = X^0 \pm X^{-1}$.

\vspace{0.5cm}
\section{De Sitter Four-Point Functions}
\label{sec:CosmoBoot}

In order to make our discussion self-contained, we begin with a brief review of relevant results from~\cite{Arkani-Hamed:2018kmz}, focusing on the bare minimum of background material required for our present purposes. All derivations and further details can be found in~\cite{Arkani-Hamed:2018kmz}.

\subsection{Boundary Correlators} 
\label{sec:boundary}

\begin{figure}[t!]
\centering
      \includegraphics[scale=0.7]{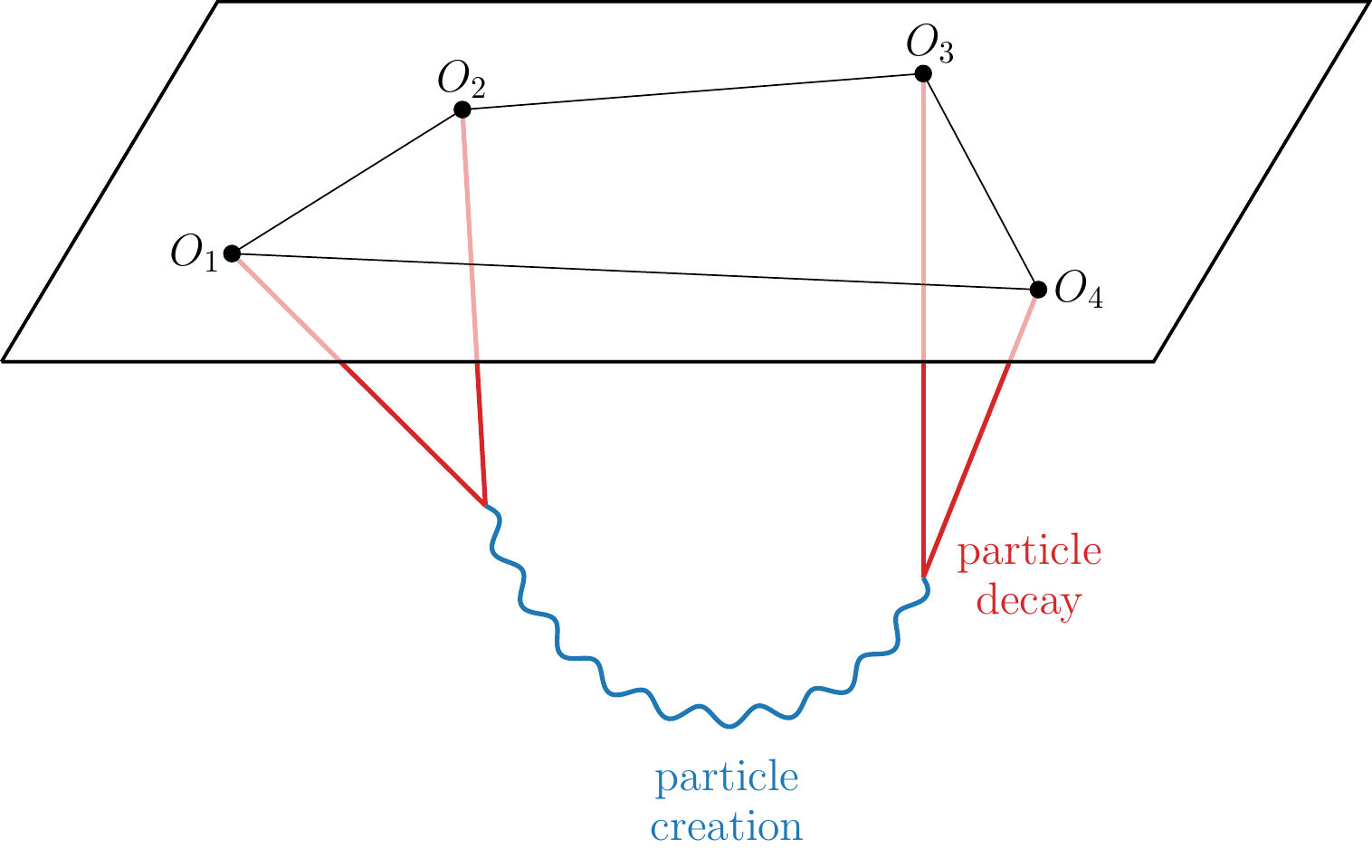}
           \caption{Illustration of particle creation and decay producing correlations on the future boundary of the de Sitter spacetime. The momentum dependence of the boundary correlators encodes the time dependence of the processes in the bulk. } 
    \label{fig:boundary}
\end{figure}

In the standard cosmological model, all cosmological correlators
can be traced back to the end of inflation (or the beginning of the hot Big Bang), where
they reside on the spacelike boundary of an approximate de Sitter spacetime  (see Fig.~\ref{fig:boundary}). The time dependence of bulk interactions is encoded in the momentum dependence of these boundary correlators. In particular, massive particles can be produced and decay during inflation, leaving their imprint in the nontrivial correlations on the boundary. 
 In the case of single-field slow-roll inflation with sufficiently weak couplings to additional massive particles, the inflationary correlators are strongly constrained by the conformal symmetry that the boundary theory inherits from the isometries of the bulk quasi-de Sitter spacetime. This is a promising arena to attempt to directly construct---or {\it bootstrap}---correlators on the future boundary by exploiting the large degree of symmetry and the expected analytic properties of tree-level processes.

\vskip 4pt
It is useful to briefly review how the symmetry constraints on late-time correlation functions arise in inflationary cosmology. The setting is de Sitter space in the inflationary slicing, which is described by the line element
\beq
{\rm d}s^2 = \frac{1}{H^2\eta^2}\left(-{\rm d}\eta^2+{\rm d}\vec x^{\hskip 1pt 2}\right).
\eeq
This spacetime has 10 Killing vectors associated with the following generators
\be
\begin{aligned}
P_i&= \partial_i\,,    & D & = -\eta \partial_\eta - x^i\partial_i\,,\\
J_{ij}&= x_{i}\partial_{j}-x_j\partial_i\,,  & \qquad K_i &=  2x_i \eta\partial_\eta +\big(2x^jx_i+(\eta^2- x^2)\delta^j_i\big)\partial_j\,.
\label{eq:conformal}
\end{aligned}
\ee
These transformations include
translations ($P_i$) and rotations ($J_{ij}$), which preserve the spatial slices, as well as dilations ($D$) and special conformal transformations ($K_i$).   We will assume that any additional matter fields only weakly break these de Sitter symmetries.  This implies that the late-time correlation functions of fields will transform like correlators in a CFT. To see how these constraints arise, let us
consider the evolution of a scalar field in de Sitter space.
At late times, a scalar field behaves as
\beq
\sigma(\vec x, \eta\to 0) = \sigma_{+}(\vec x\hskip 1pt)\, \eta^{\Delta_+} + \sigma_{-}(\vec x\hskip 1pt)\,\eta^{\Delta_-}\,.
\label{eq:dsfalloffs}
\eeq
We see that the field has two characteristic fall-offs, whose time dependence is fixed by the mass of the field, $m$, through the relation
\beq
\Delta_\pm = \frac{3}{2}\pm\sqrt{\frac{9}{4}-\frac{m^2}{H^2}} \, .
\label{equ:Delta}
\eeq
From this late-time scaling, we infer that the coefficient functions in (\ref{eq:dsfalloffs}) must transform as 
\be
\begin{aligned}
P_i\sigma_{\pm}&= \partial_i\sigma_{\pm}\,, & D\sigma_{\pm}& = -\left(\Delta_\pm + x^i\partial_i\right)\sigma_{\pm}\,,\\
J_{ij}\sigma_{\pm}&= \left(x_{i}\partial_{j}-x_j\partial_i\right)\sigma_{\pm}\,, &\qquad K_i\sigma_{\pm} &=  \left(2x_i \Delta_\pm +2x_ix^j\partial_j- x^2\partial_i\right)\sigma_{\pm}\,, 
\end{aligned}
\label{equ:Ki}
\ee
where~$\Delta_\pm$ is the weight of the operator late-time field coefficient $\sigma_{\pm}$. 
This implies that correlation functions of 
$\sigma_{\pm}$---and hence late-time correlation functions of fields in de Sitter space---obey the same kinematic constraints as conformal correlators. This allows us to leverage insights from the study of CFT to learn about inflationary correlators.

\vskip 4pt
We will be interested in the correlation functions of the late-time spatial profiles of fields in de Sitter space, focusing on the case of ``light fields," for which $m^2/H^2 \leq 9/4$. For these fields, the dominant fall-off at late times is given by $\Delta_-$, so that the physically interesting correlation functions are those of $\sigma_-$. However, in practice we will compute the correlation functions of $\sigma_+$ because these take a slightly simpler form. These correlators are related to those of  $\sigma_-$ by a shadow transform, which in Fourier space amounts to a simple multiplication by factors of power spectra.\footnote{Conformal primary operators of the same spin that are related by $\tilde\Delta = d-\Delta$ are so-called {\it shadows} of each other. These operators generate equivalent representations of the conformal group and can be mapped to each other by means of the shadow transform. For scalar operators in momentum space, the shadow transform is implemented by the map $O_{\tilde\Delta}(\vec k) = \langle O_{\tilde\Delta}(\vec k)O_{\tilde\Delta}(-\vec k)\rangle\, O_\Delta(\vec k)$.  See Appendix~A of~\cite{Anninos:2017eib} for more details.} We will denote the conformal weight of the dual boundary operators by $\Delta\equiv \Delta_+$. The cases of primary interest in this paper are conformally coupled scalars (with $m^2 = 2 H^2$ or $\Delta=2$) and massless scalars (with $\Delta=3$). Though we restrict our attention to light external fields, we will allow internal fields of arbitrary weights.


\vskip 4pt
In the cosmological context we are interested in correlation functions in Fourier space, in order to take advantage of the translational invariance of the spatial hypersurfaces. We should therefore translate the constraints coming from de Sitter/conformal invariance into momentum space.
To illustrate this, let us consider a
four-point function of scalar operators, which in momentum space takes the form
\beq
\langle O_1 O_2 O_3 O_4\rangle = F(k_1,k_2,k_3,k_4,s,t) \times (2\pi)^3 \delta^3(\k_1+\cdots +\k_4)\, , \label{equ:scalarF}
\eeq
where $O_n \equiv O_n(\k_n)$ are generic operators of scaling dimensions $\Delta_n$. Invariance under spatial rotations and translations implies that the four-point function $F$ is a function of six independent variables, which we take to be $k_n \equiv |\k_n|$, $s\equiv |\k_1+\k_2|$ and $t\equiv |\k_2+\k_3|$.
To be invariant under dilations (D) and special conformal transformations (SCTs), the function $F$ must satisfy the following {\it conformal Ward identities} 
 \begin{align}
{\rm D:} \quad 0 &= \left[9-\sum_{n=1}^4 \left(\Delta_n - k_n^j\frac{\partial}{\partial k_n^j}\right)\right] F\, ,  \label{WI_D} \\
{\rm SCT:} \quad 0&= \sum_{n=1}^4 \left[ (\Delta_n-3) \frac{\partial}{\partial k_{n,i}} -  k_n^j \frac{\partial^2}{\partial k_n^j k_{n,i}} + \frac{k_n^i}{2} \frac{\partial^2}{\partial k_n^j k_{n,j}} \right] F\, , \label{WI_SCT}
\end{align}
which are the momentum-space equivalents of the constraints in~\eqref{equ:Ki}.
 To satisfy (\ref{WI_D}) it is sufficient to write
 \beq
 F = s^{\Delta_t -  9} \F \, ,
 \label{equ:hatF}
 \eeq
 where $\Delta_t \equiv \sum_n \Delta_n$ and $\F$ is a dimensionless function. The form of $\F$ will be determined by the remaining Ward identity~\eqref{WI_SCT} and the singularities of tree-level processes.
 
\vskip 4pt
Once we have understood that cosmological correlation functions obey these symmetry constraints, we can cast aside the bulk interpretation of these correlators as arising from local causal and unitary time evolution, and attempt to reconstruct the corresponding correlation functions by directly solving~\eqref{WI_SCT}. This is the approach that we will take in this paper: The goal is to provide a systematic purely boundary derivation and interpretation of correlation functions corresponding to bulk tree-level exchange of massive particles. This is both practically useful---it will allow us to characterize slow-roll inflationary three-point functions in a completely universal way---as well as conceptually interesting, as it provides insight into how bulk time evolution is encoded in the (static) boundary correlation functions.
 
\subsection{Scalar Seed Functions}
\label{sec:seeds}

As we alluded to before, a case of special interest is the four-point function of the operator dual to conformally coupled scalars $\upvarphi$ (with $\Delta=2$), mediated by the tree-exchange of massive scalars~$\sigma$. In this case the kinematics further simplifies, and the four-point function can be written as a function of only two kinematic variables. In particular, the $s$-channel contribution takes the form 
\beq
F =  s^{-1} \F(u,v)\,,  \quad {\rm where} \quad \begin{array}{l} \displaystyle u \equiv  \frac{s}{k_1+k_2}\, , \\[0.5cm]
 \displaystyle v\equiv \frac{s}{k_3+k_4}\, .
 \end{array}
 \label{equ:ansatz}
\eeq 
The ansatz (\ref{equ:ansatz}) automatically satisfies two of the three equations contained in (\ref{WI_SCT}). 
After changing momentum variables, the remaining constraint equation can be written as~\cite{Arkani-Hamed:2018kmz}
\beq
(\Delta_u - \Delta_v) \hat F = 0\, , \label{equ:Ward}
\eeq
where $\Delta_u \equiv u^2(1-u^2) \partial_u^2 -2 u^3 \partial_u$. In general these equations have many solutions, but tree-level bulk physics is captured by solutions with a particular singularity structure.

\vskip 4pt
In~\cite{Arkani-Hamed:2018kmz}, the solutions of~\eqref{equ:Ward} were classified, for both contact interactions and tree-level exchange of massive particles.
A large class of contact solutions is extremely simple, and can be written as~\cite{Arkani-Hamed:2015bza}
\beq
 \hat C_n = \Delta_u^n \hat C_0\,, \quad{\rm where}\quad\hat C_0 = \frac{uv}{u+v}\,.\label{eq:Cncontact}
\eeq
The seed contact solution $\hat C_0$ arises from a $\upvarphi^4$ interaction in the bulk. Repeated application of the operator $\Delta_u$ produces the additional solutions $\hat C_n$ corresponding to higher-derivative interactions in the bulk.\footnote{An important caveat is that this is not the most general possible contact solution. The contact solutions shown are the ones that come from integrating out scalar particles at tree level. Integrating out higher-spin particles can produce contact solutions with dependence on additional kinematic invariants. These additional contact solutions can be generated by acting with the weight-shifting operators introduced in \S\ref{sec:InternalSpin} on the contact solutions in~\eqref{eq:Cncontact}. }

\vskip 4pt
In the case of tree exchange, the partial differential equation~\eqref{equ:Ward} can be written as two ordinary differential equations in $u$ and $v$ separately: 
\beq
\begin{aligned}
(\Delta_u + M^2) \F &= \hat C_n \,, \\
(\Delta_v + M^2) \F &= \hat C_n \,, 
\end{aligned}
\label{equ:exchange}
\eeq
where $\hat C_n$ is one of the contact solutions in \eqref{eq:Cncontact} and $M^2 \equiv (1-\Delta_\sigma)(\Delta_\sigma-2)$, which can be  
 related to the mass of the exchanged particle via (\ref{equ:Delta}).\footnote{Explicitly, we have $M^2 = m^2/H^2 -2$, where $m$ is the mass of the exchanged field in the bulk. The coupling to gravity is such that a massless scalar corresponds to $m^2 = 0$. Note that, in the limit $M\to\infty$, equation~\eqref{equ:exchange} has a formal solution as a series of contact terms of the form~\eqref{eq:Cncontact}, which is the EFT expansion arising from integrating out a heavy particle.} The replacement of the PDE in~\eqref{equ:Ward} with two ODEs in \eqref{equ:exchange} can be understood as a manifestation of locality at tree level. In particular, these differential equations can be derived from the bulk perspective using the equation of motion that the Green's function for the exchanged field obeys~\cite{Arkani-Hamed:2018kmz}.

\vskip 4pt
The equations~\eqref{equ:exchange} are second-order ordinary differential equations, so they require two boundary conditions. It is natural to impose boundary conditions at the singular points of the differential equation, $u = 0,\pm 1$. 
It is easy to see that a generic solution has the following logarithmic singularities
\be
\lim_{u\to +1}\F &\propto \log(1-u)\,,\\
\lim_{u,v\to -1}\F &\propto \log(1+u)\log(1+v)\,.
\ee
The limit $u \to 1$ corresponds to a collinear configuration where the momenta $\vec k_1$ and $\vec k_2$ align. In the standard Bunch--Davies vacuum, this limit should be regular. We therefore impose the absence of the singularity at $u= 1$ as one boundary condition. 
The limit $u,v \to -1$ cannot be reached for real momenta, but corresponds to an analytic continuation in the complex plane. In this limit, the four-point function factorizes into a product of three-particle amplitudes. The correct normalization of this factorization channel (which depends on whether we are computing the wavefunction or the correlator) provides a second boundary condition. 
These two boundary conditions uniquely fix the solutions to~\eqref{equ:exchange} given the form of the contact solutions $\hat C_n$.

\vskip 4pt 
The final singularity as $u\to0$ is physically the most interesting; in this limit the correlation function displays a characteristic non-analyticity
\beq
\lim_{u\to 0}\F \propto u^{\frac{1}{2}\pm i\mu}\,, \qquad \mu \equiv \sqrt{\frac{m^2}{H^2}-\frac{9}{4}}\, ,
\label{eq:uto0sing}
\eeq
where the parameter $\mu$ is set by the mass $m$ of the exchanged particle. 
The limit $u\to 0$ is the so-called {\it collapsed limit}---where two of the momenta nearly add to zero---and the characteristic ringing as we approach this limit is imprinted in inflationary three-point functions, providing a sharp way to test for the presence of these heavy states in observables. 
 The detailed form of the solutions $\hat F$ can be found in~\cite{Arkani-Hamed:2018kmz}, but won't be needed in our analysis below.

\subsection{Spin-Exchange Solutions} 

For scalar exchange, the correlators only depended on the variables $u,v$ (along with an overall power of $s$), but when we consider the exchange of spinning particles, the kinematics are more complicated. The four-point function will  depend on the following additional variables
\beq
\a \equiv \frac{k_1-k_2}{s} \, , \qquad \b \equiv \frac{k_3-k_4}{s}\, , \qquad \t \equiv \frac{(\k_1-\k_2)\cdot (\k_3-\k_4)}{s^2}\, .
\label{equ:abt}
\eeq
To determine the spin-exchange solutions, one approach is to re-solve~\eqref{WI_SCT}, allowing for an additional dependence of $\F$ on $\a$, $\b$ and $\t$. 
This quickly becomes very complicated and cumbersome.
Fortunately, it isn't necessary to solve the Ward identities for every case separately.  Instead, all spin-exchange solutions can be generated by acting with spin-raising operators on the scalar-exchange solutions. 
Written as a sum over helicity contributions, the spin-$S$ exchange solutions take the following form~\cite{Arkani-Hamed:2018kmz}
\beq
 \F^{(S)}  = \sum_{m=0}^S \Pi_{S,m} {\cal D}_{uv}^{(S,m)} \F^{(0)}\, ,
 \label{eq:spinldiffop}
\eeq
where we have defined $\F^{(0)}\equiv \F(u,v)$. In this paper, we use a group-theoretic approach inspired by known tools of conformal field theory~\cite{Costa:2011dw,Karateev:2017jgd,Dirac:1936fq, Weinberg:2010fx,Costa:2011mg} to provide a new and more elegant derivation of the relevant spin-raising operators ${\cal D}_{uv}^{(S,m)}$ and polarization sums $\Pi_{S,m}$ (see Section~\ref{sec:SpinExchange}).

\subsection{Inflationary Correlators} 
\label{sec:Inf}

Our real interest for applications to inflation is in massless external fields $\upphi$ (with $\Delta =3$), and not conformally coupled fields $\upvarphi$ (with $\Delta=2$).
In~\cite{Arkani-Hamed:2018kmz}, the dimensions/weights of the external fields were raised by acting with suitable differential operators on the solutions for conformally coupled scalars.
While this got the job done, the treatment in~\cite{Arkani-Hamed:2018kmz}  was unsatisfactory in a number of ways: {\it i}\hs) different weight-shifting operators had to be found for each spin-exchange solution separately and {\it ii}\hs) their derivation wasn't very systematic, so that explicit results were only shown up to spin two. 
In this paper, we will provide a much simpler and more unified derivation of a single weight-raising operator (see Section~\ref{sec:Inflation}). When this operator acts on a general spin-exchange solution (\ref{eq:spinldiffop}), it straightforwardly reproduces the results in~\cite{Arkani-Hamed:2018kmz} for low spins and automatically generalizes them to arbitrary spin.   Weakly perturbing the scaling dimension appearing in the weight-shifting operator, $\Delta_4=2 \to \Delta_4=2-\epsilon$, where $\epsilon$ is the slow-roll parameter, and taking the soft limit $k_4 \to 0$ provides an elegant way to 
 obtain inflationary three-point functions for arbitrary spin exchange.  The source functions appearing in these inflationary correlators will be given explicitly for any mass and spin of the exchange field.

\newpage
\section{CFTs in Embedding Space}
\label{sec:EmbeddingSpace}

We are interested in exploring the consequences of conformal symmetry on inflationary correlators, so it is useful to introduce some technical machinery to deal with conformal symmetry in a simple way.
Conformal transformations are complicated nonlinear transformations and become particularly involved for spinning operators. At the same time, it is easy to show that the conformal algebra on $\mathbb{R}^{d}$ is isomorphic to the algebra of Lorentz transformations on $\mathbb{R}^{1,d+1}$. This suggests
that a suitable embedding of $\mathbb{R}^{d}$ into $\mathbb{R}^{1,d+1}$, should make conformal transformations as simple as Lorentz transformations.
 The embedding space formalism of conformal field theory goes back to Dirac~\cite{Dirac:1936fq}, and has found many powerful applications in the recent CFT literature, e.g.~\cite{Costa:2011dw,Karateev:2017jgd}.  As we will see, the formalism is particularly well-suited to describe fields with spin.

\vskip 4pt 
 In this section, we provide a pedagogical review of CFTs in embedding space, based mostly on the excellent treatment in~\cite{Rychkov:2016iqz}. 
 Experts may skip directly to Section~\ref{sec:SpinExchange}, where we apply this formalism to derive the weight-shifting operators of interest in cosmology.
 
 \subsection{Projective Null Cone}\label{sec:nullcone}

We begin by describing the embedding of $d$-dimensional Euclidean space as a slice through a higher-dimensional lightcone.
 Consider $d+2$ dimensional Minkowski space, with coordinates
 \beq
  X^M, \ M=-1,0,1, \ldots, d\, , \label{equ:XM}
 \eeq
 where Lorentz transformations act as 
 \beq
 X^M \to {\Lambda^M}_N X^N\,.  \label{equ:LT}
 \eeq
 The goal is to find an embedding of $\mathbb{R}^{d}$ into $\mathbb{R}^{1,d+1}$ on which these Lorentz transformations become conformal transformations.
 We first restrict ourselves to points living on the {\it null cone} in the embedding space:
 \beq
 X^2 = 0  \, .
 \eeq
 This condition is Lorentz invariant and removes one of the coordinates in (\ref{equ:XM}).
 To remove a second coordinate and obtain a $d$-dimensional subspace, we define a {\it section} of the lightcone $X^+ = f(X^i)$, where $X^\pm \equiv X^0 \pm X^{-1} $ are lightcone coordinates and the coordinates $X^i$ are identified with the coordinates $x^i$ on $\mathbb{R}^{d}$.

  \begin{figure}[h!]
\centering
      \includegraphics[scale=1.1]{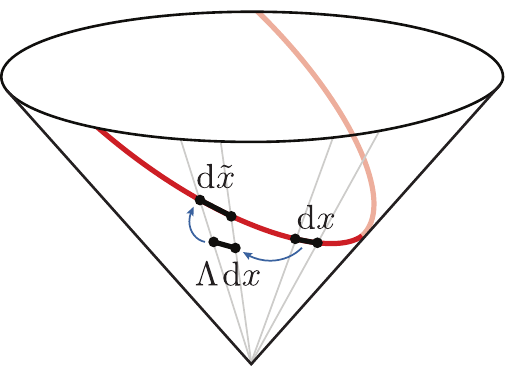}
           \caption{Illustration of the action of Lorentz transformations (and re-scalings) on points living on the Euclidean section of the lightcone in embedding space.} 
    \label{fig:Embedding}
\end{figure}

 \vskip 4pt
 We would like to understand how Lorentz transformations act on points living on the section. In particular, we want to determine for which choice of embedding function $ f(X^i)$ these transformations become conformal transformations.
 The action of the Lorentz group on the section is illustrated in Figure~\ref{fig:Embedding}. Each point on the section defines a lightray by connecting the point to the origin. Let ${\rm d} x$ be the infinitesimal interval between two nearby points on the section. The induced metric on the section relates this to the interval ${\rm d} s^2$. 
 Since the Lorentz transformation (\ref{equ:LT}) is an isometry, it will not change ${\rm d} s^2$. However, by itself, the Lorentz transformation will move the interval off the section. To map it back onto the section, we need to perform an additional rescaling 
 \beq
 X^M \to \lambda(X) X^M\, . \label{equ:ST}
 \eeq 
Under the combined action of (\ref{equ:LT}) and (\ref{equ:ST}), the induced metric on the section transforms as~\cite{Rychkov:2016iqz}
 \beq
 {\rm d} s^2 \to \Omega^2(x) \hs {\rm d} s^2\, ,  \quad {\rm with} \quad \Omega(x) = \lambda(X)\,.
 \eeq
 This corresponds to a conformal transformation on $\mathbb{R}^{d}$ if ${\rm d} s^2$ is flat. One can show that the latter requirement implies $f(X^i) = const.$ Without loss of generality, we can choose $X^+ = 1$,
 so that the {\it Euclidean section} of the lightcone is
 \beq
 X^M = (X^+, X^-, X^i) = (1,x^2,x^i)\, . \label{equ:ES}
 \eeq
As we will show next, correlators in the $d$-dimensional Euclidean space are lifted to homogeneous functions on the lightcone of the $(d+2)$-dimensional Minkowski spacetime, where the conformal group acts as the Lorentz group.

 \subsection{Tensors in Embedding Space}
 \label{sec:embeddingspacetensors}
 
 Consider a symmetric, traceless and transverse tensor $O_{M_1 \ldots M_S}$ defined on the cone $X^2=0$. Under the rescaling $X \to \lambda X$, the tensor transforms as
\beq
O_{M_1 \ldots M_S}(\lambda X) = \lambda^{-\Delta} O_{M_1 \ldots M_S}(X)\, ,  \label{equ:ST2}
\eeq
i.e.~it is a homogeneous function of degree $\Delta$.
This implies that the tensor is known everywhere on the cone if it is known on the section (\ref{equ:ES}).
The corresponding tensor on $\R^d$ is then defined through the following projection
\beq
O_{i_1 i_2 \ldots}(x) = O_{M_1 M_2 \ldots}(X)\, \frac{\partial X^{M_1}}{\partial x^{i_1}} \frac{\partial X^{M_2}}{\partial x^{i_2}} \cdots\, , \qquad \frac{\partial X^M}{\partial x^i} = (0, 2 x_i, \delta^j_i)\, .
\eeq
It is straightforward to show that the scaling transformation (\ref{equ:ST2}) for $O_{M_1 M_2 \ldots}(X)$ implies a conformal transformation for $O_{i_1 i_2 \ldots}(x)$.

\vskip 4pt
Contracting the tensors with auxiliary null polarization vectors $z^i$ and $Z^M$, we can write them in index-free notation
\be
\label{eq:index_free_vector_formalism}
O^{(S)}(x,z) &= O_{i_1\cdots i_S}(x) \hskip 2pt z^{i_1}\cdots z^{i_S}\, ,\\
O^{(S)}(X,Z) &= O_{M_1\cdots M_S}(X) \hskip 2pt Z^{M_1}\cdots Z^{M_S}\, .
\ee
In embedding space, any symmetric traceless tensor operator can therefore be written as a homogeneous function $O^{(S)}(X,Z)$ of the coordinates $X,Z\in \R^{d+1,1}$ such that $X^2 = X\cdot Z=Z^2 = 0$, with ``gauge invariance" under $Z\to Z+\beta X$. Together with the scaling~\eqref{equ:ST2}, this gauge invariance removes exactly two components per index from the tensor in embedding space, so its independent components match with those of the tensor on the Euclidean section.
Under rescalings of the embedding coordinates, we have
\be
O^{(S)}(\l X,\alpha Z) = \l^{-\De} \alpha^S O^{(S)}(X,Z)\,, \label{equ:ST3}
\ee
where $\De,S$ are the dimension and spin of $O^{(S)}$.  
Collectively, we refer to $[\hs\De,S\hs ]$ as the {\it weight} of the operator.  Conformal correlators in embedding space are simply the most general Lorentz-invariant expressions with the correct scaling behavior. The projection of these functions onto the Euclidean section defines the space of conformally-invariant correlation functions on $\mathbb{R}^d$.

\subsection{Conformal Correlators}
\label{sec:CC}

To illustrate the power of the embedding space formalism, we present a few examples of conformal correlators.

\vskip 4pt
Consider first a set of scalar primary operators $O_a \equiv O_a(X_a)$, of dimension $\Delta_a$.
Correlators of $O_a$ can only depend on the Lorentz-invariant inner products 
\beq
X_{ab} \equiv X_a \cdot X_b\, ,
\eeq
since $X_a^2=0$ on the lightcone. The scaling in (\ref{equ:ST3}) then uniquely fixes the two- and three-point functions of the operators to be
 \begin{align}
 \langle O_1 O_2 \rangle &= \frac{1}{X_{12}^{\Delta_1}} \delta_{\Delta_1 \Delta_2}\, , \label{equ:2pt} \\
 \langle O_1 O_2 O_3 \rangle &= 
 \frac{c_{123}}{X_{12}^{(\Delta_1+\Delta_2-\Delta_3)/2} X_{23}^{(\Delta_2+\Delta_3-\Delta_1)/2} X_{31}^{(\Delta_3+\Delta_1-\Delta_2)/2}}\, , \label{equ:3pt}
\end{align}
which reproduces the classic results in real space~\cite{Polyakov:1970xd}, if we use that $X_{ab} = - \frac{1}{2} (x_a - x_b)^2$.
Similarly, the four-point function of  identical scalars, of dimensions $\Delta_a \equiv \Delta$, is
 \beq
 \langle O_1 O_2 O_3 O_4 \rangle = \frac{1}{X_{12}^\Delta X_{34}^\Delta} \, g(U,V)\, , \qquad \begin{array}{l} \displaystyle U \equiv \frac{X_{12} X_{34}}{X_{13} X_{24}}\, , \\[0.5cm]
V\equiv \displaystyle \frac{X_{14} X_{32}}{X_{13} X_{24}}\, ,
 \end{array}
 \label{equ:4pt}
 \eeq
where $g$ is an arbitrary function of the cross ratios $U$ and $V$.

\vskip 4pt
The real benefit of going to embedding space becomes most manifest for operators with spin. 
For example, the two-point function of spin-$S$ operators (of dimension~$\Delta$) takes the form
\beq
\langle O_1^{(S)}  O_2^{(S)}  \rangle =  \left(Z_1 \cdot Z_2 - \frac{Z_1 \cdot X_2\, Z_2 \cdot X_1}{X_{12}} \right)^S  \langle O_1 O_2 \rangle\, ,
\eeq
where
$\langle O_1 O_2 \rangle$ is given by (\ref{equ:2pt}) and
the relative coefficient in the prefactor is fixed by transversality. 
Similarly, the three-point function of two scalars (with dimensions $\Delta_1$ and $\Delta_2$) and a spin-$S$ operator (with dimension $\Delta_3$) is
\beq
\langle O_1 O_2 O_3^{(S)} \rangle =  \left(\frac{(Z_3 \cdot X_1)(X_2 \cdot X_3) - (Z_3 \cdot X_2)(X_1\cdot X_3)}{(X_{12} X_{13} X_{23})^{1/2}}\right)^S\, \langle O_1 O_2 O_3 \rangle\, ,
 \eeq
 where $\langle O_1 O_2 O_3 \rangle$ is given by (\ref{equ:3pt}). 
 These simple examples already illustrate the power of the embedding space formalism.
 The above results are simply the most general Lorentz-invariant functions that are consistent with the scaling symmetry  \eqref{equ:ST3} and the transversality of the operators.
 More general spinning correlators are obtained in the same way and can be written in terms of a simple basis of tensor structures (see e.g.~\cite{Costa:2011dw,Costa:2011mg}), although this will not be needed in this paper. 
For specific weights, the operators become conserved and the correlators satisfy additional constraints~\cite{Costa:2011mg,Kravchuk:2016qvl}.
We will study these cases in a separate publication~\cite{CosmoBoot2}.

\vskip 4pt
In~\cite{Costa:2011dw}, conformally-invariant spin-$S_1$--spin-$S_2$--spin-$S$ three-point functions were written as derivatives of scalar--scalar--spin-$S$ three-point functions:
\be
\langle O_1^{(S_1)} O_2^{(S_2)} O_3^{(S)} \rangle = \cD\, \langle O_1 O_2 O_3^{(S)} \rangle\, ,
\ee
where the differential operator $\cD$ can be written in terms of spin-raising operators acting on~$O_1$ and $O_2$.
Relevant differential operators in this construction are
\be
D_{11} &\equiv (X_1 \cdot X_2) Z_1 \cdot \frac{\partial }{\partial X_2} -  (Z_1 \cdot X_2) X_1 \cdot \frac{\partial }{\partial X_2} -  (Z_1 \cdot Z_2) X_1 \cdot \frac{\partial }{\partial Z_2} +  (X_1 \cdot Z_2) Z_1 \cdot \frac{\partial }{\partial Z_2}\,, \label{equ:D11}\\
D_{12} &\equiv (X_1 \cdot X_2) Z_1 \cdot \frac{\partial }{\partial X_1} - (Z_1 \cdot X_2) X_1 \cdot \frac{\partial }{\partial X_1} + (Z_1 \cdot X_2) Z_1 \cdot \frac{\partial }{\partial Z_1}\, ,\label{equ:D12}
\ee
as well as two more operators with $1$ and $2$ interchanged.
Acting with $D_{ab}$ on a correlator increases the spin at point $a$ by one unit and decreases the dimension at point $b$ by one unit.\footnote{This is easy to see by counting factors of $X$ and $Z$. The detailed form of these operators is fixed by demanding that their action preserves the Euclidean section of the projective lightcone.} In Appendix~\ref{app:WeightEmbedding}, these weight-shifting operators, and others, will be discussed in more detail. We will now show how these operators can be utilized to raise the internal spin of the exchanged particles in cosmological correlation functions.

\section{Exchange of Spinning Particles}
\label{sec:SpinExchange}

A remarkable feature of the cosmological bootstrap is the fact that all spin-exchange solutions can be obtained from the scalar-exchange solution to~\eqref{equ:exchange} through the action of a {\it spin-raising operator}\hskip 1pt:
 \beq
\includegraphicsbox{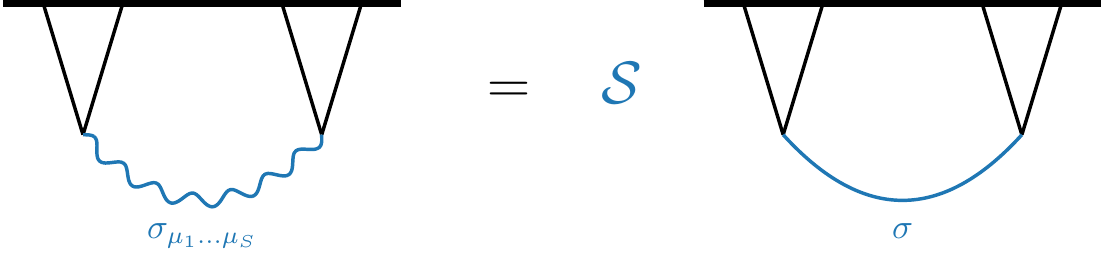}  \nonumber
  \eeq
In this section, we will use the embedding space formalism to provide a simple derivation of the relevant spin-raising operator ${\cal S}$. This operator is an example of a larger class of {\it weight-shifting operators}, which we discuss in more detail in Appendix~\ref{app:WeightEmbedding}.

\subsection{Spin-Raising Operator}

Ultimately, our goal is to raise the spin of exchanged particles in the scalar four-point functions.  To understand the origin of the relevant spin-raising operator, however, it is helpful to first consider raising the spin of an operator in a three-point function. 
Consider, for concreteness, the correlator of two scalar operators $\varphi$ (with dimension $\Delta_\varphi =2$) and a generic scalar operator $O$ (with dimension~$\Delta_3$). 
Using the expression for a general three-point function~\eqref{equ:3pt}, specialized to this case, we have
\beq
\langle \varphi \varphi O \rangle = 
 (X_{12}^{4-\Delta_3} X_{23}^{\Delta_3} X_{31}^{\Delta_3})^{-1/2}\, . \label{equ:SpinRaising}
\eeq
Acting on this correlator with the operator defined in~\eqref{equ:D12}, we find
 \beq
 \langle \varphi \tilde \varphi O^{(1)} \rangle  = - \frac{2}{\Delta_3}\hs D_{32}\,  \langle \varphi \varphi O \rangle\, ,  \label{equ:spin1}
 \eeq
where $\tilde \varphi$ is the shadow of $\varphi$, which has dimension $\Delta_{\tilde \varphi} = d - \Delta_\varphi = 1$ (for $d=3$), and $O^{(1)}$ is a spin-1 operator. The relation in~\eqref{equ:spin1} is easily confirmed using the results of \S\ref{sec:CC}. 
As advertised, the operator $D_{32}$ raises the spin of the operator at position $3$ by one unit and lowers the dimension at position $2$ by one unit. It is straightforward to translate the operator $D_{32}$ from embedding space to flat space and then write it in momentum space, where it can be applied to cosmological correlators.
Applying these transformations to the operator in~\eqref{equ:D12}, we find (see Appendix~\ref{app:WeightFourier})
\beq
D_{32} =  z_3^i \left[ (\Delta_3+S_3-1) K_{32}^i + \frac{1}{2}  k_3^i K_{32}^j K_{32}^j\right] , \quad {\rm with} \quad K_{32}^i \equiv \partial_{k_3^i} - \partial_{k_2^i}\, .
\label{eq:D32fourier} 
\eeq
Finally, we preform a shadow transform to raise the dimension of the operator $\tilde\varphi$ at position 2. In Fourier space, this simply amounts to multiplication by $k_2$, so that 
\beq
\langle \varphi \varphi O^{(1)} \rangle = k_2 \langle \varphi \tilde \varphi O^{(1)} \rangle  = - \frac{2}{\Delta_3} \, k_2 D_{32}\,  \langle \varphi \varphi O \rangle \equiv - \frac{2}{\Delta_3} \, i {\cal S}_{12}  \langle \varphi \varphi O \rangle \, , \label{equ:SR}
\eeq
where we have defined the {\it spin-raising} operator ${\cal S}_{12}$ that implements the combined action of~\eqref{eq:D32fourier} and the shadow transform. For our future uses, it is convenient to use momentum conservation to write the differential operator~\eqref{eq:D32fourier}  
in terms of $\vec k_1$ and $\vec k_2$, rather than $\k_3$. The differential operator then acts on the fields at positions $1$ and $2$, which explains the subscript of ${\cal S}_{12}$.
Repeated application of $i{\cal S}_{12} = k_2 D_{32}$ would raise the spin further.

\subsection{Raising Internal Spin}
\label{sec:InternalSpin}

We can use the operator ${\cal S}_{12}$ to raise the spin of the exchanged field using the scalar-exchange solution as a seed. We first show how this works in detail for spins 1 and 2  and then discuss the generalization to arbitrary spin.

\subsubsection{Spin-1 Exchange} 

We first consider the mapping of the scalar-exchange solution to a spin-1 exchange solution. To understand which spin-raising operator to use, it is helpful to first consider the disconnected contribution to the four-point function coming from spin-1 exchange (see e.g.~Appendix~A of~\cite{Arkani-Hamed:2018kmz}): 
\beq
\langle \varphi \varphi \varphi \varphi \rangle_{\rm d}^{(1)} = \frac{ \langle \varphi_{\vec k_1} \varphi_{\vec k_2} O_{-\vec s}^i\hskip 1pt \rangle (\Pi_1)_{ij} \langle O_{\vec s}^j \hskip 2pt\varphi_{\vec k_3} \varphi_{\vec k_4} \rangle }{\langle O_{\vec s} \hskip 2pt O_{-\vec s} \hskip 1pt\rangle}\, , \label{equ:dis}
\eeq 
where the symmetric traceless tensor $(\Pi_1)_{ij}$ contains the polarization structure of the inverse two-point function of the exchanged field:\hskip 1pt\footnote{Note that in Fourier space, the inverse of the two-point function is equivalent to the two-point function of an operator with the shadow dimension (suitably normalized).}
\beq
 \langle O^i_{\raisebox{-1pt}{\scriptsize $ \vec s$}} \,O^j_{-\vec s} \hskip 1pt\rangle^{-1} \propto (\Pi_1)^{ij} \langle O_{\vec s} \hskip 2pt O_{-\vec s}\hskip 1pt\rangle^{-1} \, .
\eeq
 Using the spin-raising operator in~\eqref{equ:SR}, the expression~\eqref{equ:dis} can be written as
\beq
\langle \varphi \varphi \varphi \varphi \rangle_{\rm d}^{(1)} = -(\Pi_1)_{ij}  \frac{ {\cal S}_{12}^i\langle \varphi_{\vec k_1}\varphi_{\vec k_2} O_{\hspace{-1pt}\raisebox{-2pt}{\scriptsize $- \vec s$}} \hskip 1pt \rangle \hskip 2pt {\cal S}_{34}^j  \langle O_{\raisebox{-2pt}{\scriptsize $\vec s$}}\hskip 2pt\varphi_{\vec k_3} \varphi_{\vec k_4} \rangle }{\langle O_{\vec s}\,O_{-\vec s}\hskip 1pt\rangle},
 \label{equ:S1E}
\eeq
where $O$ is a scalar operator of dimension $\Delta$. From this, it is clear that the operator $(\Pi_1)_{ij}\,{\cal S}_{12}^i {\cal S}_{34}^j$ acting in the numerator raises the spin of the exchanged particle. Our goal is to simplify this operator such that we can pull it  outside and have it act on the total disconnected correlator $\langle \varphi \varphi \varphi \varphi \rangle_{\rm d}^{(0)}$. 

\vskip 4pt
We start by simplifying the expression ${\cal S}_{12}\langle \varphi \varphi O\rangle$.
Since the three-point function depends only on the magnitudes of the momenta, we can write the operator in~\eqref{equ:SR} as
\be
 {\cal S}^i_{12} = \ (\Delta-1) k^i_2 \pdr{}{k_2} + k_2 s^i \bigg[ (\Delta-2)\frac{1}{s}\pdr{}{s} -\frac{1}{k_2}\pdr{}{k_2} -\frac{1}{2}\frac{\ptl^2}{\ptl k_2^2} -\frac{1}{2}\frac{\ptl^2}{\ptl s^2} -(\hat{k}_2\cdot\hat{s})\frac{\ptl^2}{\ptl k_2\ptl s} \bigg]\,, \label{equ:S12X}
\ee
where we have used $\vec{s}= \vec{k}_1+\vec{k}_2$,  
which is the momentum of the exchanged particle.
Next, we change variables from $k_1,k_2$ to \beq
u=\frac{s}{k_1+k_2} \, , \qquad \a=\frac{k_1-k_2}{s} \, .
\eeq
In these variables, the scalar three-point function has an extremely simple functional dependence~\cite{Arkani-Hamed:2018kmz}: $\langle\varphi\varphi O\rangle = s^{\Delta-2}\hat f(u)$, so we can drop any derivatives with respect to $\hat\alpha$ and derivatives with respect to $s$ act simply on the polynomial prefactor. 
In simplifying the answer it is convenient to use $\Delta_u \hat f(u) = (\Delta-1) (\Delta-2)\hat f(u)$, where $\Delta_u$ was defined below~\eqref{equ:Ward}. After some algebra, the operator~(\ref{equ:S12X}) simplifies to
\beq
\label{equ:o12}
{\cal S}^i_{12} = \frac{\Delta-1}{2s} \Big[ \, \alpha^i u^2\ptl_u-s^i \a (u\ptl_u +\Delta-2 ) \, \Big] \, , 
\eeq
where we have introduced 
$\vec\alpha\equiv \k_1-\k_2$, for later convenience.\footnote{The magnitude of this vector is not to be confused with $\hat\alpha=(k_1-k_2)/s$, i.e.~$|\vec\alpha|/s\ne \hat\alpha$.} Importantly, this operator only involves derivatives with respect to $u$. 
Similar manipulations let us simplify the operator ${\cal S}_{34}$. 
All we have to do is replace $\vec s$ by $-\vec s$, $\vec \alpha$ by $\vec\beta=\vec k_3-\vec k_4$ and $u, \hat \alpha$ by $v, \hat \beta$. 
The resulting operator will only contain derivatives with respect to $v$.

\vskip 4pt
Having simplified the differential operators, we now turn our attention to the polarization structure of the exchanged operator. For exchange of an operator of weight $\Delta$, the spin-1 polarization tensor is (see Appendix~\ref{sec:PT} for a derivation)  
\beq
(\Pi_1)_{ij}  = \pi_{ij} + \frac{1-\Delta}{\Delta - 2}\, \hat s_i \hat s_j\, ,  \label{equ:Pij}
\eeq
where $\pi_{ij} \equiv \delta_{ij} - \hat s_i \hat s_j$ is a transverse projector.
Substituting~\eqref{equ:o12} and~\eqref{equ:Pij} into~\eqref{equ:S1E}, we see that we can pull the differential operators outside to obtain 
\beq \label{equ:spin1disc}
 \langle \varphi \varphi \varphi \varphi \rangle_{\rm d}^{(1)} \, \propto\, \left( \Pi_{1,1}D_{uv}+\Pi_{1,0}\Delta_u \right) \langle \varphi \varphi \varphi \varphi \rangle_{\rm d}^{(0)} \, , 
 \eeq
where 
we have defined the differential operator $D_{uv}\equiv(uv)^2\ptl_u\ptl_v$ and the polarization sums $\Pi_{1,1} \equiv  \alpha^i \pi_{ij} \beta^j/s^2$ and  $\Pi_{1,0}\equiv \a \b$. The differential operator that raises the spin of the exchanged particle from zero to one therefore is
\begin{eBox}
\beq
 {\cal S}_{uv}^{(1)} \equiv \Pi_{1,1}D_{uv}+\Pi_{1,0}\Delta_u \, , \label{equ:S1}
\eeq
 \vskip 3pt
\end{eBox}
which is precisely the result found in~\cite{Arkani-Hamed:2018kmz}.  

\vskip 4pt
Thus far, we have only discussed the disconnected contribution to the four-point function arising from the exchange of a single scalar operator of general conformal weight $\Delta$. To relate this to the most general exchange four-point function, we note that any four-point function can be decomposed as a sum over the exchange of various states of different conformal weights, but fixed spin. 
Moreover, since the spin-raising operator ${\cal S}_{uv}^{(1)}$ in (\ref{equ:S1}) does {\it not} depend on conformal dimension, acting with it on the complete scalar exchange solution $\F^{(0)}$ produces the complete spin-1 exchange solution~$\F^{(1)}$:
\begin{eBox}
\beq
\F^{(1)} = {\cal S}_{uv}^{(1)} \F^{(0)}\, .
\eeq
 \vskip 3pt
\end{eBox}
Indeed, this relation is precisely what was found by other means in~\cite{Arkani-Hamed:2018kmz}.

\subsubsection{Spin-2 Exchange} 

We can repeat the same exercise for spin-2 exchange, which is  algebraically more involved, but conceptually the same. In this case, the disconnected part of the four-point function is
\beq
\langle \varphi \varphi \varphi \varphi \rangle_{\rm d}^{(2)} = \frac{ \langle \varphi_{\vec k_1} \varphi_{\vec k_2} O_{-\vec s}^{ij} \hskip 1pt\rangle (\Pi_2)_{ij,lm}(\hat s) \langle O_{\vec s}^{lm}\hskip 2pt \varphi_{\vec k_3} \varphi_{\vec k_4} \rangle }{\langle O_{\vec s} \hskip 2pt O_{-\vec s}\hskip 1pt\rangle} \label{equ:diss2} \, ,
\eeq
where $(\Pi_2)_{ij,lm}$ is the spin-2 polarization tensor. Like in the spin-1 case, we can write this in terms of the spin-raising operator~\eqref{equ:SR} acting on scalar three-point functions
\beq
\langle \varphi \varphi \varphi \varphi \rangle_{\rm d}^{(2)} = (\Pi_2)_{ij,lm}  \frac{\left( {\cal S}_{12}^2\right)^{ij} \langle \varphi_{\vec k_1} \varphi_{\vec k_2} O_{\hspace{-1pt}\raisebox{-2pt}{\scriptsize $- \vec s$}} \hskip 1pt \rangle  \left( {\cal S}_{34}^2\right)^{lm}\langle O_{\raisebox{-2pt}{\scriptsize $\vec s$}}\hskip 2pt \varphi_{\vec k_3} \varphi_{\vec k_4} \rangle }{\langle O_{\vec s} \hskip 2pt O_{-\vec s} \hskip 1pt\rangle}.
 \label{equ:S2E} 
\eeq
As before, we want to write the spin-raising operators in such way that we can act with them on the entire disconnected correlator. This proceeds very similarly to the spin-1 case. We first focus on $ {\cal S}_{12}^2 \langle \varphi\varphi O\rangle$. Changing variables to $u,\hat \alpha, \vec \alpha$ and eliminating $s$-derivatives, we can write the spin-raising operator as
\be
\begin{aligned}
\left( {\cal S}_{12}^2\right)^{ij} \propto \,\, & \frac{\Delta (\Delta-1)}{s^2} \bigg[ \delta^{ij} s^2 \left[ 3(1-{\a}^2)u\ptl_u +(\Delta-2)(\Delta-3{\a}^2) \right] \\
& - 3\alpha^i\alpha^j u^2\ptl_u(u^2\ptl_u) + 3(\alpha^is^j+s^i\alpha^j) u \a \left[ \Delta-3+\ptl_u(u^2\ptl_u) \right] \\
& - 3s^is^j \Big( \Delta-2+u\ptl_u +{\a}^2(\Delta-2)(\Delta-4)+ {\a}^2 (2\Delta-5)u\ptl_u +{\a}^2u^2\ptl_u^2 \Big) \bigg]\,,
\end{aligned}
\label{equ:o12X}
\ee
where again all derivatives are only with respect to $u$.
As in the spin-1 case, we can get ${\cal S}_{34}^2$ from the replacement $\{\vec s, u,  \hat\alpha, \vec\alpha\}\mapsto \{-\vec s, v,\hat\beta, \vec \beta\}$. Importantly, these differential operators again have no $s$-derivatives, so we can pull them outside to act on the full disconnected correlator $\langle \varphi \varphi \varphi \varphi \rangle_{\rm d}^{(0)}$.

\vskip 4pt
Next, we consider the spin-2 polarization structure, which is given by (see Appendix~\ref{sec:PT})
\beq
(\Pi_2)_{ij,lm}  = (\Pi_{2,2})_{ij,lm}-\frac{\Delta}{\Delta-3}(\Pi_{2,1})_{ij,lm}+ \frac{\Delta(\Delta-1)}{(\Delta-2)(\Delta-3)}(\Pi_{2,0})_{ij,lm} \,, \label{equ:Pijrs}
\eeq
where we have defined the individual helicity components as
\be
\begin{aligned}
(\Pi_{2,2})^{ij}{}_{lm} &= \pi^{(i}_{(l} \pi^{j)}_{m)}-\frac{1}{2}\pi^{ij}\pi_{lm}\,,\\[4pt]
(\Pi_{2,1})^{ij}{}_{lm} &= 2 \hskip 1pt \hat s^{(i} \hat s_{(l}\pi^{j)}_{m)}\,,\\
(\Pi_{2,0})^{ij}{}_{lm} &= \frac{3}{2}\left (\hat s^{i} \hat s^{j}-\frac{1}{3}\delta^{ij}\right )\left (\hat s_l \hat s_m -\frac{1}{3}\delta_{lm}\right) .
\end{aligned}
\ee
Combining these expressions together with the simplified expressions for ${\cal S}_{12}^2$ and ${\cal S}_{34}^2$, and performing some algebra, then gives
\beq
\begin{aligned}
 \langle \varphi \varphi \varphi \varphi \rangle_{\rm d}^{(2)} &\, \propto\, \left( \Pi_{2,2}D_{uv}^2 + \Pi_{2,1} D_{uv}(\Delta_u-2) +\Pi_{2,0}\Delta_u(\Delta_u-2) \right) \langle \varphi \varphi \varphi \varphi \rangle_{\rm d}^{(0)}  \, ,
 \end{aligned}
\eeq
where we have defined the following polarization sums 
\be
\begin{aligned}
\Pi_{2,2} &\equiv \frac{3}{2s^4} \alpha_i \alpha_j {(\Pi_{2,2})^{ij}}_{lm}  \beta^l \beta^m\,, \\[4pt]
\Pi_{2,1} &\equiv \frac{3}{s^2}\hskip 1pt \hat \alpha\hat \beta\hs  \alpha^i \pi_{ij}  \beta^j\, , \\[4pt]
\Pi_{2,0} &\equiv \frac{1}{4}(1-3{\a}^2)(1-3{\b}^2)\,. 
\end{aligned}
\ee
We have therefore found an operator that raises the spin of the exchanged particle from zero to two:
\begin{eBox}
\beq
{\cal S}_{uv}^{(2)}\equiv \Pi_{2,2}D_{uv}^2 + \Pi_{2,1} D_{uv}(\Delta_u-2) +\Pi_{2,0}\Delta_u(\Delta_u-2) \, .
\eeq
\vskip 3pt
\end{eBox}
For the same reason as before, we can take this operator and apply it to the full connected four-point function to obtain the massive spin-2 exchange solution
\begin{eBox}
\beq
\F^{(2)} = {\cal S}_{uv}^{(2)} \F^{(0)}\, ,
\eeq
\vskip 3pt
\end{eBox}
which is the same result as in~\cite{Arkani-Hamed:2018kmz}.

\subsubsection{Higher-Spin Exchange}

The spin-raising procedure we have described is
completely algorithmic and can therefore easily be extended to arbitrary spin.
In particular, the spin-$S$ exchange solution can be written schematically as
\be
\F^{(S)} = (\Pi_S)_{i_1 \ldots i_S \hskip 1pt,\hskip 1pt j_1 \ldots j_S} \left( {\cal S}_{12}^S\right)^{i_1 \ldots i_S} \left( {\cal S}_{34}^S\right)^{j_1 \ldots j_S} \F^{(0)}\, ,
\ee
where ${\cal S}_{ab}^S$ are the spin-$S$ analogues of~\eqref{equ:o12} and~\eqref{equ:o12X}, $(\Pi_S)_{i_1 \ldots,\hskip 1pt j_1 \ldots }$ is the polarization structure of the spin-$S$ two-point function, and $\F^{(0)}$ is the same seed function as before. 

\vskip 4pt
It is instructive to write the spin-raising operator as a sum over the different helicity components
\begin{eBox}
\beq
 \F^{(S)}  = \sum_{m=0}^S \Pi_{S,m} {\cal D}_{uv}^{(S,m)} \F^{(0)}\, , \label{equ:POL}
\eeq
\end{eBox}
where the differential operators at each helicity $m$ are given by
\begin{eBox}
\beq
 {\cal D}_{uv}^{(S,m)} =  D_{uv}^m \prod_{j=1}^{S-m} \Big(\Delta_u - (S-j)(S-j+1)\Big)\, . \label{equ:Duvellm}
 \eeq
 \end{eBox}
 The polarization sums appearing in~\eqref{equ:POL} are presented in detail in Appendix~\ref{sec:PT}. Specializing to $d=3$ dimensions, they can be written as
\begin{eBox}
\beq
	\Pi_{S,m} = (2-\delta_{m0})(-\hat L)^m \cos(m\psi) \tilde P_S^m(\a)\hskip 1pt \tilde P_S^{-m}(\b)\, ,
	\label{equ:Piellm}
\eeq
\vskip -10pt
\end{eBox}
where $\tilde P_S^m(x) \equiv (1-x^2)^{-|m|/2} P_S^m(x)$ is a modified version of the associated Legendre polynomial  $P_S^m(x)$, and $\psi$ is the angle between $\hat k_1$ and $\hat k_3$ on the plane perpendicular to $\hat s$, given by 
\begin{align}
	\cos\psi = \frac{\cos\gamma-\cos\theta_1\cos\theta_3}{\sin\theta_1\sin\theta_3} = \frac{\hat T}{\hat L}\,.\label{eq:psi}
\end{align}
In this expression, we have defined the angles $\cos\gamma\equiv \hat k_1\cdot\hat k_3$, $\cos\theta_a\equiv \hat k_a\cdot \hat s$, and the kinematic invariants: 
\begin{align}
\hat T &\equiv  \frac{\alpha_i \pi^{ij} \beta_j}{s^2}= \t +\frac{\a\b}{uv}\, , \label{equ:T}\\
\hat L^2 &\equiv \frac{ \alpha_i \pi^{ij} \alpha_j \hs  \beta_k \pi^{kl} \beta_l}{s^4}= \frac{(1-u^2)(1-v^2)}{u^2v^2} (1-\a^2)(1-\b^2)\, . \label{equ:L}
\end{align}
The overall normalization was chosen such that $\Pi_{0,0}=1$. 

\vskip 4pt
Note that some of the ``angles'' appearing in \eqref{equ:Piellm} are slightly unusual. However, taking the collapsed limit, $s\to 0$, the angular dependence simply becomes
\begin{eBox}
\beq
	\lim_{s\to 0}\Pi_{S,m} = (2-\delta_{m0})(-uv)^m\cos( m\psi)P_S^m(\cos\theta_1)P_S^{-m}(\cos\theta_3)\, .\label{eq:PolCollapsed}
\eeq
\end{eBox}
Along with the momentum dependence~\eqref{eq:uto0sing}, this angular dependence in the collapsed limit is one of the hallmarks of the exchange of massive particles with spin. The characteristic angular structure allows  both mass and spin spectroscopy to be performed using the collapsed limit of the four-point function (or the squeezed limit of the three-point function).

\section{Inflationary Three-Point Functions}
\label{sec:Inflation}

So far, we have presented results for the four-point functions of $\Delta = 2$ operators in de Sitter space.  To describe inflationary correlators, however, we further need to raise the weight of the external operators to $\Delta=3$, corresponding to a massless field in the bulk. This field plays the role of the inflaton  in conventional models of slow-roll inflation. 

\vskip 4pt
In~\cite{Arkani-Hamed:2018kmz}, a set of weight-raising operators was introduced, relating the four-point functions of conformally-coupled and massless scalars.  The derivation of these operators, however, was 
somewhat unsatisfactory because they had to be found separately for each spin-exchange solution $\F^{(S)}$. Explicit results were therefore only presented for low spins. In this section, we will show that the embedding space formalism allows for very simple derivation of a {\it single} weight-raising operator:
\beq
\includegraphicsbox{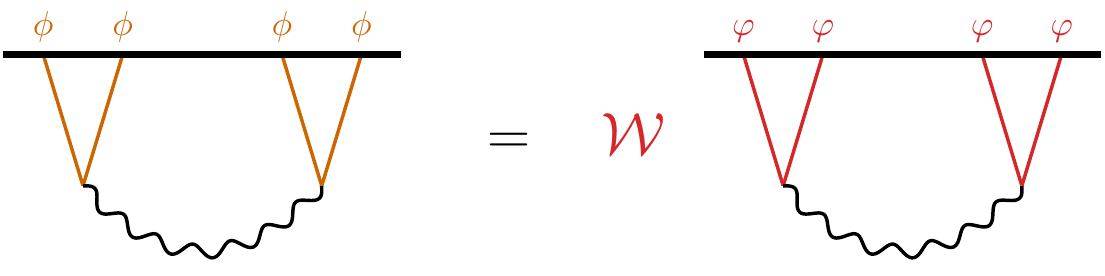}  \nonumber
\eeq
Acting with this operator ${\cal W}$ on the solutions $\F^{(S)}$ straightforwardly reproduces the results in~\cite{Arkani-Hamed:2018kmz} for low spin, and automatically generalizes them to arbitrary spin.

\vskip 4pt   
Finally, we will show how to perturb our de Sitter four-point functions to obtain inflationary three-point functions. 
We will present explicit formulas for the
source functions appearing in these inflationary correlators for any mass and spin of the exchanged field.

\subsection{Weight-Raising Operator}
\label{sec:WR}

Recall from~\eqref{equ:4pt} that, in embedding space, the most general four-point functions of a $\Delta = 2$ scalar operator, $\varphi$, and a $\Delta= 3$ scalar operator, $\phi$, can be expressed as\hs\footnote{Recently, the momentum space versions of these formulas have been presented in \cite{Bzowski:2019kwd}.}
\begin{align}
 \langle \varphi \varphi \varphi \varphi \rangle &= \frac{1}{X_{12}^2 X_{34}^2} \, f(U,V)\, , \\
 \langle \phi \phi \phi \phi \rangle &= \frac{1}{X_{12}^3 X_{34}^3} \, h(U,V)\, ,
 \end{align}
 where $f$ and $h$ are arbitrary functions of the cross ratios $U$ and $V$. 
It is then straightforward to show that the following relation holds
\beq
\langle \phi \phi \phi \phi \rangle  = {\cal W}_{12} {\cal W}_{34} \hskip 1pt  \langle \varphi \varphi \varphi \varphi \rangle\, , \label{equ:WR}
\eeq
where we have defined
 \beq
{\cal W}_{ab} \equiv \eta^{MN} \left(\frac{\partial}{\partial X_a^M} + \frac{X_{a,M}}{3}  \frac{\partial^2 }{\partial X_a^2}\right)  \left( \frac{\partial}{\partial X_b^N} + \frac{X_{b,N}}{3}  \frac{\partial^2 }{\partial X_b^2}\right) . \label{equ:W}
 \eeq
We see that the operator ${\cal W}_{12} {\cal W}_{34}$ acts as a {\it weight-raising operator}. The structure of the operators in~\eqref{equ:W} acting on the points $a$ and $b$ can be understood as follows: to lower  the conformal weight by one unit, one must reduce the overall power of $X$ by one unit. This is done by taking derivatives with respect to~$X$. 
On the projective lightcone, there are only two possible operators that reduce the weight by one, namely $\partial/\partial X$ and $X\partial^2/\partial X^2$.  The specific linear combination of these two operators in~\eqref{equ:W} is the unique combination that
preserves the Euclidean section. More details of the construction of these weight-shifting operators can be found in Appendix~\ref{app:WeightEmbedding}.

\subsection{Raising External Weight}
\label{sec:RaiseW}

Our task now is simply to project~\eqref{equ:W} to position space and then transform the result to momentum space, so that we can act on our solutions for  $\langle \varphi \varphi \varphi \varphi \rangle$  to produce the corresponding correlators $\langle \phi \phi \phi \phi \rangle$. For  $\Delta_1=\Delta_2= 2$, we find (see Appendix~\ref{app:WeightFourier})
\beq
\cW_{12} = \frac{(k_1k_2)^2}{2}\vec K_{12}\cdot \vec K_{12} -\k_1\cdot \k_2 - \Big(k_2^2\, \k_1\cdot \vec K_{12}+ 1\leftrightarrow 2\Big) \, ,\label{eq:W12Fourier}
\eeq
where $\vec K_{12}$ was defined as in~\eqref{eq:D32fourier}. 
To apply the above weight-shifting operator to the spin-exchange four-point function~\eqref{equ:POL}, it is helpful to express it in terms of the dimensionless kinematic variables $\{s,u,v,\hat\alpha,\hat\beta,\hat\tau\}$. 
Although this is conceptually straightforward, it is algebraically somewhat involved. 
After some work, we can express \eqref{eq:W12Fourier} as  
\begin{align}
	\cW_{12} &=  \frac{s^2}{8u^2}\bigg[2(1-u^2(2-\a^2)+2u(2-3u^2+u^4\a^2)\partial_u+u^2(1-u^2)(1-u^2\a^2)\partial_u^2 \\
	&\quad -2\a(1-u^2(2- \a^2))\partial_{\a}+(1- \a^2)(1-u^2\a^2)\partial_{\a}^2+\frac{ v\big(1- \a^2 u^2\big)^2 \big(1-v^2(1-\b^2)\big)}{u^2 v^2}\partial_{\t}^2 \nonumber\\[7pt]
	&\quad -\frac{4 u (\a  u (2 \b + \a  \t  u v)+v\hat\tau )+2 (1-\a^2 u^2) \big(u^2 (\a \b  u+v\t )\partial_u+(\b +\a \t  u v)\partial_{\a}\big)}{u v}\partial_{\t}\bigg]\, , \nonumber
\end{align}
where $\a$, $\b$ and $\t$ were defined in~\eqref{equ:abt}.

\vskip 4pt
When acting on spinning correlators in momentum space, it will be useful to decompose ${\cal W}_{12}$ into helicity components. We do this by commuting it through the polarization structure as 
\begin{align}
	\cW_{12} \sum_{m=0}^S \Pi_{S,m} {\cal D}_{uv}^{(S,m)} \F^{(0)} =  \frac{s^2}{2}\sum_{m=0}^S \Pi_{S,m} U_{12}^{(S,m)}{\cal D}_{uv}^{(S,m)} \F^{(0)}\,,
\end{align}
where $U_{12}^{(S,m)}$ is a dimensionless, helicity-decomposed weight-shifting operator. Using the detailed form of the polarization structure \eqref{equ:Piellm}, we find that this operator is given by
\begin{eBox}
\vskip 5pt
\beq
\begin{aligned}
	U_{12}^{(S,m)} &\equiv \left(1+\frac{u(1-\a^2u^2)}{4} \hs\ptl_u \right)\frac{(1-u^2)(1+u\partial_u) - 2S}{u^2}+(S+m)\a \,\frac{P_{S-1}^m(\a)}{P_S^m(\a)} \\[5pt]
	&\quad\  +\frac{(S-m)[(S+m+1)u^2 \a^2-S-m+3+2u(1-u^2\a^2)\partial_u]}{4u^2}\, ,\label{U12SmX}
\end{aligned}
\eeq
\vskip 2pt
\end{eBox}
where $P_{-1}^0(\a)= 1$. This agrees with the results of~\cite{Arkani-Hamed:2018kmz} for $S =1,2$, but now holds for all~$S$. The corresponding operator $U_{34}^{(S,m)}$ is obtained from this expression by swapping the kinematic variables $\{u,\hat\alpha\}\mapsto \{v,\hat\beta\}$.

\vskip 4pt
Using the operators $U_{12}^{(S,m)}$ and $U_{34}^{(S,m)}$, the general spin-exchange four-point function of massless scalars can then be written in the following form
\begin{eBox}
\vskip 1pt
\begin{equation}
	F_{\Delta=3}^{(S)} = s^3 \sum_{m=0}^S \Pi_{S,m} U_{12}^{(S,m)}U_{34}^{(S,m)} \,\cD_{uv}^{(S,m)}\F^{(0)}_{\Delta=2}\, .
\end{equation}
\end{eBox}

As we will show in the next section, only the longitudinal $(m=0)$ component contributes to inflationary three-point functions. In that case, the operator in (\ref{U12SmX}) can be written in a slightly simplified form as
\beq
U_{12}^{(S,0)} = \frac{\big(1-u^2\a^2\big)\big(\Delta_u-(S-1)(S+2)\big)}{4u^2}+\frac{1-u^2}{u}\partial_u+ S\a\, \frac{ P_{S -1}(\a)}{ P_{S}(\a)} - 1 +(1-S)\a^2\, .  \label{U12long}
\eeq
This operator acts on the longitudinal part of the $\Delta=2$ solution, 
\beq
\F_L^{(S)} \equiv {\cal D}_{uv}^{(S,0)} \F^{(0)}  = \prod_{j=1}^{S} \big(\Delta_u - (S-j)(S-j+1)\big) \F^{(0)}\, , \label{equ:FL}
\eeq
where the operator ${\cal D}_{uv}^{(S,0)} $ was defined in~\eqref{equ:Duvellm}.

\subsection{From de Sitter to Inflation}
\label{sec:toInf}

Using the operator $U_{12}^{(S,m)}$ defined in (\ref{U12SmX}), we are able to efficiently generate $\Delta = 3$ scalar solutions to the conformal Ward identities. To apply these results to inflation, we must take into account that the inflaton field is not exactly massless and has a time-dependent expectation value~$\bar \upphi(t)$.
We assume that the associated breaking of the de Sitter symmetries is weak and can be treated perturbatively. As explained in~\cite{Creminelli:2003iq, Kundu:2014gxa}, 
 inflationary three-point functions to leading order in slow-roll can then be obtained from our de Sitter four-point functions by the following procedure:
\begin{itemize}
\item First, perturb the scaling dimensions of the external fields, $\Delta = 3 \mapsto 3-\epsilon$, where $\epsilon \ll 1$ is the slow-roll parameter.

\item Second, take the soft limit of one of the external momenta (which we take to be $k_4 \to 0$), and expand the result in powers of $\epsilon$.

\end{itemize}
From the bulk point of view, this amounts to evaluating one of the external fields $\upphi$ on its time-dependent background value $\bar\upphi(t)$. For shift-symmetric inflaton interactions,\footnote{For non-derivatively coupled interactions, like $\upphi^4$, the expectation values will generically contain logarithms and not be de Sitter invariant. However, the breaking of de Sitter symmetry is mild and the violation of de Sitter symmetry is given by local terms. These cases have been analyzed carefully in \cite{Maldacena:2011nz, Bzowski:2015pba,Pajer:2016ieg,Arkani-Hamed:2018kmz}.  } this gives a three-point vertex with a coupling proportional to $\dot{\bar\upphi}$. In slow-roll inflation, this coupling is almost constant and can be related to the slow-roll parameter $\epsilon$. 
This allows us to identify the soft limit of the perturbed de Sitter four-point function with the corresponding three-point function in slow-roll inflation.

\vskip 4pt
Since the weight-shifting operator changes the weight by precisely one unit, it makes sense to perturb the conformal dimensions of the seed to $\Delta = 2-\epsilon$. To apply the weight-raising operator to the perturbed seed correlator, we need the generalization of~\eqref{equ:W} to general $\Delta$:
 \beq
{\cal W}_{ab} \equiv \eta^{MN} \left(\frac{\partial}{\partial X_a^M} + \frac{X_{a,M}}{2\Delta-1}  \frac{\partial^2 }{\partial X_a^2}\right)  \left( \frac{\partial}{\partial X_b^N} + \frac{X_{b,N}}{2\Delta-1}  \frac{\partial^2 }{\partial X_b^2}\right) . \label{equ:W2}
 \eeq
Note that, due to the transversality of the polarization tensors, all of the polarization sums \eqref{equ:Piellm} except the longitudinal component, $m=0$, vanish in the soft limit. This is easily seen from the $\hat L$ dependence of \eqref{equ:Piellm}, which vanishes as $u,\hat\alpha\to 1$ (or $v,\hat\beta\to 1$). Repeating the derivation in \S\ref{sec:RaiseW}
 for $\Delta = 2 -\epsilon$, and expanding for small $\epsilon$, we obtain the following weight-shifting operator for the longitudinal mode of the four-point function
\beq
	U_{34}^{(S,0)} \,\to\, \bar U_{34}^{(S,0)} - \epsilon \left[\frac{3-2(S-1)v\b-(4+(2S-3)\b^2)v^2}{2v^2}+\frac{S(1+v\b)}{2v} \frac{P_{S-1}(\b)}{P_S(\b)}\right] ,
\eeq
where $\bar U_{34}^{(S,0)}$ is given by \eqref{U12long}, and we have only kept the correction at linear order in $\epsilon$.

\vskip 4pt
In the soft limit $k_4\to 0$ (or $v,\b\to 1$), the unperturbed weight-raising operator $\bar U_{34}^{(S,0)}$ vanishes identically (i.e.~independent of the correlator it acts on) and we simply get
\beq
	U_{34}^{(S,0)}\, \xrightarrow{\  k_4\hs\to\hs 0\ }\, -\epsilon \, ,
\eeq
independent of spin. This means that, at order $\epsilon$, we can simply take the seed function to be that of the unperturbed dimension $\Delta=2$ operators, $\F_L^{(S)}(u,1)$, evaluated at $v=1$, and multiply it by $-\epsilon$.  
The corresponding inflationary bispectrum is then obtained by applying the operator $U_{12}^{(S,0)}$ and summing over permutations (to account for the fact that the four-point function was evaluated for the $s$-channel exchange). 
Putting everything together, the inflationary bispectrum for spin-$S$ exchange can then be written as\hskip 1pt\footnote{The bispectrum function, $B^{(S)}$, is the correlator of the sub-leading fall-off dual to $\zeta$ and is hence related to the observed bispectrum by
\beq
\langle\zeta_{\vec k_1}\zeta_{\vec k_2}\zeta_{\vec k_3}\rangle = \prod_{a=1}^3\left(\frac{H^2}{2M_{\rm pl}^2\epsilon k_a^3}\right)B^{(S)}(k_1,k_2,k_3)\,,
\eeq
where $H$ is the Hubble scale during inflation and $M_{\rm pl}$ is the (reduced) Planck mass. The prefactor arises from the shadow transform.
}
\begin{eBox}
\vskip 1pt
\beq
B^{(S)}(k_1,k_2,k_3) = -\epsilon\hs k_3^3\hs P_S(\a)\hs U_{12}^{(S,0)} \F_L^{(S)}(u,1)\, +\, {\rm perms}. \, ,
\label{equ:InfB}
\eeq
\vskip 3pt
\end{eBox}
where $U_{12}^{(S,0)}$ and $\F_L^{(S)}$ were defined in (\ref{U12long}) and (\ref{equ:FL}), respectively, and $P_S$ is a Legendre polynomial that arises from the polarization sum $\Pi_{S,0}$. This formula holds for any~$S$, generalizing the result of \cite{Arkani-Hamed:2018kmz} to all spins.

\subsection{Partially Massless Exchange}
\label{sec:pm}

As a new application of the weight-shifting technology developed in this paper, we construct the inflationary bispectra arising from the exchange of  partially massless (PM) fields. 
These PM fields are unitary spinning representations that exist on (anti) de Sitter space, but have no flat-space counterparts~\cite{Deser:1983mm,Brink:2000ag,Deser:2001us,Deser:2003gw,Deser:2013uy,deRham:2013wv,Baumann:2017jvh}. They occur at special values 
of the mass-to-Hubble ratio, 
\beq
 \frac{m^2}{H^2} = S(S-1)-T(T+1)   \, ,\label{eq:depth}
\eeq
where $S$ is the spin and $T\in\{0,\cdots\hskip -1pt,S-1\}$ is the ``depth" of the field. 
The corresponding dual operators have integer dimensions $\Delta_\sigma = 2 + T$.
At these special points, the theory possesses an additional gauge invariance that projects out modes with helicity $\leq T$.   
The tight structure of PM representations is reflected in the simplicity of their mode functions, which leads to simple analytic expressions for the associated correlation functions.

\vskip 4pt
We are interested in the inflationary three-point functions arising from the exchange of PM fields.  As we have seen above, these bispectra only depend on the longitudinal modes of the exchanged particle, which for a PM field isn't a propagating degree of freedom. This does not mean, however, that the corresponding bispectra are trivial. Higher-spin particles have a nontrivial constraint structure, where the non-propagating degrees of freedom are fixed in terms of the propagating ones. In particular, the longitudinal modes mediate Coulomb-like potentials, which lead to distinct imprints in the inflationary bispectra. In order to probe all degrees of freedom of the PM field, we would have to measure inflationary trispectra. Nonetheless, as we will see below, the Coulomb potentials generate bispectrum shapes with striking features that uniquely characterize the presence of a PM field in the spectrum.

\vskip 4pt
The general features of PM exchange are substantially simpler to describe than those of massive exchange. Recall that the longitudinal part of the $\Delta = 2$ correlator coming from spin-$S$ exchange is given by~\eqref{equ:FL}. We can write this in a slightly more illuminating form by redefining $T\equiv S-j$ to obtain 
\beq
\F_L^{(S)} = \prod_{T=0}^{S-1} \big[\Delta_u-T(T+1) \big]\hat F_{\Delta_\sigma}^{(0)}\,,
\label{eq:pmlongop}
\eeq
where we have added the weight dependence of the seed as a subscript to avoid confusion.
Notice that the differential operators appearing on the right-hand side are precisely those appearing in the differential equation describing the exchange of a PM particle of depth $T$:
\beq
 \big[\Delta_u-T(T+1)\big] \hat F_{\Delta_\sigma=2+T}^{(0)} = \hat C_0 \, .
 \label{eq:pmODE}
\eeq
It is important to emphasize that a given PM seed $\hat F_{\Delta_\sigma}^{(0)}$ will only have the right conformal weight to satisfy~\eqref{eq:pmODE} for a particular depth. Since~\eqref{eq:pmlongop} includes the left-hand-side of~\eqref{eq:pmODE} for all depths though, this means that at every PM point, one of the operators in~\eqref{eq:pmlongop} will reduce the seed exchange function to a contact solution through~\eqref{eq:pmODE}. This contact solution is then acted on by the operators in \eqref{equ:InfB}. It is for this reason that the PM-induced bispectra are so simple.

\vskip 4pt
Cleanly disentangling the effects of PM exchange is somewhat subtle. In the case of massive particle exchange, there are characteristic non-analytic features that cannot be mimicked by contact interactions and therefore are unambiguously attributed to the exchange. In the present case, however, the resulting bispectra will be rational functions of momenta, and so it is less obvious which parts to ascribe to particle exchange. In practice, this subtlety can be managed by explicitly subtracting off all possible contact solutions. After performing this subtraction, the leftover shape is a sharp signature of the exchanged PM particle.

\subsubsection*{Spin-2 exchange}

Let us first consider the special case $S=2$.
There are then two (partially) massless points with corresponding dual operators having $\Delta_\sigma = 3$ and $\Delta_\sigma=2$. The former is the ordinary graviton ($m^2=0$) and the latter is the PM state that saturates the Higuchi bound ($m^2 = 2 H^2$).
We will discuss these two cases in turn.

\begin{itemize}
\item {\it Graviton.}---For a $\Delta_\sigma =3$ state, the exchange equation~\eqref{equ:exchange} becomes $(\Delta_u-2)\F^{(0)}=\hat C_0$,
with the simplest contact term on the right-hand side.  Comparing this to the longitudinal coefficient function in (\ref{equ:FL}), we find
\beq
\F_L^{(2)}= \Delta_u(\Delta_u-2)\F^{(0)} =\hat C_1\, .
\eeq
Plugging this solution into \eqref{equ:InfB}, we obtain the corresponding bispectrum.\footnote{In \cite{Arkani-Hamed:2018kmz}, a different exchange solution was utilized, with an unphysical contact term. The goal there was to mimic the number of derivatives of the bulk Lagrangian for graviton exchange. We see here that such a choice was unnecessary; one can just as well use a more physical exchange solution as a seed.} 
We wish to isolate the part of this result that can unambiguously be attributed to massless spin-2 exchange. In other words, we want to see whether any parts of this correlator can be mimicked by contact terms, and then subtract off these pieces.

The relevant contact contributions to the bispectrum are
\beq
B_{c}= - \frac{\epsilon k_3^3}{3}\, \sum_{n} a_n U_{12}^{(0,0)} \hat C_n(u,1)  + {\rm perms.}  
\label{eq:pmcontactterms}
\eeq
To identify the parts that are degenerate with the PM-induced bispectrum, we consider the limit $k_t\equiv \sum_a k_a \to 0$. In this limit, the exchange bispectrum has a leading singularity scaling as $k_t^{-3}$. This singularity can be removed by adding the contact interaction $\hat C_1$, with $a_1 = 15$ in~\eqref{eq:pmcontactterms} (and all other coefficients zero). 
After this subtraction, the bispectrum still has a $k_t^{-1}$ singularity, which cannot be removed by another contact term. 
The part of the bispectrum that is cleanly associated with the exchange of a graviton therefore is
\beq
B_{\rm inf}= 3 \epsilon \left[\sum_{a \ne b} k_a k_b^2 +\frac{8}{k_t} \sum_{a>b}k_a^2k_b^2-3\sum_a k_a^3\right] . \label{equ:Binf}
\eeq
Up to a local term, $\sum_a k_a^3$, arising from the gauge transformation from spatially flat gauge to comoving gauge, this is precisely the famous bispectrum of single-field slow-roll inflation~\cite{Maldacena:2002vr}.

\item {\it PM graviton.}---The exchange equation for a $\Delta_\sigma = 2$ state is  
\beq
\Delta_u \F^{(0)}=\hat C_0\, .
\eeq 
 Although the exchange solution is slightly different than for the graviton, after plugging it into~\eqref{equ:InfB}, we still obtain a bispectrum with a $k_t^{-3}$ singularity. As before, we can subtract this singularity, together with the subleading $k_t^{-2}$ term, by adding a suitable choice of contact terms. In this case, we must take $a_0 = 24$ and $a_1 = 15$ in~\eqref{eq:pmcontactterms}, with all the other coefficients vanishing. 
 
 This time, however, there is no leftover bispectrum! This means that the bispectrum due to the exchange of a PM graviton can be represented completely by a certain mixture of equilateral non-Gaussianity coming from inflaton self-interactions. This, however, does not rule out the possibility that inflaton correlators with PM exchange involve transcendental functions and therefore give shapes that are non-degenerate with contact diagrams. This is because our seed functions---the four-point functions of conformally coupled scalars with PM exchange---were rational functions.
 \end{itemize}

\subsubsection*{Higher-spin exchange}

The story becomes richer for higher spins, especially because there are now multiple PM points.
Let us first describe a few qualitative features of these bispectra, before presenting details for the special cases $S=4$ and~$6$.\footnote{Since we require a coupling to two identical $\Delta = 2$ scalars in the relevant seed function, only particles of even spin can contribute to the bispectrum of an uncharged inflaton.} Acting on the PM seed function $\hat F_{\Delta_\sigma}^{(0)}$ as in~\eqref{eq:pmlongop} yields the longitudinal part of the exchange correlator, which we then feed into~\eqref{equ:InfB} to produce the inflationary bispectrum. To isolate the part which is an unambiguous signature of PM exchange, we adopt the following procedure:
\begin{itemize} 
\item The bispectrum will have a leading singularity for $k_t \to 0$ that can be removed by adding a contact interaction. In other words, part of the bispectrum shape is indistinguishable from equilateral non-Gaussianity. After removing those terms, the resulting shape will have a singularity scaling as $k_t^{1-S}$ (for any depth).

\item The exchange of a PM field of depth $T=S-2$ can be absorbed completely by a sum of contact terms, with no exchange contribution remaining.\footnote{We do not have a deep explanation for this fact, but it is interesting to speculate that it has something to do with the fact that the corresponding operator obeys a double-conservation condition~\cite{Dolan:2001ih,Brust:2016gjy,Brust:2016zns}, which somehow behaves differently from the other multiple-conservation conditions that dual PM operators satisfy.}
 These fields therefore do not produce a distinct bispectrum shape. Their imprint will only appear cleanly in the four-point function. 

\item After fixing the singularity for $k_t\to 0$, we still have the freedom to choose some contact terms to remove the least soft pieces of the bispectrum in the squeezed limit, so that
\beq
\lim_{k_3 \to 0} \langle \upphi_{\k_1}\upphi_{\k_2}\upphi_{\k_3}\rangle\simeq \frac{1}{(k_1k_3)^3} \left(\frac{k_3}{k_1}\right)^p  P_{J}(\cos \theta)+\cdots\,,
\label{eq:squeezedscaling}
\eeq
where $\theta$ is the angle between $\vec{k}_1$ and $\vec{k}_3$, and the power $p$ depends on $S$ and $T$ in a nontrivial way. This subtraction can be done without changing the overall singularity in $k_t$. If there is any freedom left, we try to remove the highest possible Legendre polynomial $P_J$. 
\end{itemize}
We now illustrate this procedure for PM fields of spin 4 and 6.

\begin{itemize}
\item {\bf Spin-4:} 
 A spin-4 field has four PM points. 
 Carrying out the procedure described above at each of them, we are able to pick various contact terms in~\eqref{eq:pmcontactterms} to isolate the parts of the inflationary bispectra due to the exchange of PM fields. We tabulate the necessary contact term coefficients and the resulting scalings in the squeezed limit~\eqref{eq:squeezedscaling} below:
\begin{center}
\begin{tabular}{cccccc}
\toprule
$T$&$a_0$&$a_1$&$a_2$&$a_3$&$(p\hskip 1pt, J)$\\
\midrule
$0$ & 0 & 0 & 2394 & 63& (2\hskip 1pt, \hskip -2pt 4) \\
$1$ & $6880$ & $9128$ & 2178 & 63 & (3\hskip 1pt, \hskip -2pt 4)\\
$2$ & 4320 & 7128 & 1746 & 63  & --- \\
$3$ & 2880 & 3708 & 1098 & 63  & (3\hskip 1pt, \hskip -2pt 0)\\
\bottomrule
\end{tabular}
\end{center}
We see that there is a rich structure of scalings and angular dependences of the final inflationary bispectra. In particular, the depth-2 PM point has {\it no} unambiguous signature in the inflationary bispectrum. The other PM points lead to distinct behaviors in the squeezed limit.

\item {\bf Spin-6:} Finally, we consider the exchange of a spin-6 field, with five PM points. 
A novelty of this situation is that, to perform the subtraction, we must introduce  contact terms arising from integrating out intermediate spin-2 particles:
\beq
B_{c}= - \epsilon k_3^3\, \sum_{n}  b_n P_2 (\hat \alpha) \hskip 1pt U_{12}^{(2,0)} \Big[(\Delta_u-2)\Delta_u\hat C_n(u,1)\Big] + {\rm perms.} \, , 
\label{eq:pmcontactterms2}
\eeq
where $\hat \alpha = (k_1-k_2)/k_3$.  
The coefficients  $a_n$ and $b_n$ in~\eqref{eq:pmcontactterms} and \eqref{eq:pmcontactterms2} are fixed by the same requirements as above. 
Their precise values are not very illuminating, so we don't display them.
Instead, we present the final squeezed-limit scalings and angular dependences of the inflationary bispectra:
\begin{center}
\begin{tabular}{cccccccc}
\toprule
$T$ & \vrule &$0$ & $1$ & $2$ & $3$ & $4$ & $5$\\
\midrule
$(p\hskip 1pt, \hskip -1pt J)$ & \vrule &$(2\hskip 1pt, \hskip -1pt 6)$ & $(3\hskip 1pt, \hskip -1pt 6)$ & $(4\hskip 1pt, \hskip -1pt 6)$ & $(4\hskip 1pt, \hskip -1pt 2)$ & --- & $(4\hskip 1pt, \hskip -1pt 2)$\\
\bottomrule
\end{tabular}
\end{center}
Again, this displays an interesting range of scalings, with the depth-4 point being degenerate with a set of contact interactions.
\end{itemize}
In summary, just like the famous graviton-induced bispectrum~\eqref{equ:Binf}, PM fields generate new, distinct shapes of non-Gaussianity, which can be written as polynomials in the momenta divided by some overall power of the total energy $k_t$. These shapes uniquely characterize the presence of PM fields of various depths in the early universe. Whether consistent interacting theories of those PM fields (beyond gravity and gauge theory) exist remains an open problem (but see \cite{Joung:2019wwf, Boulanger:2019zic}).

\newpage
\section{Conclusions}
\label{sec:Conclusions}

In this paper, we have presented a highly streamlined and much more universal derivation of the spin-raising and weight-shifting operators that appeared in the original work on the cosmological bootstrap~\cite{Arkani-Hamed:2018kmz}.  This was facilitated by the existence of a corresponding set of weight-shifting operators in conformal field theory~\cite{Dirac:1936fq, Costa:2011mg, Karateev:2017jgd}, which when Fourier transformed can be applied in the cosmological context.  Our treatment highlights the power and elegance of the bootstrap method by making manifest that a single seed function (corresponding to the correlator of conformally coupled scalars) can be transformed into all correlators of interest through the application of only two simple spin-raising and weight-raising operators. This provides explicit results for inflationary bispectra arising from tree-level exchange of massive particles of arbitrary spin. The more systematic approach has also allowed us to characterize the effects from the exchange of (partially) massless fields of arbitrary spin, whose signatures have a rich structure with some surprising features. 

\vskip 4pt
While we have restricted our application of the weight-shifting technology to reproducing and generalizing the operators found in~\cite{Arkani-Hamed:2018kmz}, many new applications have now opened up:
\begin{itemize}

\item It is straightforward to use our formalism to raise the spin of the external fields. In that case, the different tensor structures arise from the unique scalar seed because there is more than one way to raise the spin and weight of the external fields. These different ways lead to distinct answers which can be combined into the known tensor structures of spinning correlators.

\item An interesting special case is correlators involving conserved tensors. For example, the stress tensor is dual to a bulk graviton and therefore part of any inflationary model. Correlation functions of conserved tensors a further constrained, because in addition to the conformal Ward identities discussed here, they must obey the Ward--Takahashi identities associated with current conservation. The interplay between these two differential constraints underlies the rich structure present for correlation functions of conserved operators. The weight-shifting formalism is a powerful way to  study these correlation functions, and allows for a systematic classification similar to the one provided here for scalars.

\item When the exchanged particles are massless, the spin-raised correlators are not guaranteed to be local. In flat space, the requirement of local four-point interactions is a powerful constraint on the space of consistent interactions between massless particles with spin~\cite{Benincasa:2007xk, McGady:2013sga}, making almost all theories, other than the familiar gauge theories and gravity, inconsistent.  Similar considerations should restrict the theory space of viable interactions of massless fields in de Sitter space.
\end{itemize} 
The tools that we have developed in this paper provide the first steps toward unraveling the intricate web of relations between these directions, and we will
present our findings on these issues in a separate publication~\cite{CosmoBoot2}.

\vspace{0.5cm}
\paragraph{Acknowledgements} 
We are grateful to Hiroshi Isono, Denis Karateev, Petr Kravchuk, Toshifumi Noumi, Charlotte Sleight, Massimo Taronna, and Sasha Zhiboedov for discussions, and to David Simmons-Duffin for his patient tutoring in the weight-shifting formalism. We are also especially grateful to Nima Arkani-Hamed for encouragement, many fruitful discussions and collaboration. DB~and CDP are supported by a Vidi grant of the Netherlands Organisation for Scientific Research~(NWO) that is funded by the Dutch Ministry of Education, Culture and Science~(OCW). AJ is supported in part by NASA grant NNX16AB27G. HL is supported by NSF grant AST-1813694 and DOE grant DE-SC0019018. GP is supported by the European Union's Horizon 2020 research and innovation programme under the Marie-Sklodowska Curie grant agreement number 751778. The work of DB and GP is part of the Delta-ITP consortium.
DB, AJ, and GP thank the theory group at Harvard for its hospitality while this work was being completed. AJ thanks the Delta-ITP and the University of Amsterdam for hospitality while part of this work was completed.

\newpage
\appendix
\section{Weight Shifting in Embedding Space}
\label{app:WeightEmbedding}

In the main text, we have utilized the spin-raising operator ${\cal S}_{12}$,~cf.~\eqref{equ:o12}, and the weight-raising operator ${\cal W}_{12}$,~cf.~\eqref{equ:W}.
In this appendix, we wish to place these operators in a broader context and describe how they (and other useful operators) arise naturally from more formal conformal representation theory considerations. Our discussion is meant to be self-contained, although we do assume some familiarity with basic aspects of conformal field theory (see e.g.~\cite{Rychkov:2016iqz,Simmons-Duffin:2016gjk}). For more details, readers are encouraged to consult~\cite{Karateev:2017jgd}.

\subsection{General Preliminaries}
\label{sec:generalWSstuff}

The general philosophy is fairly simple to state: given a solution to the conformal Ward identities~\eqref{WI_D} and \eqref{WI_SCT}, we would like to find differential operators that act on this solution to generate new solutions with different quantum numbers (either conformal weight or spin). 

\vskip 4pt
The most natural thing to look for would be a {\it conformally-invariant} operator that accomplishes this. Such operators do exist, but only in very special situations~\cite{Erramilli:2019njx}. We can see this by considering the action of a putative operator, ${\cal D}$, on a conformal primary of weight $\Delta$ and spin~$S$, which would be of the form
\beq
{\cal D} O_\Delta^{(S)} = \tilde O_{\Delta'}^{(S')}\,,
\label{eq:confinfD}
\eeq
where the operator $\tilde O_{\Delta'}^{(S')}$ transforms in some new representation of the conformal group, with weight $\Delta'$ and spin $S'$. Asking ${\cal D}$ to be conformally invariant is a strong constraint. First of all, it implies that the operator commutes with all conformal generators. In particular, this means that the operator is translationally invariant. It therefore cannot depend on coordinates and a differential operator with $n$ derivatives must then change the weight as follows
\beq
\Delta \mapsto\Delta' = \Delta +n\,.
\eeq
Second, if the operator is conformally invariant it cannot change the quadratic Casimir eigenvalue of the representation, and therefore must satisfy
\beq
\Delta (\Delta -d) +{\cal C}_2(S) = (\Delta +n)(\Delta +n-d)+{\cal C}_2(S')\,,
\label{eq:casimirmatch}
\eeq
where we have substituted for $\Delta'$ on the right-hand side and ${\cal C}_2(S)$ denotes the ${\rm SO}(d)$ quadratic Casimir of the spin-$S$ representation.\footnote{More generally, any SO$(d)$ representations can appear in~\eqref{eq:confinfD}. This does not change the argument. In what follows, we will often restrict to spin-$S$ representations for notational simplicity, but nothing depends on this choice.} Their precise values are not important; the point is that if we fix the spin representations that the operator ${\cal D}$ maps between, the constraint~\eqref{eq:casimirmatch} becomes a linear equation for $\Delta$, which is only solved for one specific value.\footnote{An example is that $\partial_\mu J^\mu$ transforms like a scalar, but only for $\Delta_J = d-1$.} This means that---given a particular target weight and spin representation---we are generically unable to find a conformally-invariant differential operator that maps us there from a given starting point.

\vskip 4pt
The loophole to this argument is fairly intuitive, we must relax our requirements and search instead for differential operators that themselves transform in some representation of the conformal group. We therefore are in search of {\it conformally-covariant} (as opposed to invariant) differential operators. The construction of such weight-shifting operators was performed systematically in~\cite{Karateev:2017jgd}, and here we wish to review their construction. Special cases of weight-shifting operators were constructed earlier in~\cite{Costa:2011dw}.

\subsection{Some Group Theory}
Some representation-theoretic considerations are helpful in order to understand what weight-shifting operators should exist. These details are not essential and readers who are willing to take the existence and transformation properties of weight-shifting operators on faith can skip to the next subsection to see them constructed explicitly.

\vskip 4pt
It is useful to consider conformally-covariant differential operators that transform in finite-dimensional representations of the conformal algebra.\footnote{Note that these finite-dimensional representations are not unitary---as they represent a non-compact algebra---but this does not affect their usefulness. These representations should be thought of as a tool to generate operators with the kinematic transformation properties we desire. Imposing unitarity of the final results will require additional information beyond kinematics that will have to be input at a later time in some situations.} The content of these finite-dimensional representations, $W$, can be understood by thinking of them as analytically continued ${\rm SO}(d+2)$ representations; they decompose under dilations\,$\times\,{\rm SO}(d)$ as a direct sum
\beq
W = \bigoplus_{i=-j}^j W_i \,,
\label{eq:degenstates}
\eeq
where the dilation eigenvalue of the subspaces $W_i$ runs from $-j$ to $j$  and is analogous to the spin quantum number, $J_z$, for SO(3) representations.

\vskip 4pt
We now consider the tensor product of $W$ with another irreducible conformal representation, which we denote by $V_{\Delta, S}$, where $\Delta, S$ label the weight and spin of the lowest-weight state. The goal is to decompose the tensor product $W\otimes V_{\Delta, S}$ into irreducible representations of the conformal group. This is equivalent to finding the conformal primary operators that appear in the product representation. A straightforward way to accomplish this is to use the state-operator correspondence and think of the representation~$W$ as being generated by the conformal primary operator $w_{-j}^{(S_w)}(0)$ with weight $\Delta_w = -j$ and spin $S_w$ and think of the representation $V_{\Delta, S}$ as being generated by the conformal primary $O_\Delta^{(S)}(0)$. Since $W$ is finite-dimensional, $w_{-j}^{(S_w)}(0)$ will have a finite number of descendants---we can only take so many derivatives before the resulting states are guaranteed to be null.

\vskip 4pt
Primary operators appearing in $W\otimes V_{\Delta, S}$ can be constructed by considering products of $w$ and $O$ of the form~\cite{Karateev:2017jgd}
\beq
\tilde O_{\Delta'}^{S'}= \partial_{i_1}\cdots\partial_{i_m}w_{-j}^{(S_w)}(0) \otimes O_\Delta^{(S)}(0) +c_1\,\partial_{i_1}\cdots\partial_{i_{m-1}}w_{-j}^{(S_w)}(0) \otimes \partial_{i_{m}}O_\Delta^{(S)}(0)+\cdots\, ,
\label{eq:tensordecomp}
\eeq
where $m$ can range from $0$ to $2j$ and the ellipses denote all other ways of distributing the $m$ derivatives on both operators. The various coefficients can be fixed uniquely by demanding that the expression~\eqref{eq:tensordecomp} is a primary operator, i.e.~that it is annihilated by the special conformal generator. We are being somewhat schematic about the spin representations, but the spin representations appearing in~\eqref{eq:tensordecomp} should also be decomposed into irreducible components. Note that this involves both the indices carried by the derivatives as well as possible ${\rm SO}(d)$ indices of the operators $w$ and $O$.

\vskip 4pt
Each of the resulting primary operators constructed in this way will have weight $\Delta' = \Delta - j+m$ and will transform in some definite ${\rm SO}(d)$ representation. Once we have decomposed~\eqref{eq:tensordecomp} like this, we can interpret the result as a differential operator acting on $O_\Delta^{(S)}$ by recalling that $w$ transforms in a finite-dimensional representation of the conformal group, which therefore has a finite basis, $e^A$. We can write~\eqref{eq:tensordecomp} as
\beq
\begin{aligned}
\tilde O_{\Delta'}^{S'} &= e^A\otimes\left(\partial_{i_1}\cdots\partial_{i_m}w_A(0) \otimes O_\Delta^{(S)}(0) +c_1\,\partial_{i_1}\cdots\partial_{i_{m-1}}w_A(0) \otimes \partial_{i_{m}}O_\Delta^{(S)}(0)+\cdots\right)\\
& \equiv e^A \otimes {\cal D}_A O_\Delta^{(S)}\,.
\end{aligned}
\label{eq:weightshiftingopdef}
\eeq
The operator ${\cal D}_A$ is a weight-shifting operator: it changes the conformal weight and/or spin representation of the operator $O$ and transforms in a finite-dimensional representation of the conformal group.

\vskip 4pt
The preceding discussion was very abstract, so we now proceed to construct some particularly useful weight-shifting operators explicitly. It should be noted, however, that there are, in principle, an infinite number of such operators. Any finite-dimensional representation of the conformal group can be used to construct a set of weight-shifting operators. It could be that some more exotic possibilities are also of use in cosmology.

\subsection{Vector Representation}\label{app:WSvec}
One of the most important sets of weight-shifting operators arises from the finite-dimensional vector representation of the conformal group, $W_M$ (which has $j=1$), where $M$ is an embedding space index with $d+2$ components. Decomposing this ${\rm SO}(d+1,1)$ representation into ${\rm SO}(1,1)\times{\rm SO}(d)$ representations as in~\eqref{eq:degenstates}, we find 
\beq
\gyoung(;)
\xrightarrow[{\scriptscriptstyle {\rm SO}(1,1)\times{\rm SO}(d)}]{}
\big(\bullet\big)_{-1}\raisebox{0.325ex}{\medoplus}~
\big(\,
\gyoung(;)
\,\big)_{0}\,
\raisebox{0.325ex}{\medoplus} ~\big(\bullet\big)_{1} \,,
\eeq
where $\bullet$ denotes the trivial spin representation, and the subscripts are the dilation weight.\footnote{The dilation weights of the various operators can be obtained by noting that the representation must shorten at the second level of descendents, with the spin-2 state becoming null. For a discussion of the shortening of conformal representations, see~\cite{Penedones:2015aga}.} This is just a pictorial way of representing the information that a $d+1+1$ split of the vector $W_M$
contains two scalars, $W_{-1}$ and $W_{0}$, and one $d$-dimensional vector, $W_i$. Equivalently, the lowest-weight state in the representation is a scalar, $w_{-1}(0)$, with weight $\Delta_w = -1$. This scalar  satisfies the differential equation 
\beq
\partial_{(i}\partial_{j)_T} w_{-1}(0) = 0\,,
\label{eq:vecprimdiffeq}
\eeq
where the notation $(\cdots)_T$ denotes the symmetric trace-free part.

\vskip 4pt
By considering products of these operators with a primary of interest, like in~\eqref{eq:weightshiftingopdef},
we can see what kind of weight-shifting operators they correspond to.

\begin{itemize}
\item {\bf Weight $\boldsymbol{-1}$:}
The lowest-weight state has weight $\Delta=-1$. The corresponding weight-shifting operator will shift the weight of the representation by $\Delta\mapsto \Delta-1$. This state is a scalar under SO$(d)$, so it does not change the spin representation.

\item {\bf Weight $\boldsymbol{0}$:}
The first descendent state transforms as a vector under SO$(d)$ and has weight $\Delta = 0$. The weight-shifting operators  associated with this state leave the weight invariant, but change the spin representation. To see how the spin for example of a spin-$S$ representation is changed, we decompose the tensor product
\beq
\gyoung(_4S)^{\,T}\,\raisebox{0.35ex}{\medotimes}~
\gyoung(;) \,=\, 
\gyoung(_3{S-1})^{\,T}\,\raisebox{0.35ex}{\medoplus}~
\gyoung(_5{S+1})^{\,T}
\,\raisebox{0.35ex}{\medoplus}~
\raisebox{1ex}{\gyoung(_4{S},;)}^{\,T} \, .
\label{eq:spinplusboxyoung}
\eeq
Acting on spin-$S$ representations there are therefore three weight-shifting operators: one which shifts the spin down by one unit, one that shifts the spin up by one unit, and a third operator that projects onto a mixed-symmetry representation.\footnote{We will not utilize weight-shifting operators that generate mixed-symmetry representations, because our principal interest is in $d=3$ dimensions, where mixed-symmetry representations can all be dualized to symmetric tensor representations. However, these operators may be useful for some calculations.} (For more general spin representations the story is similar, but the details can be more complicated.)

\item {\bf Weight $\boldsymbol{1}$:} Finally, the highest-weight state is a scalar with weight $\Delta =1$ and the corresponding weight-shifting operator will
shift the weight of the representation by $\Delta\mapsto \Delta+1$. 
\end{itemize}
\subsubsection*{Operators in embedding space}
The vector representation therefore gives rise to a set of differential operators that
shift the dimension and spin of  CFT operators $O(X,Z)$, which we denote by
\beq
\cD^{\alpha \beta}_M : [\De,S] \mapsto [\De+\alpha,S+\beta]\,.
\eeq
In principle, these operators can be constructed by solving the differential equation~\eqref{eq:vecprimdiffeq} and then building states of the form~\eqref{eq:weightshiftingopdef}. However, in practice, the formal representation-theoretic arguments we have just reviewed are more useful to catalog which weight-shifting operators should exist, and then construct them directly in embedding space.

\vskip 4pt
The benefit of working in embedding space is that it is easy to implement the correct conformal transformation properties of the weight-shifting operators. For example, weight-shifting operators coming from the vector representation will carry an uncontracted embedding space index. It is then algorithmic to construct weight-shifting operators: the above list completely catalogs the possible operators, our task is to find embedding space expressions with the correct conformal dimensions and spin weights. 

\vskip 4pt
One additional subtlety needs to be addressed: we must ensure that the resulting expressions preserve the Euclidean section of the projective lightcone. This is not automatic because embedding space tensors are invariant under the shift $Z\mapsto Z+\beta X$ and subject to the constraints
\beq
X^2 = X\cdot Z=Z^2 = 0\,.
\label{eq:lightconeconstraints} 
\eeq
We must then make sure that these combined constraints are preserved by the differential operators we construct.
This turns out to be a strong enough requirement to uniquely fix the embedding space expressions for each weight-shifting operator.

\vskip 4pt
The weight-shifting operators in the vector representation were constructed explicitly in Appendix C of~\cite{Karateev:2017jgd}. Here, we summarize their results.  The algorithm involves first making an ansatz that has the correct weights:\hs\footnote{Note that $Z$ raises the spin by one, $X$ lowers the weight by one and derivatives do the opposite. The action of the operators $X\cdot\partial_X = -\Delta$ and $Z\cdot\partial_Z = S$ can be absorbed into the constants, while $X\cdot\partial_Z= 0$ is one of the constraints we will impose, so this operator does not appear.}
\be
\label{eq:vectoroperators}
\cD_M^{-0} &= X_M\,,\\
\cD_M^{0-} &= a_1\frac{\partial}{\partial Z^M}+a_2Z_M \frac{\partial^2}{\partial Z^2}+a_3X_M\frac{\partial^2}{\partial X\cdot \partial Z}+a_4 X_M Z\cdot \frac{\partial}{\partial X}\frac{\partial^2}{\partial Z^2}\, ,
\\
\cD_M^{0+} &= b_1Z_M+ b_2X_M Z\cdot\pdr{}{X}\,, \label{eq:vectoroperators3}\\
\cD_M^{+0} &= c_1\pdr{}{X^M} + c_2 X_M \frac{\ptl^2}{\ptl X^2} + c_3 Z_M \frac{\ptl^2}{\ptl Z\cdot\ptl X} + c_4 Z\cdot\pdr{}{X} \pdr{}{Z^M} \\
&\quad + c_5 X_M Z\cdot \pdr{}{X}\frac{\ptl^2}{\ptl Z\cdot\ptl X} + c_6 Z_M Z\cdot\pdr{}{X} \frac{\ptl^2}{\ptl Z^2} + c_7 X_M \p{Z\cdot\pdr{}{X}}^2 \frac{\ptl^2}{\ptl Z^2} \, . \nn
\ee
We now wish to fix the various coefficients in the operators above. First, we require that the action of the weight-shifting operators preserves invariance under the shift $Z \mapsto Z+\beta X$.
This implies
\beq
\left(X\cdot\frac{\partial}{\partial Z}\right){\cal D}_M f_{\Delta, S}(X,Z)= 0\,,
\label{eq:gaugepreservation}
\eeq
where $ f_{\Delta, S}$ is any homogeneous polynomial of spin $S$ and weight $\Delta$ that is invariant under the shift of $Z$~\cite{Costa:2011mg,Karateev:2017jgd}. The latter can be written as
\beq
 f_{\Delta, S}(X,Z) = (X\cdot Y)^{-\Delta-S}\left( P^MQ^NC_{MN}\right)^S \,,
\eeq
where $C_{MN} = Z_M X_N - Z_N X_M$ and $Y, P, Q$ are arbitrary constant vectors. 
Second, we also have to ensure that the constraints~\eqref{eq:lightconeconstraints} are preserved. This is somewhat tricker to enforce because there are many possible ways to extend the operators away from the lightcone. In practice, we only need to impose this constraint on a small number of polynomials in order to completely fix all of the free coefficients. 
 The following set of polynomials, which all have weight $\Delta$ and spin $S$, does the job~\cite{Karateev:2017jgd}
\be
g_1(X,Z) &= X^2 f_{\Delta+2,S}(X,Z)\,,\\
g_2(X,Z) &= S^MX^N C_{MN} f_{\Delta+2,S-1}(X,Z)\,,\\
g_3(X,Z) &= Z^MX^N C_{MN} f_{\Delta+2,S-2}(X,Z)\,,\\
g_4(X,Z) &= S^MS_N C_{MO}C^{ON} f_{\Delta+2,S-2}(X,Z)\,,
\ee
where $S$ is an arbitrary constant vector.
It is straightforward to check that these polynomials all vanish after imposing~\eqref{eq:lightconeconstraints}. Finally, we act with the weight-shifting operators {\it before} imposing the constraints, then impose the constraints and demand that the result vanishes. This completely fixes all the remaining free coefficients and we find 
\be
\label{eq:vectoroperators}
\cD_M^{-0} &= X_M\,,\\
\cD_M^{0-} &=
\p{(\De-d + 2 - S)\de_M^N + X_M \pdr{}{X_N}} \p{(d - 4 + 2 S)  \pdr{}{Z^N}
-Z_N \frac{\ptl^2}{\ptl Z^2}},
\\
\cD_M^{0+} &= (S+\De)Z_M+ X_M Z\cdot\pdr{}{X}\,, \label{eq:vectoroperators3}\\
\cD_M^{+0} &= c_1\pdr{}{X^M} + c_2 X_M \frac{\ptl^2}{\ptl X^2} + c_3 Z_M \frac{\ptl^2}{\ptl Z\cdot\ptl X} + c_4 Z\cdot\pdr{}{X} \pdr{}{Z^M} \label{eq:vectoroperators4} \\
&\quad + c_5 X_M Z\cdot \pdr{}{X}\frac{\ptl^2}{\ptl Z\cdot\ptl X} + c_6 Z_M Z\cdot\pdr{}{X} \frac{\ptl^2}{\ptl Z^2} + c_7 X_M \p{Z\cdot\pdr{}{X}}^2 \frac{\ptl^2}{\ptl Z^2} \,, \nn
\ee
where the coefficients $c_n$ in (\ref{eq:vectoroperators4}) are given by
\beq
\begin{aligned}
c_1 &= \left(\frac{d}{2}-\Delta - 1\right)(\Delta+S-1)(d-\Delta+S-2)\, , \qquad & c_5 &=\frac{d}{2}+S-2\, ,\\
c_2 &=-\frac{1}{2}(\Delta+S-1)(d-\Delta+S-2) \, , & c_6 &=\frac{d}{2}-\Delta-1\,,\\
c_3 &=-\left(\frac{d}{2}-\Delta-1\right)(\Delta+S-2)\, , & c_7 &=-\frac{1}{2}\,. \\
c_4 &=-\left(\frac{d}{2}-\Delta-1\right)(d-\Delta+S-2) \,,
\end{aligned}
\eeq
The operators defined in this way preserve the Euclidean section of the projective lightcone and  shift the weights of the operators they act on (as indicated by the superscripts of ${\cal D}_M^{\alpha \beta}$).

\subsubsection*{Bi-local operators}

The weight-shifting operators constructed in this way satisfy all of our requirements---they are conformally covariant and change the quantum numbers of representations---but it is natural to ask if we can do better. It would be preferable to have some objects that are actually conformally invariant. From the arguments in \S\ref{sec:generalWSstuff}, we don't expect to be able to accomplish this with operators that act at a single point, but nothing prevents us from combining pairs of weight-shifting operators into singlets that are bi-local. In particular, it is useful to introduce the following combinations of operators that act at two points (labeled by $1$ and $2$) by contracting their embedding space indices
\be
{\cal S}_{12}^{++} \equiv \cD_{1}^{0+}  \cdot \cD_{2}^{0+}\, , \label{equ:S+}\\
{\cal S}_{12}^{--} \equiv \cD_{1}^{0-}  \cdot \cD_{2}^{0-} \, , \label{equ:S-} \\
{\cal W}_{12}^{++} \equiv \cD_{1}^{+0}  \cdot \cD_{2}^{+0}\, , \label{equ:W+} \\
{\cal W}_{12}^{--} \equiv \cD_{1}^{-0}  \cdot \cD_{2}^{-0} \, . \label{equ:W-}
\ee
Defined in this way, the operators ${\cal S}_{12}^{\pm \pm}$ raise (lower) the spins at points 1 and 2, while the operators ${\cal W}_{12}^{\pm \pm}$ raise (lower) the dimensions. In the main text, we have used the operator ${\cal W}_{12}^{++} \equiv {\cal W}_{12}$ to raise the weight of external fields from $\Delta =2$ to $\Delta =3$.

\vskip 4pt
The operators ${\cal S}_{12}^{\pm \pm}$ and ${\cal W}_{12}^{\pm \pm}$ act in the same way at both points, but nothing requires us to combine the operators in this way. In fact, an extremely useful combination is the operator
\beq
\begin{aligned}
\label{eq:D12op}
D_{12} \equiv {\cal D}_1^{0+}\cdot {\cal D}_2^{-0} &= (\Delta_1+S_1) Z_1\cdot X_2+(X_1\cdot X_2) Z_1\cdot\frac{\partial}{\partial X_1}\\
&=-(Z_1\cdot X_2)\,X_a\cdot\frac{\partial}{\partial X_1}+(Z_1\cdot X_2)\,Z_1\cdot\frac{\partial}{\partial Z_1}+(X_1\cdot X_2) Z_1\cdot\frac{\partial}{\partial X_1},
\end{aligned}
\eeq
which raises the spin by one unit at point $1$ and lowers the dimension by one at point $2$. This operator was first derived in~\cite{Costa:2011dw}. They also constructed an operator that raises the spin and lowers the weight at the same point, but to construct this operator we have to consider the adjoint representation, which we do next.

\subsection{Adjoint Representation}
We can repeat the same construction with any finite-dimensional representation of the conformal group. Another particular useful example is the adjoint representation, which is the anti-symmetric tensor representation of ${\rm SO}(d+1,1)$; under ${\rm SO}(1,1)\!\times{\rm SO}(d)$, it decomposes as
\beq
\raisebox{1ex}{\gyoung(;,;)}
\xrightarrow[{\scriptscriptstyle {\rm SO}(1,1)\times{\rm SO}(d)}]{}
\big(\,\gyoung(;)\,\big)_{-1}\,\raisebox{0.3ex}{\medoplus}~
\left(\,\bullet~\raisebox{0.3ex}{\medoplus}~
\raisebox{1ex}{\gyoung(;,;)}
~\right)_{0}\,
\raisebox{0.3ex}{\medoplus}~
\big(\,\gyoung(;)\,\big)_{1}\, .
\eeq
We see that the adjoint representation has $j=-1$ and is generated by a vector primary state, $w_{-1}^i(0)$. In this case we can again employ representation-theoretic arguments to count the weight-shifting operators that should exist. We will not be quite as systematic as we were for the vector representation, but rather we will just construct some useful operators.
\begin{itemize}
\item {\bf Weight $\boldsymbol{-1}$:} The lowest-weight state is a vector, so the degenerate primary, $w^i$, carries a SO$(d)$ vector index. Acting on a symmetric tensor representation, the spin decomposition is the same as in~\eqref{eq:spinplusboxyoung} and we should be able to find weight-shifting operators that lower the weight by one unit and either raise or lower the spin by one unit. In addition, there is an operator that projects onto a mixed-symmetry representation, but we will not utilize it.

\item {\bf Weight $\boldsymbol{0}$:} At the first level of descendants there are two kinds of weight-shifting operators. The first kind is associated to the trivial spin representation and does nothing to either the spin or the conformal weight. These operators are actually exactly the conformal generators: they are conformally-covariant differential operators that transform in the adjoint and keep us in the same representation.
The second kind of weight-shifting operator at this level doesn't change the conformal dimension, but can map us to spin representations appearing in the decomposition
\beq
\gyoung(_4S)~\raisebox{0.325ex}{\medotimes}~
\raisebox{1ex}{\gyoung(;,;)}\ .
\eeq
We will not use these operators because only mixed-symmetry tensors appear in this decomposition.

\item {\bf Weight $\boldsymbol{1}$:} The story at weight 1 is similar to the case at weight $-1$: there are three kinds of operators. Two of them raise the dimension by one unit and either spin-up or spin-down a symmetric tensor operator. The third operator again maps a symmetric tensor to a mixed-symmetry representation.
\end{itemize}

\subsubsection*{Operators in embedding space}
Now that we know what type of weight-shifting operators should exist, we can search for them in embedding space like we did for the vector representation. The difference is that in this case the operators will carry a pair of anti-symmetric embedding space indices, as they must transform in the adjoint. In practice, this means that we can construct the adjoint embedding space operators by multiplying and anti-symmetrizing the vector embedding space operators.

\vskip 4pt
Rather than systematically derive all possible operators, like we did for the vector representation, we instead only quote some operators which are particularly useful.
For example, the operator that lowers the weight by one and raises the spin by one is 
\beq
{\cal D}_{MN}^{-+} = X_{[M}Z_{N]}\,.
\eeq
Another semi-trivial example is provided by the conformal generators
\beq
{\cal D}_{MN}^{00} = J_{MN} = X_{[M}\frac{\partial}{\partial X^{N]}}+ Z_{[M}\frac{\partial}{\partial Z^{N]}}\,,
\eeq
which by itself do not do anything to correlation functions if summed over all points, but can be useful in tandem with one of the other weight-shifting operators. 

\vskip 4pt
There are three other weight-shifting operators that can be built from the adjoint representation, which we do not construct explicitly because they are not needed in our analysis. The operators act in the following way: one operator lowers both the weight and spin at a point by one unit, one operator raises both the weight and spin at a point by one unit and finally another operator raises the weight by one unit and lowers the spin by one unit. If desired, these operators can be constructed by antisymmetrizing various combinations of the operators in the vector representation.

\subsubsection*{Bi-local operators}

As before, these operators are most useful when paired together into conformally-invariant combinations. For example, a convenient combination is
\beq
H_{12} \equiv {\cal D}_1^{-+}\cdot {\cal D}_2^{-+} = (Z_1\cdot Z_2)(X_1\cdot X_2)-(Z_1\cdot X_2)(Z_2\cdot X_1)\,,
\label{equ:H12X}
\eeq
which lowers the weight and raises the spin by one unit at both points $1$ and $2$.
Another very useful operator is 
\be
D_{11} &\equiv {\cal D}_1^{-+}\cdot J_2  \label{equ:D11X} \\\nn
&= (X_1\cdot X_2) Z_1\cdot\frac{\partial}{\partial X_2}-(X_2\cdot Z_1)\,X_1\cdot\frac{\partial}{\partial X_2}+(X_1\cdot Z_2) Z_1\cdot\frac{\partial}{\partial Z_2}-(Z_2\cdot Z_1)\,X_1\cdot\frac{\partial}{\partial Z_2}\,,
\ee
which raises the spin and lowers the weight by one unit at point $1$, but does nothing at point $2$. This operator was also considered in~\cite{Costa:2011dw}.

\newpage
\section{Weight Shifting in Fourier Space}
\label{app:WeightFourier}
Cosmological correlation functions naturally live in Fourier space, so we would like to understand the action of the bi-local weight-shifting operators in Fourier space. This involves first projecting the weight-shifting operators from embedding space to position space, and then transforming the result to Fourier space. In this appendix, we will give details of the weight-shifting operators in Fourier space. We will present results that are valid for general dimensions.

\subsection{Projection to Position Space}
We first consider projecting the embedding space weight-shifting operators to the Euclidean section of the lightcone.
Using $X^M=(1,x^2,x^{i})$ and $Z^M=(0,2\hskip 1pt \x\cdot \z,z^{i})$, we can write the derivatives with respect to the embedding coordinates as
\be
\pdr{}{X^M} &=\frac{\partial x^{i}}{\partial X^M}\frac{\partial}{\partial x^{i}}+\frac{\partial z^{i}}{\partial X^M}\frac{\partial}{\partial z^{i}}= \left( -\Delta -x^{j}\frac{\partial}{\partial x^{j}}\,,\,0\,,\,\frac{\partial}{\partial x^{i}} \right) , \\
\pdr{}{Z^M} &=\frac{\partial x^{i}}{\partial Z^M}\frac{\partial}{\partial x^{i}}+\frac{\partial z^{i}}{\partial Z^M}\frac{\partial}{\partial z^{i}}\, = \left( -x^{j}\pdr{}{z^{j}}\,,\,0\,,\,\pdr{}{z^{i}} \right) .
\ee
This allows us to write all relevant scalar products in 
 embedding space and scalar products in position space:
\beq
\begin{aligned}
Z_a \cdot Z_b &= \z_a\cdot \z_b \, ,&\qquad\quad
Z_a\cdot \pdr{}{Z_b} &= \z_a\cdot \pdr{}{\z_b}\, , \\ 
X_a\cdot Z_b &=(\x_a- \x_b)\cdot \z_b \, ,&
Z_a\cdot\pdr{}{X_b} &= \z_a\cdot\pdr{}{\x_b}  \, ,\\
 X_a\cdot X_b &= - \frac{(x_a-x_b)^2}{2}  \, , &
X_a\cdot\pdr{}{X_b} &=-\Delta_b +(\x_a- \x_b)\cdot\pdr{}{\x_b}\, , \\
\pdr{}{Z_a}\cdot\pdr{}{Z_b} &= \pdr{}{\z_a}\cdot\pdr{}{\z_b}\, , &
X_a\cdot \pdr{}{Z_b} &= (\x_a-\x_b)\cdot \pdr{}{\z_b} \, , \\
\pdr{}{X_a}\cdot\pdr{}{X_b}&=
\pdr{}{\x_a}\cdot\pdr{}{\x_b} \, ,
\end{aligned}
\eeq
where $a,b$ label different positions. We will not write out the full position space expressions for the weight-shifting operators, but they can be obtained straightforwardly by substituting these scalar products into the embedding space expressions given in Appendix~\ref{app:WeightEmbedding}.

\subsection{Fourier-Transformed Operators}\label{app:WSfourier}
Once the weight-shifting operators have been written in position space, it is an algorithmic (though tedious!) task to transform these operators to Fourier space. In the following, we record the Fourier space expressions for a variety of useful weight-shifting operators that act on pairs of points.

\begin{itemize}
\item The operator ${\cal W}_{12}^{--}$, defined in~\eqref{equ:W-}, lowers weights at points $1$ and $2$ by one unit. In Fourier space, it takes the form 
\begin{eBox}
\beq
{\cal W}_{12}^{--} = \frac{1}{2} \left( \pdr{}{\k_1}-\pdr{}{\k_2} \right) \cdot \left( \pdr{}{\k_1}-\pdr{}{\k_2} \right) \equiv \frac{1}{2} \vec K_{12}\cdot \vec K_{12}  \,,
\label{eq:Wlower}
\eeq
\end{eBox}
where we have defined the differential operator 
\beq
\vec K_{12} \equiv \frac{\partial}{\partial \k_1}-\frac{\partial}{\partial \k_2}\,. \label{equ:K12}
\eeq
\item Similarly, the operator ${\cal W}_{12}^{++}$ in (\ref{equ:W+}) raises the weights by one at both points $1$ and $2$. It was used in \S\ref{sec:WR} to change the weight of the external fields from $\Delta=2$ to $\Delta=3$.
In general, this operator takes a very complicated form in Fourier space,
\begin{align}
	\cW_{12}^{++} \propto\ &  D_1^{(A)}D_2^{(A)} \cW_{12}^{--}-\kza\kzb \Big(\delta_1\delta_2(\z_1\cdot\z_2)\ddza\ddzb+\eta_{12}^2\eta_{21}^2(\partial_{\z_1}\cdot\partial_{\z_2})\Big) \nn\\[5pt]
	&+\Big[\delta_1 D_2^{(a)}(\vec K_{12}\cdot \vec z_1)\big(\kza\ddza-\tau_1\kdza \big)- \eta_{11}^2\kza D_1^{(A)}(\vec K_{12}\cdot \partial_{\vec z_1}) \nn\\[5pt]
	&+\delta_2\kza\Big(D_{21}^{(B)}+\kza\big(\sigma_2(1+\Delta_2-d)-\tau_2(\z_2\cdot\partial_{\z_2})\big)\Big)\ddza -\delta_1\eta_{12}\sigma_2\kza \nn\\[5pt]
	& -\delta_2\Big[\sigma_2(-4+d+2S_1)(1-d+\Delta_2)\kza +\tau_1 D_{21}^{(B)}\Big]\kdza \nonumber \\[5pt]
	&+(1\leftrightarrow 2) \Big]\, ,
	\label{eq:Wraise}
\end{align}
where we have defined 
\beq
\begin{aligned}
	D_a^{(A)} &\equiv (\k_a\cdot \z_a)^2\nabla_{\z_1}^2-(d-4+2S_a)(\k_a\cdot \z_a)(\k_a\cdot \partial_{\z_a})-k_a^2\sigma_a^2 \, ,\\
	D_{ab}^{(B)} &\equiv \delta_a\big(\tau_a(\z_1\cdot\z_2)(\k_a\cdot\partial_{\z_a})+\sigma_b(\k_a\cdot\z_b)\big)+\eta_{ab}(\k_a\cdot\z_a)(\z_b\cdot\partial_{\z_a})\, ,
\end{aligned}
\eeq
and 
\beq
\begin{aligned}
	\sigma_a^2 &\equiv (\Delta_a+S_a-1)(d-2-\Delta_a+S_a)\, ,\\
	\eta_{ab}^2 &\equiv(d-3-\Delta_a+S_a)(d-2\Delta_b)\, ,\\
	\tau_a &\equiv \Delta_a+S_a-1\, ,\\
	\delta_a &\equiv d-\Delta_a\, .
\end{aligned}
\eeq
Fortunately, this becomes somewhat more manageable when acting on scalars: \begin{eBox}
\beq
\begin{aligned}
{\cal W}_{12}^{++} =  &\,(k_1k_2)^2\hs{\cal W}_{12}^{--} -(d-2\Delta_1)(d-2\Delta_2)\,\k_1\cdot \k_2 \\[3pt]
&+ \Big(k_2^2(d-2\Delta_1)\big(d-1-\Delta_1+ \k_1\cdot \vec K_{12}\big)+ (1\leftrightarrow 2)\Big) \, .\label{W12Fourier}
\end{aligned}
\eeq
\end{eBox}
In the main text, this operator played an important role and was denoted ${\cal W}_{12}$, i.e.~we dropped the superscripts to avoid clutter.

\item The spin-raising operator ${\cal S}_{12}^{++}$, defined in~\eqref{equ:S+},  has the following Fourier representation
\vspace{-.4cm}
\begin{eBox}
\beq
\begin{aligned}
{\cal S}_{12}^{++} = &\ (S_1+\Delta_1-1)(S_2+\Delta_2-1) \,{\z}_1\cdot {\z}_2  -({\z}_1\cdot \k_1)({\z}_2\cdot  \k_2) \,{\cal W}_{12}^{--}\\[3pt]
& + \Big[(S_1+\Delta_1-1)(\k_2\cdot {\z}_2)({\z}_1 \cdot \vec K_{12}) + (1 \leftrightarrow 2) \Big]\,.
\end{aligned}
\label{eq:Spinraise}
\eeq
\end{eBox}
This operator raises the spin by one unit at both points $1$ and $2$, which is useful for generating spinning correlators~\cite{CosmoBoot2}.

\item The operator $D_{12}$, defined in (\ref{eq:D12op}), raises the spin at point $1$ and lowers the weight at point $2$. In Fourier space, it reads
\begin{eBox}
\beq
D_{12} = (\Delta_1+S_1-1)\,{\z}_1\cdot \vec K_{12}- ({\z}_1\cdot \k_1) \cW_{12}^{--}\,,
\eeq
\vskip 5pt
\end{eBox}
where we have dropped an overall factor of $i$. We have used this operator in the main text to raise the spin of exchanged fields.

\item The operator $D_{11}$ in (\ref{equ:D11X}) both raises the spin and lowers the weight at the point $1$ and is useful for raising the spin of external fields~\cite{CosmoBoot2}. In Fourier space, it becomes
\begin{eBox}
\vspace{-.3cm}
\beq
\label{equ:D11k}
D_{11} =  \big(\Delta_2-d+ \k_2\cdot \vec K_{12} \big)\, {\z}_1\cdot \vec K_{12} -(\k_2\cdot {\z}_1) \cW_{12}^{--} - {\z}_2\cdot \vec K_{12}\,{\z}_1\cdot \partial_{\vec z_2} + ({\z}_1\cdot {\z}_2) \partial_{\vec z_2}\cdot \vec K_{12} \,,
\eeq
\vspace{-.5cm}
\end{eBox}
where we have again dropped an overall factor of $i$. 

\item It is often helpful to use the operator $H_{12}$, defined in (\ref{equ:H12X}), which raises the spin and lowers the weight by one unit at both points $1$ and $2$. Its Fourier representation is 
\begin{eBox}
\beq
 \label{equ:H12Fourier}
H_{12} = 2\,{\z}_1\cdot \vec K_{12}\,{\z}_2\cdot \vec K_{12}-({\z}_1\cdot {\z}_2) \vec K_{12}\cdot \vec K_{12}\,.
\eeq
\vskip 5pt
\end{eBox}
\item Although we have not made use of it in this work, for completeness let us also give the expression for the spin-lowering operator $\cS_{12}^{--}$, defined in (\ref{equ:S-}), which lowers the spin by one unit at both points 1 and 2:
\begin{align}
	\cS_{12}^{--} =\ & D_1^{(C)}D_2^{(C)} \cW_{12}^{--} - \rho_1\rho_2(\z_1\cdot\z_2)\ddza\ddzb \nn\\
	&+\Big[ \rho_1 D_2^{(C)} K_{12}\cdot(\z_1\ddza-\lambda_1\partial_{\z_1}\big)+\rho_1\rho_2\lambda_1\partial_{\z_1}\cdot(\lambda_2 \partial_{\z_2} - \z_2)+(1\leftrightarrow 2)\Big]\, ,
	\label{eq:Spinlower}
\end{align}
where 
\beq
\begin{aligned}
	D_{a}^{(C)}&\equiv (\k_a\cdot\z_a)\nabla_{z_a}^2 -\lambda_a(\k_a\cdot\partial_{\z_a})\, ,\\
	\rho_a &\equiv d-1-\Delta_a+S_a\, ,\\
	\lambda_a &\equiv 2S_a+d-4\, .
\end{aligned}
\eeq 
\end{itemize}
Finally, we need to highlight  that, in momentum space, correlation functions take the form
\be
\langle O_{\k_1}\cdots O_{\k_n}\rangle = (2\pi)^d\delta^d(\k_1+\cdots+\k_n)\langle O_{\k_1}\cdots O_{\k_n}\rangle' \, .
\ee
We often want to act with weight-shifting operators directly on the primed correlator, with the delta function removed. In that case, one might be concerned that the momentum space operators would also act on the delta function, leading to extra terms after integration-by-parts. This situation is familiar from the Fourier space action of the dilation operator. However, inspection of the weight-shifting operators presented in this section reveals that derivatives always appear in the combination $\vec K_{12}$ defined in (\ref{equ:K12}), so that they depend on momentum differences and will therefore pass through the delta function.

\subsection{A Few Simple Examples}

To illustrate the power of these spinning weight-shifting operators, we give a couple simple applications. We will present a much more systematic study of spinning correlation functions in~\cite{CosmoBoot2}.

\begin{itemize}
\item As a first example, let us consider spinning up the two-point function of scalar operators to obtain the two-point function of operators with spin. The scalar two-point function is given by
\beq
\langle O O \rangle = k^{2\Delta-d}\, ,
\eeq
where the normalization is arbitrary. We want to act on this object with the spin-raising operator~\eqref{eq:Spinraise}. Momentum conservation implies that $\k_1 = -\k_2$, so that the spin-raising operator simplifies to
\beq
\begin{aligned}
{\cal S}^{++}_{12} = &\ (1-S-\Delta)^2 \,{\z}_1\cdot{\z}_2+(1-S-\Delta)\Big[ (\k\cdot{\z}_1)\,{\z}_2\cdot\partial_{\k}+(\k\cdot{\z}_2)\,{\z}_1\cdot\partial_{\k}\Big]  \\
&+\frac{(\k\cdot{\z}_1)(\k\cdot{\z}_2)}{2} \,\partial_\k\cdot\partial_\k\, .
\end{aligned}
\eeq
Acting with this operator repeatedly on $\langle OO\rangle$ generates the two-point function of spinning operators
\be
\langle O^{(S)}\,  O^{(S)}\rangle &=({\cal S}^{++})^S\langle OO\rangle  \nn \\
&\propto \frac{[(\hat k\cdot {\z}_1)(\hat k\cdot {\z}_2)]^S}{k^{d-2\Delta}}\, P_S^{(\Delta-S-d/2,d/2-2)}\left(1-\frac{{\z}_1\cdot {\z}_2}{(\hat k\cdot {\z}_1)(\hat k\cdot {\z}_2)}\right) , \label{equ:OJ}
\ee
where $P_S^{(a,b)}$ is the Jacobi polynomial. 

\item As a simple higher-point example, we construct the three-point correlation function between the stress tensor and two $\Delta = 2$ scalars in $d=3$ dimensions from a scalar seed. It is well-known that this correlation function is completely fixed by conformal invariance, and here we indeed reproduce this result. We start with the 
three-point function of $\Delta = 3$ scalar fields:
\beq
\langle \phi \phi \phi \rangle = \log(k_t/\mu) \sum_a k_a^3 - \sum_{a \ne b} k_a^2 k_b + k_1k_2k_3 \, ,\label{equ:GP}
\eeq
where $k_t \equiv k_1+k_2+k_3$. By applying the $D_{12}$ operator we can lower the weights of two of the scalars to $\Delta = 2$, while spinning up the third one to $S  =2$. Explicitly, we find
\begin{align}
\label{eq:t223pt}
\langle T \varphi \varphi\rangle &= D_{13}D_{12}\langle\phi \phi\phi\rangle \nn \\[6pt]
& \begin{aligned} = \ &
\frac{9(k_1-k_2+k_3)^2}{2k_1 k_t^2}( \k_1\cdot {\z}_1)^2+\frac{36(k_1+k_3)}{k_t^2}(\k_1\cdot {\z}_1)(\k_2\cdot {\z}_1) \\
&+\frac{18(2k_1+k_2+k_3)}{k_t^2}(\k_2\cdot {\z}_1)^2\, .
\end{aligned}
\end{align}
Since the correlation function of any spinning operator with two scalars is uniquely fixed by conformal invariance, this answer is guaranteed to be conserved, because the spin-2 operator has $\Delta = 3$. It can be checked that this agrees with the results of~\cite{Bzowski:2013sza}, who derived this correlation function in momentum space by directly solving the conformal Ward identity differential equations. Finally, we can use the weight-raising operator ${\cal W}_{23}^{++}$ to raise the scalar weights to $\Delta = 3$, which also yields a result that agrees with~\cite{Mata:2012bx,Bzowski:2013sza}.
\end{itemize}

\newpage
 \section{Polarization Tensors}
 \label{sec:PT}

 Our derivation of the spin-exchange solutions to the conformal Ward identities in \S\ref{sec:InternalSpin} required explicit expressions for the polarization tensors of spinning operators in a conformal field theory. Fortunately, these can be obtained from the two-point functions of spinning operators~\eqref{equ:OJ} in a relatively straightforward way. In this appendix, we describe this construction and provide explicit formulas for the relevant polarization sums. 
 
 \vskip 4pt
 We being by considering the two-point function of spin-$S$ operators
\beq
\langle O^{(S)}\,  O^{(S)}\rangle= \frac{(-2)^S S!}{(\tilde\Delta -1)_S}\frac{[(\hat k\cdot {\z}_1)(\hat k\cdot {\z}_2)]^S}{k^{d-2\tilde \Delta}}\, P_S^{(\tilde \Delta-S-d/2,d/2-2)}(\omega)\,,
\eeq
where the weight appearing, $\tilde \Delta=d-\Delta$, is the shadow dimension to that of the exchanged operator---because we are interested in the inverse of the two-point function---and the argument of the Jacobi polynomial is 
\beq
\omega \equiv 1-\frac{\vec z_1\cdot \vec z_2}{(\hat k\cdot \vec z_1 )(\hat k\cdot \vec z_2)}\, .
\eeq
The polarization tensor used in the main text is then defined as
\beq
(\Pi_S)^{i_1\cdots i_S}_{j_1\cdots j_S} = \frac{1}{(S!(\frac{d-2}{2})_S)^2}D_{z_1}^{i_1} \cdots D_{z_1}^{i_S}  D^{z_2}_{j_1} \cdots D^{z_2}_{j_S} \left(k^{d-2\tilde\Delta}\langle O^{(S)}\,  O^{(S)}\rangle\right), 
\label{equ:PT}
\eeq
where $D_{z}^i$ is the Todorov operator that strips off the null vectors $z^i$ from the index-free form of the two-point function:
\beq
D_{z}^i = \left(\frac{d}{2}-1+ \vec z\cdot\frac{\partial}{\partial \vec z}\right)\frac{\partial}{\partial z_i}-\frac{1}{2}z^i \frac{\partial^2}{\partial \vec z\cdot\partial \vec z}\,.
\label{eq:Todorov}
\eeq
To see the equivalence of the projector (\ref{equ:PT}) with the more familiar expressions, we should work in something closer to the helicity basis. This amounts to decomposing $\Pi_S$ into a set of irreducible components. We first show how this works for spins 1 and 2, before turning to the general case.
\begin{itemize}
\item {\bf Spin 1:} Specializing (\ref{equ:PT}) to the case of spin one, we get 
\beq
(\Pi_1)^{i}_{j} = \delta^i_j - \frac{(d-2\Delta)}{(d-\Delta-1)}\hat k^i\hat k_j\,.
\eeq
Introducing the projector
\beq
\pi^i_j \equiv \delta^i_j -\hat k^i\hat k_j\,, \label{equ:pij}
\eeq
this can be split into transverse and longitudinal components
\beq 
(\Pi_1)^{i}_{j}  = \pi^i_j + \frac{(1-\Delta)}{(1-d+\Delta)}\hat k^i\hat k_j\, ,
\eeq
which is~\eqref{equ:Pij}.
\item {\bf Spin 2:} In the case of spin two, we decompose the projector in the orthonormal basis of projectors for traceless two-index tensors 
\begin{subequations}
\begin{align}
(\Pi_{2,2})^{ij}{}_{lm} &= \pi^{(i}_{(l} \pi^{j)}_{k)}-\frac{1}{d-1}\pi^{ij}\pi_{lm}\,,\\
(\Pi_{2,1})^{ij}{}_{lm} &= 2\hat k^{(i}\hat k_{(l}\pi^{j)}_{m)}\,,\\
(\Pi_{2,0})^{ij}{}_{lm} &= \frac{d}{d-1}\left (\hat k^i\hat k^j-\frac{1}{d}\delta_{ij}\right )\left (k_lk_m-\frac{1}{d}\delta^{lm}\right) .\label{piS}
\end{align}
\label{Projectors}
\end{subequations}
Using this, the spin-2 version of (\ref{equ:PT}) becomes 
\beq
(\Pi_2)^{ij}_{lm}  = (\Pi_{2,2})^{ij}{}_{lm}+\frac{\Delta}{d-\Delta}(\Pi_{2,1})^{ij}{}_{lm}+ \frac{\Delta(\Delta-1)}{(d-\Delta)(d-\Delta-1)}(\Pi_{2,0})^{ij}{}_{lm}\,,
\eeq
which is (\ref{equ:Pijrs}). 

\item {\bf Spin $\boldsymbol S$:} 
For general spins, the polarization tensors can be constructed by methods of harmonic analysis~\cite{Dobrev:1977qv}, but we can get an intuitive understanding of the answer as follows: After normalizing the highest-helicity projector to 1, for each lower-helicity state there will be a pole at each of the partially-massless weights that projects out that mode. The numerator for each helicity mode is given by the same polynomial with $\Delta \to d-\Delta$. The unitary/non-unitary states between the partially-massless points fixes the relative signs between adjacent helicity components.
 The final answer is 
\begin{align}
	(\Pi_{S})^{i_1\cdots i_S}_{j_1\cdots j_S} = \sum_{m=0}^S \frac{(\Delta-1+m)_{S-m}}{(d-\Delta-1+m)_{S-m}}(\Pi_{S,m})^{i_1\cdots i_S}_{j_1\cdots j_S}\, .\label{piStensor1}
\end{align}
The polarization tensor contracted with null vectors $\z_1$ and $\z_2$ takes the form~\cite{Dobrev:1977qv}
\begin{align}
\Pi_{S,m}(\z_1, \z_2) &\equiv z_1^{i_1}\cdots z_1^{i_S}(\Pi_{S,m})^{i_1\cdots i_S}_{j_1\cdots j_S}z_2^{j_1}\cdots z_2^{j_S}\nn\\[3pt]
	&=\frac{(d-3+2m)(-1)^mS!}{(S-m)!}\frac{\left(\frac{d-2}{2}\right)_S}{(d-3)_{S+m+1}}\big[2(\hat k\cdot \z_1)(\hat k\cdot \z_2)\big]^S C_m^\frac{d-3}{2}(\omega)\, ,
\label{eq:bookprojectors}
\end{align}
where $C_m^\frac{d-3}{2}$ is the Gegenbauer polynomial. The explicit polarization tensors can be obtained by stripping off the auxiliary null vectors with the help of the operator~\eqref{eq:Todorov}. These polarization tensors satisfy 
\be
&{\rm orthonormality}:& & \hspace{-0.75cm} (\Pi_{S,m})^{i_1\cdots i_S}_{j_1\cdots j_S}(\Pi_{S,m'})^{j_1\cdots j_S}_{l_1\cdots l_S} = \delta_{mm'}(\Pi_{S,m})^{i_1\cdots i_S}_{l_1\cdots l_S}\, , \\
&{\rm completeness}: & & \hspace{-0.75cm}  1= \sum_{m=0}^S \Pi_{S,m} \, , \\
&{\rm transversality}: && \hspace{-0.75cm}  0=\hat k^{j_m}\cdots \hat k^{j_S}(\Pi_{S,m})^{i_1\cdots i_S}_{j_1\cdots j_S} \, .
\ee
It is also useful to know the harmonic extension of~\eqref{eq:bookprojectors}, which can be contracted with {\it non-null} vectors. Contracting with generic vectors $\vec w_1$  and $\vec w_2$ leads to~\cite{Dobrev:1977qv}
\beq
\Pi_{S,m}(\vec w_1,\vec w_2) = 2^{S-m}\frac{(m+d/2-1)_{S-m}}{(2m+d-2)_{S-m}}{S\choose m} \,L_{S,m}(\vec w_1)\, \Pi_{m,m}(\vec w_1,\vec w_2)\,L_{S,m}(\vec w_2)\, , \label{PiPi}
\eeq
where we have defined 
\be
L_{S,m}(\vec w\hskip 1pt) &= 2^{m-S}{d/2+S-1 \choose S-m}^{-1} (w^2)^\frac{S-m}{2} C_{S-m}^{d/2+m-1}(\hat k\cdot \hat w)\, , \\
\Pi_{m,m}(\vec w_1,\vec w_2) &= (-2)^m m! \frac{ (d/2-1)_m}{(d-3)_{2m}}(\pi^{(11)}\pi^{(22)})^{\frac{m}{2}} C_m^{\frac{d-3}{2}}\left(-\frac{\pi^{(12)}}{\sqrt{\pi^{(11)}\pi^{(22)}}}\right) . \label{equ:PiPi}
\ee
By design, $L_{S,m}(\vec w\hskip 1pt) = (\hat k\cdot \vec w)^{S-m}$ when $w^2=0$.
The tensors $\pi^{(ab)}$ in (\ref{equ:PiPi}) correspond to~$\pi_{ij}$ of (\ref{equ:pij}) contracted with the auxiliary vectors:
\beq
\pi^{(ab)} 
= \vec w_a\cdot \vec w_b -(\hat k\cdot \vec w_a)(\hat k\cdot \vec w_b)\, .
\eeq
In the $s$-channel exchange studied in Section~\ref{sec:SpinExchange},
the polarization tensor was associated to the internal momentum $\vec s = \k_1+\k_2$. 
To obtain the polarization sums used in the main text, we therefore let $\vec w_1 \to \k_1$ and $\vec w_2 \to \k_2$, and write the result in terms of $\cos\theta_i\equiv \hat k_i\cdot\hat s$ and $\psi$ (the angle between $\hat k_1$ and $\hat k_3$ projected on the plane perpendicular to $\hat s$). To take the limit $d \to 3$, we make use of the following limits of the Gegenbauer polynomials
\begin{align}
	\lim_{d\to 3} C_{S-m}^{d/2-m-1}(z) &= \frac{2^{1-m}}{(\frac{3}{2})_{m-1}}\frac{(-1)^m}{(1-z^2)^{m/2}}\, P_S^m(z)\, ,\\
	\lim_{d\to 3} \frac{m}{d-3}C_{m}^{\frac{d-3}{2}}(\cos\psi) &= \cos (m\psi) \times \begin{cases}
		1 & m=1,2,\cdots\\ \frac{1}{2} & m=0
	\end{cases}\, .\label{Pidim3}
\end{align}
This gives
\begin{eBox}
\vspace{-0.3cm}
\beq
	\lim_{d\to 3} \Pi_{S,m}(\hat k_1,\hat k_3) = \frac{ S!}{(2S-1)!!}(2-\delta_{m0})(-1)^m\cos(m\psi)P_S^m(\cos\theta_1)P_S^{-m}(\cos\theta_3)\, ,\label{Pidim3ang}
\eeq
\vspace{-0.4cm}
\end{eBox}
which is the polarization sum used in (\ref{eq:PolCollapsed}).

\end{itemize}

\newpage
\section{Notation and Conventions}
\label{app:Notation}

\begin{center}
\renewcommand*{\arraystretch}{1.08}
\begin{longtable}{c p{10cm} c}
\toprule
\multicolumn{1}{c}{\textbf{Symbol}} &
\multicolumn{1}{l}{\textbf{Meaning}} &
\multicolumn{1}{c}{\textbf{Reference}} \\
\midrule
\endfirsthead
\multicolumn{3}{c}
{} \\
\toprule
\multicolumn{1}{c}{\textbf{Symbol}} &
\multicolumn{1}{l}{\textbf{Meaning}} &
\multicolumn{1}{c}{\textbf{Reference}} \\
\midrule
\endhead
\bottomrule
\endfoot
\bottomrule
\endlastfoot

$\k$ & Three-momentum vector & \S\ref{sec:boundary} \\
$k_i$	  & 	Spatial component of $\k$	&	\S\ref{sec:boundary}	\\ 
$\k_a$	 	&Momentum of the $a$-th leg	&	\S\ref{sec:boundary}	\\ 
$k_a$	 	& Magnitude of $\k_a$, $k_a \equiv |\k_a|$	&	\S\ref{sec:boundary}	\\ 
$\hat k_a$ & Unit vector, $\hat k_a \equiv \k_a/ k_a$ & \S\ref{sec:InternalSpin} \\
$k_t$	 	& Sum of momentum magnitudes, $k_t \equiv \sum_{a=1}^N k_a$	&	\S\ref{sec:pm}	\\ 
$s$	  &  Exchange momentum, $s \equiv |\k_1+\k_2|$  	&	\S\ref{sec:boundary}		\\ 
$t$	 	& Exchange momentum, $t \equiv |\k_2+\k_3|$  	&\S\ref{sec:boundary}		\\ 
$u$ & Momentum ratio, $u \equiv s/(k_1+k_2)$& \eqref{equ:ansatz} \\ 
$v$ & Momentum ratio, $v \equiv s/(k_3+k_4)$& \eqref{equ:ansatz} \\   
$\vec\alpha$ & Difference of momentum vectors, $\vec \alpha \equiv \vec k_1-\vec k_2$ & \S\ref{sec:InternalSpin} \\ 
$\vec\beta$ & Difference of momentum vectors, $\vec\beta \equiv \vec k_3-\vec k_4$  & \S\ref{sec:InternalSpin}\\ 
$\hat\alpha$ & Dimensionless difference of momenta, $\hat\alpha \equiv (k_1-k_2)/s$ & \eqref{equ:abt} \\ 
$\hat\beta$ & Dimensionless  difference of momenta, $\hat\beta \equiv (k_3-k_4)/s$  & \eqref{equ:abt} \\ 
$\hat\tau$	  & 	Angular variable, $\hat \tau \equiv \vec \alpha \cdot \vec \beta/s^2$ 	&	(\ref{equ:abt})	\\ 
$\hat T$ & Angular variable, $\hat T \equiv \hat \tau + \hat \alpha \hat \beta/(uv)$  & \eqref{equ:T} \\ 
$\hat L$ & Angular variable & \eqref{equ:L}\\ 
$\gamma$ & Angle between $\hat k_1$ and $\hat k_3$, $\cos\gamma\equiv \hat k_1\cdot\hat k_3$ & \eqref{eq:psi}\\ 
$\theta_a$ & Angle between $\hat k_a$ and $\hat s$,  $\cos\theta_a\equiv \hat k_a\cdot \hat s$ & \eqref{eq:psi}\\ 
$\psi$ & Projected angle between $\hat k_1$ and $\hat k_3$, $\cos \psi \equiv \hat T/\hat L^2$ & \eqref{eq:psi}\\ 
$\Pi_{S,m}$ & Polarization sum & \eqref{equ:Piellm} \\
\midrule
$X$ & Embedding space coordinate & \S\ref{sec:nullcone} \\
$Z$ & Embedding space null vector & \S\ref{sec:embeddingspacetensors} \\
$X^M$ & Component of $X$ & \S\ref{sec:nullcone} \\
$Z^M$ & Component of $Z$ & \S\ref{sec:embeddingspacetensors} \\
$X_{ab}$ & Dot product, $X_{ab}\equiv X_a\cdot X_b$ & \S\ref{sec:embeddingspacetensors} \\
$U$ & Cross ratio, $U\equiv X_{12}X_{34}/X_{13}X_{24}$ & \S\ref{sec:CC} \\
$V$ & Cross ratio, $V\equiv X_{14}X_{32}/X_{13}X_{24}$ & \S\ref{sec:CC} \\
 $O^{(S)}$ & Index-free spin-$S$ operator, $O^{(S)} \equiv O_{M_1\cdots M_S} \hskip 2pt Z^{M_1}\cdots Z^{M_S}$ &  \S\ref{sec:embeddingspacetensors}\\ 
\midrule
$\sigma$ & Generic bulk scalar field & \S\ref{sec:boundary} \\
$\upvarphi$ & Conformally-coupled scalar field & \S\ref{sec:seeds} \\
$\upphi$ & Massless scalar field & \S\ref{sec:Inf} \\
 $O$ & Operator dual to $\sigma$ & \S\ref{sec:boundary}\\ 
 $\Delta$	  & Scaling dimension (conformal weight)	&	 \S\ref{sec:boundary}	\\  
$\varphi$ & Operator dual to $\upvarphi$ $(\Delta = 2)$ & \S\ref{sec:seeds}\\ 
$\phi$	  & Operator dual to $\upphi$ ($\Delta = 3$)	&	 \S\ref{sec:Inf}	\\  
$\Delta_t$	  & Total conformal weight, $\Delta_t \equiv \sum_n \Delta_n$ &	  \S\ref{sec:boundary}		\\ 
$M$ & Mass parameter  & \S\ref{sec:seeds} \\ 
$S$ & Spin of exchanged particle & \S\ref{sec:InternalSpin}\\ 
$m$ & Helicity of exchanged particle & \S\ref{sec:InternalSpin}\\ 
$T$ & Depth of partially massless field & \S\ref{sec:pm}\\ 
\midrule
$F$ & Scalar four-point function & (\ref{equ:scalarF}) \\ 
$\F$ & Dimensionless four-point function, $\F \equiv s^{9-\Delta_t} F$ & (\ref{equ:hatF})	\\ 
$\hat C$ & Contact four-point function & \eqref{eq:Cncontact} \\ 
$F^{(S)}$ & Four-point function from spin-$S$ exchange & (\ref{equ:POL})\\ 
$F_L^{(S)}$ & Longitudinal part of four-point function & (\ref{equ:FL})\\ 
$B^{(S)}$ & Bispectrum from spin-$S$ exchange & \eqref{equ:InfB}\\ 
$B_c$ & Contact contributions to the bispectrum & \eqref{eq:pmcontactterms}\\ 
$B_{\rm inf}$ & Bispectrum of slow-roll inflation & \eqref{equ:Binf}\\
$\epsilon$ & Slow-roll parameter &  \S\ref{sec:toInf} \\ 
$\theta$ & Angle in the squeezed limit & \eqref{eq:squeezedscaling}\\  
\midrule
$\Delta_u$	  & Differential operator, $\Delta_u \equiv u^2(1-u^2) \partial_u^2 - 2 u^3 \partial_u$	&(\ref{equ:Ward})	\\
$D_{uv}$	  & Differential operator, $D_{uv} \equiv (uv)^2 \partial_u\partial_v$	&(\ref{equ:spin1disc})	\\
$\vec{K}_{ab}$	  & Vector differential operator, $\vec{K}_{ab} \equiv \partial_{\vec{k}_a}-\partial_{\vec{k}_b}$	&(\ref{equ:K12})	\\
${\cal W}_{ab}^{++}$ & Weight-raising operator & \eqref{eq:Wraise} \\
${\cal W}_{ab}^{--}$ & Weight-lowering operator & \eqref{eq:Wlower} \\
$\cW_{ab}$	  & Weight-raising operator, $\cW_{ab} \equiv \cW_{ab}^{++}$ & 	\eqref{eq:W12Fourier} \\ 
$U_{ab}^{(S,m)}$	  & Helicity-decomposed weight-shifting operator & 	\eqref{U12SmX} \\ 
${\cal S}_{ab}^{++}$ & Spin-raising operator & \eqref{eq:Spinraise} \\
${\cal S}_{ab}^{--}$ & Spin-lowering operator & \eqref{eq:Spinlower} \\
${\cal S}_{ab}$ & Spin-raising operator, $ {\cal S}_{ab} \equiv {\cal S}_{ab}^{++}$ & (\ref{equ:SR})\\ 
${\cal D}_{uv}^{(S,m)}$	  &  Helicity-decomposed spin-raising operator & \eqref{equ:Duvellm}	\\  
$D_{12}$ & Operator that raises spin at 1 and lowers weight at 2& \eqref{equ:D12}\\ 
$D_{11}$ &  Operator that raises spin at 1 and lowers weight at 1 &\eqref{equ:D11}\\ 
$H_{12}$ & Operator that raises spin and lowers weight at  1 and 2&  \eqref{equ:H12Fourier}\\ 
\midrule
$d$ & Boundary space dimension & \S\ref{app:WSfourier}\\
$z^i$ & Auxiliary null vector, $z^2=0$ & \S\ref{sec:embeddingspacetensors} \\
$(\Pi_S)^{i_1\cdots i_S}_{j_1\cdots j_S}$ & Polarization tensor & \eqref{equ:PT} \\
$D_{z}^i$ & Todorov operator for $z^i$ & \eqref{eq:Todorov} \\
$\pi_{ij}$ & Spin-1 projector & (\ref{equ:pij}) \\
\newpage
$\delta_{ab}$ & Kronecker delta & \cite{MathWorld} \\ 
$P_S$ & Legendre polynomial & \cite{MathWorld} \\ 
$P_S^m$ & Associated Legendre polynomial & \cite{MathWorld} \\ 
$C_m^\lambda$ & Gegenbauer polynomial & \cite{MathWorld} \\ 
$P_S^{(a,b)}$ & Jacobi polynomial & \cite{MathWorld} \\ 
$(\cdot)_n$ & Pochhammer symbol & \cite{MathWorld} \\ 
\end{longtable}
\end{center}

\clearpage
\phantomsection
\addcontentsline{toc}{section}{References}
\bibliographystyle{utphys}
\bibliography{WeightShifting-Refs}

\providecommand{\href}[2]{#2}\begingroup\raggedright\begin{thebibliography}{10}

\bibitem{Arkani-Hamed:2018kmz}
N.~Arkani-Hamed, D.~Baumann, H.~Lee, and G.~L. Pimentel, ``{The Cosmological
  Bootstrap: Inflationary Correlators from Symmetries and Singularities},''
\href{http://arxiv.org/abs/1811.00024}{{\ttfamily arXiv:1811.00024 [hep-th]}}.

\bibitem{Arkani-Hamed:2017fdk}
N.~Arkani-Hamed, P.~Benincasa, and A.~Postnikov, ``{Cosmological Polytopes and
  the Wavefunction of the Universe},''
\href{http://arxiv.org/abs/1709.02813}{{\ttfamily arXiv:1709.02813 [hep-th]}}.

\bibitem{Arkani-Hamed:2018bjr}
N.~Arkani-Hamed and P.~Benincasa, ``{On the Emergence of Lorentz Invariance and
  Unitarity from the Scattering Facet of Cosmological Polytopes},''
\href{http://arxiv.org/abs/1811.01125}{{\ttfamily arXiv:1811.01125 [hep-th]}}.

\bibitem{Benincasa:2018ssx}
P.~Benincasa, ``{From the Flat-Space S-matrix to the Wavefunction of the
  Universe},''
\href{http://arxiv.org/abs/1811.02515}{{\ttfamily arXiv:1811.02515 [hep-th]}}.

\bibitem{Sleight:2019mgd}
C.~Sleight, ``{A Mellin Space Approach to Cosmological Correlators},''
\href{http://arxiv.org/abs/1906.12302}{{\ttfamily arXiv:1906.12302 [hep-th]}}.

\bibitem{Sleight:2019hfp}
C.~Sleight and M.~Taronna, ``{Bootstrapping Inflationary Correlators in Mellin
  Space},''
\href{http://arxiv.org/abs/1907.01143}{{\ttfamily arXiv:1907.01143 [hep-th]}}.

\bibitem{Benincasa:2019vqr}
P.~Benincasa, ``{Cosmological Polytopes and the Wavefuncton of the Universe for
  Light States},''
\href{http://arxiv.org/abs/1909.02517}{{\ttfamily arXiv:1909.02517 [hep-th]}}.

\bibitem{Maldacena:2011nz}
J.~Maldacena and G.~L. Pimentel, ``{On Graviton Non-Gaussianities During
  Inflation},'' \href{http://dx.doi.org/10.1007/JHEP09(2011)045}{{\em JHEP}
  {\bfseries 09} (2011) 045},
\href{http://arxiv.org/abs/1104.2846}{{\ttfamily arXiv:1104.2846 [hep-th]}}.

\bibitem{Antoniadis:2011ib}
I.~Antoniadis, P.~Mazur, and E.~Mottola, ``{Conformal Invariance, Dark Energy,
  and CMB Non-Gaussianity},''
  \href{http://dx.doi.org/10.1088/1475-7516/2012/09/024}{{\em JCAP} {\bfseries
  1209} (2012) 024},
\href{http://arxiv.org/abs/1103.4164}{{\ttfamily arXiv:1103.4164 [gr-qc]}}.

\bibitem{Creminelli:2011mw}
P.~Creminelli, ``{Conformal Invariance of Scalar Perturbations in Inflation},''
  \href{http://dx.doi.org/10.1103/PhysRevD.85.041302}{{\em Phys. Rev.}
  {\bfseries D85} (2012) 041302},
\href{http://arxiv.org/abs/1108.0874}{{\ttfamily arXiv:1108.0874 [hep-th]}}.

\bibitem{Kehagias:2012pd}
A.~Kehagias and A.~Riotto, ``{Operator Product Expansion of Inflationary
  Correlators and Conformal Symmetry of de Sitter},''
  \href{http://dx.doi.org/10.1016/j.nuclphysb.2012.07.004}{{\em Nucl. Phys.}
  {\bfseries B864} (2012) 492--529},
\href{http://arxiv.org/abs/1205.1523}{{\ttfamily arXiv:1205.1523 [hep-th]}}.

\bibitem{Arkani-Hamed:2015bza}
N.~Arkani-Hamed and J.~Maldacena, ``{Cosmological Collider Physics},''
\href{http://arxiv.org/abs/1503.08043}{{\ttfamily arXiv:1503.08043 [hep-th]}}.

\bibitem{Raju:2012zr}
S.~Raju, ``{New Recursion Relations and a Flat Space Limit for AdS/CFT
  Correlators},'' \href{http://dx.doi.org/10.1103/PhysRevD.85.126009}{{\em
  Phys. Rev.} {\bfseries D85} (2012) 126009},
\href{http://arxiv.org/abs/1201.6449}{{\ttfamily arXiv:1201.6449 [hep-th]}}.

\bibitem{Raju:2012zs}
S.~Raju, ``{Four-Point Functions of the Stress Tensor and Conserved Currents in
  AdS$_4$/CFT$_3$},'' \href{http://dx.doi.org/10.1103/PhysRevD.85.126008}{{\em
  Phys. Rev.} {\bfseries D85} (2012) 126008},
\href{http://arxiv.org/abs/1201.6452}{{\ttfamily arXiv:1201.6452 [hep-th]}}.

\bibitem{Mata:2012bx}
I.~Mata, S.~Raju, and S.~Trivedi, ``{CMB from CFT},''
  \href{http://dx.doi.org/10.1007/JHEP07(2013)015}{{\em JHEP} {\bfseries 07}
  (2013) 015},
\href{http://arxiv.org/abs/1211.5482}{{\ttfamily arXiv:1211.5482 [hep-th]}}.

\bibitem{Ghosh:2014kba}
A.~Ghosh, N.~Kundu, S.~Raju, and S.~Trivedi, ``{Conformal Invariance and the
  Four-Point Scalar Correlator in Slow-Roll Inflation},''
  \href{http://dx.doi.org/10.1007/JHEP07(2014)011}{{\em JHEP} {\bfseries 1407}
  (2014) 011},
\href{http://arxiv.org/abs/1401.1426}{{\ttfamily arXiv:1401.1426 [hep-th]}}.

\bibitem{Kundu:2014gxa}
N.~Kundu, A.~Shukla, and S.~Trivedi, ``{Constraints from Conformal Symmetry on
  the Three-Point Scalar Correlator in Inflation},''
  \href{http://dx.doi.org/10.1007/JHEP04(2015)061}{{\em JHEP} {\bfseries 04}
  (2015) 061},
\href{http://arxiv.org/abs/1410.2606}{{\ttfamily arXiv:1410.2606 [hep-th]}}.

\bibitem{Kundu:2015xta}
N.~Kundu, A.~Shukla, and S.~Trivedi, ``{Ward Identities for Scale and Special
  Conformal Transformations in Inflation},''
  \href{http://dx.doi.org/10.1007/JHEP01(2016)046}{{\em JHEP} {\bfseries 01}
  (2016) 046},
\href{http://arxiv.org/abs/1507.06017}{{\ttfamily arXiv:1507.06017 [hep-th]}}.

\bibitem{Coriano:2013jba}
C.~Corian\`o, L.~Delle~Rose, E.~Mottola, and M.~Serino, ``{Solving the
  Conformal Constraints for Scalar Operators in Momentum Space and the
  Evaluation of Feynman's Master Integrals},''
  \href{http://dx.doi.org/10.1007/JHEP07(2013)011}{{\em JHEP} {\bfseries 07}
  (2013) 011},
\href{http://arxiv.org/abs/1304.6944}{{\ttfamily arXiv:1304.6944 [hep-th]}}.

\bibitem{Coriano:2018bbe}
C.~Corian\`o and M.~Maglio, ``{Exact Correlators from Conformal Ward Identities
  in Momentum Space and the Perturbative $TJJ$ Vertex},''
\href{http://arxiv.org/abs/1802.07675}{{\ttfamily arXiv:1802.07675 [hep-th]}}.

\bibitem{Maglio:2019grh}
C.~Corian\`o and M.~Maglio, ``{On Some Hypergeometric Solutions of the
  Conformal Ward Identities of Scalar Four-Point Functions in Momentum
  Space},'' \href{http://dx.doi.org/10.1007/JHEP09(2019)107}{{\em JHEP}
  {\bfseries 09} (2019) 107},
\href{http://arxiv.org/abs/1903.05047}{{\ttfamily arXiv:1903.05047 [hep-th]}}.

\bibitem{Bzowski:2013sza}
A.~Bzowski, P.~McFadden, and K.~Skenderis, ``{Implications of Conformal
  Invariance in Momentum Space},''
  \href{http://dx.doi.org/10.1007/JHEP03(2014)111}{{\em JHEP} {\bfseries 03}
  (2014) 111},
\href{http://arxiv.org/abs/1304.7760}{{\ttfamily arXiv:1304.7760 [hep-th]}}.

\bibitem{Anninos:2014lwa}
D.~Anninos, T.~Anous, D.~Freedman, and G.~Konstantinidis, ``{Late-Time
  Structure of the Bunch-Davies De Sitter Wavefunction},''
  \href{http://dx.doi.org/10.1088/1475-7516/2015/11/048}{{\em JCAP} {\bfseries
  1511} no.~11, (2015) 048},
\href{http://arxiv.org/abs/1406.5490}{{\ttfamily arXiv:1406.5490 [hep-th]}}.

\bibitem{Bzowski:2015pba}
A.~Bzowski, P.~McFadden, and K.~Skenderis, ``{Scalar Three-Point Functions in
  CFT: Renormalisation, Beta Functions and Anomalies},''
  \href{http://dx.doi.org/10.1007/JHEP03(2016)066}{{\em JHEP} {\bfseries 03}
  (2016) 066},
\href{http://arxiv.org/abs/1510.08442}{{\ttfamily arXiv:1510.08442 [hep-th]}}.

\bibitem{Bzowski:2015yxv}
A.~Bzowski, P.~McFadden, and K.~Skenderis, ``{Evaluation of Conformal
  Integrals},'' \href{http://dx.doi.org/10.1007/JHEP02(2016)068}{{\em JHEP}
  {\bfseries 02} (2016) 068},
\href{http://arxiv.org/abs/1511.02357}{{\ttfamily arXiv:1511.02357 [hep-th]}}.

\bibitem{Bzowski:2017poo}
A.~Bzowski, P.~McFadden, and K.~Skenderis, ``{Renormalised Three-Point
  Functions of Stress Tensors and Conserved Currents in CFT},''
\href{http://arxiv.org/abs/1711.09105}{{\ttfamily arXiv:1711.09105 [hep-th]}}.

\bibitem{Isono:2018rrb}
H.~Isono, T.~Noumi, and G.~Shiu, ``{Momentum Space Approach to Crossing
  Symmetric CFT Correlators},''
  \href{http://dx.doi.org/10.1007/JHEP07(2018)136}{{\em JHEP} {\bfseries 07}
  (2018) 136},
\href{http://arxiv.org/abs/1805.11107}{{\ttfamily arXiv:1805.11107 [hep-th]}}.

\bibitem{Isono:2019ihz}
H.~Isono, T.~Noumi, and T.~Takeuchi, ``{Momentum Space Conformal Three-Point
  Functions of Conserved Currents and a General Spinning Operator},''
  \href{http://dx.doi.org/10.1007/JHEP05(2019)057}{{\em JHEP} {\bfseries 05}
  (2019) 057},
\href{http://arxiv.org/abs/1903.01110}{{\ttfamily arXiv:1903.01110 [hep-th]}}.

\bibitem{Isono:2019wex}
H.~Isono, T.~Noumi, and G.~Shiu, ``{Momentum Space Approach to Crossing
  Symmetric CFT Correlators II: General Spacetime Dimension},''
\href{http://arxiv.org/abs/1908.04572}{{\ttfamily arXiv:1908.04572 [hep-th]}}.

\bibitem{Albayrak:2018tam}
S.~Albayrak and S.~Kharel, ``{Towards the Higher-Point Holographic Momentum
  Space Amplitudes},'' \href{http://dx.doi.org/10.1007/JHEP02(2019)040}{{\em
  JHEP} {\bfseries 02} (2019) 040},
\href{http://arxiv.org/abs/1810.12459}{{\ttfamily arXiv:1810.12459 [hep-th]}}.

\bibitem{Albayrak:2019asr}
S.~Albayrak, C.~Chowdhury, and S.~Kharel, ``{New Relation for AdS
  Amplitudes},''
\href{http://arxiv.org/abs/1904.10043}{{\ttfamily arXiv:1904.10043 [hep-th]}}.

\bibitem{Albayrak:2019yve}
S.~Albayrak and S.~Kharel, ``{Towards the Higher-Point Holographic Momentum
  Space Amplitudes II: Gravitons},''
\href{http://arxiv.org/abs/1908.01835}{{\ttfamily arXiv:1908.01835 [hep-th]}}.

\bibitem{Bzowski:2019kwd}
A.~Bzowski, P.~McFadden, and K.~Skenderis, ``{Conformal Four-Point Functions in
  Momentum Space},''
\href{http://arxiv.org/abs/1910.10162}{{\ttfamily arXiv:1910.10162 [hep-th]}}.

\bibitem{Pajer:2016ieg}
E.~Pajer, G.~L. Pimentel, and J.~V.~S. Van~Wijck, ``{The Conformal Limit of
  Inflation in the Era of CMB Polarimetry},''
  \href{http://dx.doi.org/10.1088/1475-7516/2017/06/009}{{\em JCAP} {\bfseries
  1706} no.~06, (2017) 009},
\href{http://arxiv.org/abs/1609.06993}{{\ttfamily arXiv:1609.06993 [hep-th]}}.

\bibitem{Farrow:2018yni}
J.~Farrow, A.~Lipstein, and P.~McFadden, ``{Double Copy Structure of CFT
  Correlators},'' \href{http://dx.doi.org/10.1007/JHEP02(2019)130}{{\em JHEP}
  {\bfseries 02} (2019) 130},
\href{http://arxiv.org/abs/1812.11129}{{\ttfamily arXiv:1812.11129 [hep-th]}}.

\bibitem{Chen:2009zp}
X.~Chen and Y.~Wang, ``{Quasi-Single-Field Inflation and Non-Gaussianities},''
  \href{http://dx.doi.org/10.1088/1475-7516/2010/04/027}{{\em JCAP} {\bfseries
  1004} (2010) 027},
\href{http://arxiv.org/abs/0911.3380}{{\ttfamily arXiv:0911.3380 [hep-th]}}.

\bibitem{Baumann:2011nk}
D.~Baumann and D.~Green, ``{Signatures of Supersymmetry from the Early
  Universe},'' \href{http://dx.doi.org/10.1103/PhysRevD.85.103520}{{\em Phys.
  Rev.} {\bfseries D85} (2012) 103520},
\href{http://arxiv.org/abs/1109.0292}{{\ttfamily arXiv:1109.0292 [hep-th]}}.

\bibitem{Assassi:2012zq}
V.~Assassi, D.~Baumann, and D.~Green, ``{On Soft Limits of Inflationary
  Correlation Functions},''
  \href{http://dx.doi.org/10.1088/1475-7516/2012/11/047}{{\em JCAP} {\bfseries
  1211} (2012) 047},
\href{http://arxiv.org/abs/1204.4207}{{\ttfamily arXiv:1204.4207 [hep-th]}}.

\bibitem{Chen:2012ge}
X.~Chen and Y.~Wang, ``{Quasi-Single-Field Inflation with Large Mass},''
  \href{http://dx.doi.org/10.1088/1475-7516/2012/09/021}{{\em JCAP} {\bfseries
  1209} (2012) 021},
\href{http://arxiv.org/abs/1205.0160}{{\ttfamily arXiv:1205.0160 [hep-th]}}.

\bibitem{Pi:2012gf}
S.~Pi and M.~Sasaki, ``{Curvature Perturbation Spectrum in Two-field Inflation
  with a Turning Trajectory},''
  \href{http://dx.doi.org/10.1088/1475-7516/2012/10/051}{{\em JCAP} {\bfseries
  1210} (2012) 051},
\href{http://arxiv.org/abs/1205.0161}{{\ttfamily arXiv:1205.0161 [hep-th]}}.

\bibitem{Noumi:2012vr}
T.~Noumi, M.~Yamaguchi, and D.~Yokoyama, ``{EFT Approach to Quasi-Single-Field
  Inflation and Effects of Heavy Fields},''
  \href{http://dx.doi.org/10.1007/JHEP06(2013)051}{{\em JHEP} {\bfseries 06}
  (2013) 051},
\href{http://arxiv.org/abs/1211.1624}{{\ttfamily arXiv:1211.1624 [hep-th]}}.

\bibitem{Baumann:2012bc}
D.~Baumann, S.~Ferraro, D.~Green, and K.~Smith, ``{Stochastic Bias from
  Non-Gaussian Initial Conditions},''
  \href{http://dx.doi.org/10.1088/1475-7516/2013/05/001}{{\em JCAP} {\bfseries
  1305} (2013) 001},
\href{http://arxiv.org/abs/1209.2173}{{\ttfamily arXiv:1209.2173
  [astro-ph.CO]}}.

\bibitem{Assassi:2013gxa}
V.~Assassi, D.~Baumann, D.~Green, and L.~McAllister, ``{Planck-Suppressed
  Operators},'' \href{http://dx.doi.org/10.1088/1475-7516/2014/01/033}{{\em
  JCAP} {\bfseries 1401} (2014) 033},
\href{http://arxiv.org/abs/1304.5226}{{\ttfamily arXiv:1304.5226 [hep-th]}}.

\bibitem{Gong:2013sma}
J.-O. Gong, S.~Pi, and M.~Sasaki, ``{Equilateral Non-Gaussianity from Heavy
  Fields},'' \href{http://dx.doi.org/10.1088/1475-7516/2013/11/043}{{\em JCAP}
  {\bfseries 1311} (2013) 043},
\href{http://arxiv.org/abs/1306.3691}{{\ttfamily arXiv:1306.3691 [hep-th]}}.

\bibitem{Lee:2016vti}
H.~Lee, D.~Baumann, and G.~L. Pimentel, ``{Non-Gaussianity as a Particle
  Detector},'' \href{http://dx.doi.org/10.1007/JHEP12(2016)040}{{\em JHEP}
  {\bfseries 12} (2016) 040},
\href{http://arxiv.org/abs/1607.03735}{{\ttfamily arXiv:1607.03735 [hep-th]}}.

\bibitem{Kehagias:2017cym}
A.~Kehagias and A.~Riotto, ``{On the Inflationary Perturbations of Massive
  Higher-Spin Fields},''
  \href{http://dx.doi.org/10.1088/1475-7516/2017/07/046}{{\em JCAP} {\bfseries
  1707} no.~07, (2017) 046},
\href{http://arxiv.org/abs/1705.05834}{{\ttfamily arXiv:1705.05834 [hep-th]}}.

\bibitem{Kumar:2017ecc}
S.~Kumar and R.~Sundrum, ``{Heavy-Lifting of Gauge Theories by Cosmic
  Inflation},'' \href{http://dx.doi.org/10.1007/JHEP05(2018)011}{{\em JHEP}
  {\bfseries 05} (2018) 011},
\href{http://arxiv.org/abs/1711.03988}{{\ttfamily arXiv:1711.03988 [hep-ph]}}.

\bibitem{An:2017hlx}
H.~An, M.~McAneny, A.~Ridgway, and M.~Wise, ``{Quasi-Single-Field Inflation in
  the Nonperturbative Regime},''
  \href{http://dx.doi.org/10.1007/JHEP06(2018)105}{{\em JHEP} {\bfseries 06}
  (2018) 105},
\href{http://arxiv.org/abs/1706.09971}{{\ttfamily arXiv:1706.09971 [hep-ph]}}.

\bibitem{An:2017rwo}
H.~An, M.~McAneny, A.~Ridgway, and M.~Wise, ``{Non-Gaussian Enhancements of
  Galactic Halo Correlations in Quasi-Single-Field Inflation},''
  \href{http://dx.doi.org/10.1103/PhysRevD.97.123528}{{\em Phys. Rev.}
  {\bfseries D97} no.~12, (2018) 123528},
\href{http://arxiv.org/abs/1711.02667}{{\ttfamily arXiv:1711.02667 [hep-ph]}}.

\bibitem{Baumann:2017jvh}
D.~Baumann, G.~Goon, H.~Lee, and G.~L. Pimentel, ``{Partially Massless Fields
  During Inflation},'' \href{http://dx.doi.org/10.1007/JHEP04(2018)140}{{\em
  JHEP} {\bfseries 04} (2018) 140},
\href{http://arxiv.org/abs/1712.06624}{{\ttfamily arXiv:1712.06624 [hep-th]}}.

\bibitem{Goon:2018fyu}
G.~Goon, K.~Hinterbichler, A.~Joyce, and M.~Trodden, ``{Shapes of Gravity:
  Tensor Non-Gaussianity and Massive Spin-2 Fields},''
\href{http://arxiv.org/abs/1812.07571}{{\ttfamily arXiv:1812.07571 [hep-th]}}.

\bibitem{Kumar:2018jxz}
S.~Kumar and R.~Sundrum, ``{Seeing Higher-Dimensional Grand Unification In
  Primordial Non-Gaussianities},''
  \href{http://dx.doi.org/10.1007/JHEP04(2019)120}{{\em JHEP} {\bfseries 04}
  (2019) 120},
\href{http://arxiv.org/abs/1811.11200}{{\ttfamily arXiv:1811.11200 [hep-ph]}}.

\bibitem{Liu:2019fag}
T.~Liu, X.~Tong, Y.~Wang, and Z.-Z. Xianyu, ``{Probing P and CP Violations on
  the Cosmological Collider},''
\href{http://arxiv.org/abs/1909.01819}{{\ttfamily arXiv:1909.01819 [hep-ph]}}.

\bibitem{Kumar:2019ebj}
S.~Kumar and R.~Sundrum, ``{Cosmological Collider Physics and the Curvaton},''
\href{http://arxiv.org/abs/1908.11378}{{\ttfamily arXiv:1908.11378 [hep-ph]}}.

\bibitem{Alexander:2019vtb}
S.~Alexander, S.~J. Gates, L.~Jenks, K.~Koutrolikos, and E.~McDonough,
  ``{Higher-Spin Supersymmetry at the Cosmological Collider: Sculpting SUSY
  Rilles in the CMB},''
\href{http://arxiv.org/abs/1907.05829}{{\ttfamily arXiv:1907.05829 [hep-th]}}.

\bibitem{Kim:2019wjo}
S.~Kim, T.~Noumi, K.~Takeuchi, and S.~Zhou, ``{Heavy Spinning Particles from
  Signs of Primordial Non-Gaussianities: Beyond the Positivity Bounds},''
\href{http://arxiv.org/abs/1906.11840}{{\ttfamily arXiv:1906.11840 [hep-th]}}.

\bibitem{CosmoBoot2}
D.~Baumann, C.~Duaso~Pueyo, A.~Joyce, H.~Lee, and G.~L. Pimentel, {\em {to
  appear.}}

\bibitem{Gillioz:2016jnn}
M.~Gillioz, X.~Lu, and M.~Luty, ``{Scale Anomalies, States, and Rates in
  Conformal Field Theory},''
  \href{http://dx.doi.org/10.1007/JHEP04(2017)171}{{\em JHEP} {\bfseries 04}
  (2017) 171},
\href{http://arxiv.org/abs/1612.07800}{{\ttfamily arXiv:1612.07800 [hep-th]}}.

\bibitem{Gillioz:2018kwh}
M.~Gillioz, X.~Lu, and M.~Luty, ``{Graviton Scattering and a Sum Rule for the c
  Anomaly in 4D CFT},'' \href{http://dx.doi.org/10.1007/JHEP09(2018)025}{{\em
  JHEP} {\bfseries 09} (2018) 025},
\href{http://arxiv.org/abs/1801.05807}{{\ttfamily arXiv:1801.05807 [hep-th]}}.

\bibitem{Gillioz:2018mto}
M.~Gillioz, ``{Momentum Space Conformal Blocks on the Light Cone},''
  \href{http://dx.doi.org/10.1007/JHEP10(2018)125}{{\em JHEP} {\bfseries 10}
  (2018) 125},
\href{http://arxiv.org/abs/1807.07003}{{\ttfamily arXiv:1807.07003 [hep-th]}}.

\bibitem{Bautista:2019qxj}
T.~Bautista and H.~Godazgar, ``{Lorentzian CFT Three-Point Functions in
  Momentum Space},''
\href{http://arxiv.org/abs/1908.04733}{{\ttfamily arXiv:1908.04733 [hep-th]}}.

\bibitem{Gillioz:2019lgs}
M.~Gillioz, ``{Conformal Three-Point Functions and the Lorentzian OPE in
  Momentum Space},''
\href{http://arxiv.org/abs/1909.00878}{{\ttfamily arXiv:1909.00878 [hep-th]}}.

\bibitem{Costa:2011dw}
M.~Costa, J.~Penedones, D.~Poland, and S.~Rychkov, ``{Spinning Conformal
  Blocks},'' \href{http://dx.doi.org/10.1007/JHEP11(2011)154}{{\em JHEP}
  {\bfseries 11} (2011) 154},
\href{http://arxiv.org/abs/1109.6321}{{\ttfamily arXiv:1109.6321 [hep-th]}}.

\bibitem{Karateev:2017jgd}
D.~Karateev, P.~Kravchuk, and D.~Simmons-Duffin, ``{Weight-Shifting Operators
  and Conformal Blocks},''
  \href{http://dx.doi.org/10.1007/JHEP02(2018)081}{{\em JHEP} {\bfseries 02}
  (2018) 081},
\href{http://arxiv.org/abs/1706.07813}{{\ttfamily arXiv:1706.07813 [hep-th]}}.

\bibitem{Costa:2018mcg}
M.~Costa and T.~Hansen, ``{AdS Weight-Shifting Operators},''
  \href{http://dx.doi.org/10.1007/JHEP09(2018)040}{{\em JHEP} {\bfseries 09}
  (2018) 040},
\href{http://arxiv.org/abs/1805.01492}{{\ttfamily arXiv:1805.01492 [hep-th]}}.

\bibitem{Dirac:1936fq}
P.~A.~M. Dirac, ``{Wave Equations in Conformal Space},''
\href{http://dx.doi.org/10.2307/1968455}{{\em Annals Math.} {\bfseries 37}
  (1936) 429--442}.

\bibitem{Weinberg:2010fx}
S.~Weinberg, ``{Six-dimensional Methods for Four-dimensional Conformal Field
  Theories},'' \href{http://dx.doi.org/10.1103/PhysRevD.82.045031}{{\em Phys.
  Rev.} {\bfseries D82} (2010) 045031},
\href{http://arxiv.org/abs/1006.3480}{{\ttfamily arXiv:1006.3480 [hep-th]}}.

\bibitem{Costa:2011mg}
M.~Costa, J.~Penedones, D.~Poland, and S.~Rychkov, ``{Spinning Conformal
  Correlators},'' \href{http://dx.doi.org/10.1007/JHEP11(2011)071}{{\em JHEP}
  {\bfseries 11} (2011) 071},
\href{http://arxiv.org/abs/1107.3554}{{\ttfamily arXiv:1107.3554 [hep-th]}}.

\bibitem{Rychkov:2016iqz}
S.~Rychkov, ``{EPFL Lectures on Conformal Field Theory in $D\ge 3$
  Dimensions},''
\href{http://arxiv.org/abs/1601.05000}{{\ttfamily arXiv:1601.05000 [hep-th]}}.

\bibitem{Polyakov:1970xd}
A.~Polyakov, ``{Conformal Symmetry of Critical Fluctuations},'' {\em JETP
  Lett.} {\bfseries 12} (1970) 381--383.
[Pisma Zh. Eksp. Teor. Fiz.12,538(1970)].

\bibitem{Kravchuk:2016qvl}
P.~Kravchuk and D.~Simmons-Duffin, ``{Counting Conformal Correlators},''
  \href{http://dx.doi.org/10.1007/JHEP02(2018)096}{{\em JHEP} {\bfseries 02}
  (2018) 096},
\href{http://arxiv.org/abs/1612.08987}{{\ttfamily arXiv:1612.08987 [hep-th]}}.

\bibitem{Anninos:2017eib}
D.~Anninos, F.~Denef, R.~Monten, and Z.~Sun, ``{Higher-Spin de Sitter Hilbert
  Space},''
\href{http://arxiv.org/abs/1711.10037}{{\ttfamily arXiv:1711.10037 [hep-th]}}.

\bibitem{Creminelli:2003iq}
P.~Creminelli, ``{On Non-Gaussianities in Single-Field Inflation},''
  \href{http://dx.doi.org/10.1088/1475-7516/2003/10/003}{{\em JCAP} {\bfseries
  0310} (2003) 003},
\href{http://arxiv.org/abs/astro-ph/0306122}{{\ttfamily arXiv:astro-ph/0306122
  [astro-ph]}}.

\bibitem{Deser:1983mm}
S.~Deser and R.~Nepomechie, ``{Gauge Invariance Versus Masslessness in De
  Sitter Space},''
\href{http://dx.doi.org/10.1016/0003-4916(84)90156-8}{{\em Annals Phys.}
  {\bfseries 154} (1984) 396}.

\bibitem{Brink:2000ag}
L.~Brink, R.~Metsaev, and M.~Vasiliev, ``{How Massless are Massless Fields in
  AdS(d)},'' \href{http://dx.doi.org/10.1016/S0550-3213(00)00402-8}{{\em Nucl.
  Phys.} {\bfseries B586} (2000) 183--205},
\href{http://arxiv.org/abs/hep-th/0005136}{{\ttfamily arXiv:hep-th/0005136
  [hep-th]}}.

\bibitem{Deser:2001us}
S.~Deser and A.~Waldron, ``{Partial Masslessness of Higher Spins in (A)dS},''
  \href{http://dx.doi.org/10.1016/S0550-3213(01)00212-7}{{\em Nucl. Phys.}
  {\bfseries B607} (2001) 577--604},
\href{http://arxiv.org/abs/hep-th/0103198}{{\ttfamily arXiv:hep-th/0103198
  [hep-th]}}.

\bibitem{Deser:2003gw}
S.~Deser and A.~Waldron, ``{Arbitrary Spin Representations in de Sitter from
  dS/CFT with Applications to dS Supergravity},''
\href{http://dx.doi.org/10.1016/S0550-3213(03)00348-1}{{\em Nucl. Phys.}
  {\bfseries B662} (2003) 379--392}.

\bibitem{Deser:2013uy}
S.~Deser, M.~Sandora, and A.~Waldron, ``{Nonlinear Partially Massless from
  Massive Gravity?},'' \href{http://dx.doi.org/10.1103/PhysRevD.87.101501}{{\em
  Phys. Rev.} {\bfseries D87} no.~10, (2013) 101501},
\href{http://arxiv.org/abs/1301.5621}{{\ttfamily arXiv:1301.5621 [hep-th]}}.

\bibitem{deRham:2013wv}
C.~de~Rham, K.~Hinterbichler, R.~Rosen, and A.~Tolley, ``{Evidence for and
  Obstructions to Nonlinear Partially Massless Gravity},''
  \href{http://dx.doi.org/10.1103/PhysRevD.88.024003}{{\em Phys. Rev.}
  {\bfseries D88} no.~2, (2013) 024003},
\href{http://arxiv.org/abs/1302.0025}{{\ttfamily arXiv:1302.0025 [hep-th]}}.

\bibitem{Maldacena:2002vr}
J.~Maldacena, ``{Non-Gaussian Features of Primordial Fluctuations in
  Single-Field Inflationary Models},''
  \href{http://dx.doi.org/10.1088/1126-6708/2003/05/013}{{\em JHEP} {\bfseries
  05} (2003) 013},
\href{http://arxiv.org/abs/astro-ph/0210603}{{\ttfamily arXiv:astro-ph/0210603
  [astro-ph]}}.

\bibitem{Dolan:2001ih}
L.~Dolan, C.~R. Nappi, and E.~Witten, ``{Conformal operators for partially
  massless states},''
  \href{http://dx.doi.org/10.1088/1126-6708/2001/10/016}{{\em JHEP} {\bfseries
  10} (2001) 016},
\href{http://arxiv.org/abs/hep-th/0109096}{{\ttfamily arXiv:hep-th/0109096
  [hep-th]}}.

\bibitem{Brust:2016gjy}
C.~Brust and K.~Hinterbichler, ``{Free $\Box^{k}$ Scalar Conformal Field
  Theory},'' \href{http://dx.doi.org/10.1007/JHEP02(2017)066}{{\em JHEP}
  {\bfseries 02} (2017) 066},
\href{http://arxiv.org/abs/1607.07439}{{\ttfamily arXiv:1607.07439 [hep-th]}}.

\bibitem{Brust:2016zns}
C.~Brust and K.~Hinterbichler, ``{Partially Massless Higher-Spin Theory},''
  \href{http://dx.doi.org/10.1007/JHEP02(2017)086}{{\em JHEP} {\bfseries 02}
  (2017) 086},
\href{http://arxiv.org/abs/1610.08510}{{\ttfamily arXiv:1610.08510 [hep-th]}}.

\bibitem{Joung:2019wwf}
E.~Joung, K.~Mkrtchyan, and G.~Poghosyan, ``{Looking for Partially Massless
  Gravity},'' \href{http://dx.doi.org/10.1007/JHEP07(2019)116}{{\em JHEP}
  {\bfseries 07} (2019) 116},
\href{http://arxiv.org/abs/1904.05915}{{\ttfamily arXiv:1904.05915 [hep-th]}}.

\bibitem{Boulanger:2019zic}
N.~Boulanger, C.~Deffayet, S.~Garcia-Saenz, and L.~Traina, ``{A Theory for
  Multiple Partially Massless Spin-2 Fields},''
\href{http://arxiv.org/abs/1906.03868}{{\ttfamily arXiv:1906.03868 [hep-th]}}.

\bibitem{Benincasa:2007xk}
P.~Benincasa and F.~Cachazo, ``{Consistency Conditions on the S-Matrix of
  Massless Particles},''
\href{http://arxiv.org/abs/0705.4305}{{\ttfamily arXiv:0705.4305 [hep-th]}}.

\bibitem{McGady:2013sga}
D.~McGady and L.~Rodina, ``{Higher-Spin Massless S-Matrices in Four
  Dimensions},'' \href{http://dx.doi.org/10.1103/PhysRevD.90.084048}{{\em Phys.
  Rev.} {\bfseries D90} no.~8, (2014) 084048},
\href{http://arxiv.org/abs/1311.2938}{{\ttfamily arXiv:1311.2938 [hep-th]}}.

\bibitem{Simmons-Duffin:2016gjk}
D.~Simmons-Duffin, ``{TASI Lectures on The Conformal Bootstrap},''
\href{http://arxiv.org/abs/1602.07982}{{\ttfamily arXiv:1602.07982 [hep-th]}}.

\bibitem{Erramilli:2019njx}
R.~Erramilli, L.~Iliesiu, and P.~Kravchuk, ``{Recursion Relation for General 3d
  Blocks},''
\href{http://arxiv.org/abs/1907.11247}{{\ttfamily arXiv:1907.11247 [hep-th]}}.

\bibitem{Penedones:2015aga}
J.~Penedones, E.~Trevisani, and M.~Yamazaki, ``{Recursion Relations for
  Conformal Blocks},'' \href{http://dx.doi.org/10.1007/JHEP09(2016)070}{{\em
  JHEP} {\bfseries 09} (2016) 070},
\href{http://arxiv.org/abs/1509.00428}{{\ttfamily arXiv:1509.00428 [hep-th]}}.

\bibitem{Dobrev:1977qv}
V.~Dobrev, G.~Mack, V.~Petkova, S.~Petrova, and I.~Todorov, ``{Harmonic
  Analysis on the $n$-Dimensional Lorentz Group and Its Application to
  Conformal Quantum Field Theory},''
\href{http://dx.doi.org/10.1007/BFb0009678}{{\em Lect. Notes Phys.} {\bfseries
  63} (1977) 1--280}.

\bibitem{MathWorld}
{\em Wolfram MathWorld}.
\newblock \url{http://mathworld.wolfram.com/}.

\end{thebibliography}\endgroup

\end{document}